\documentclass{emulateapj} 
\usepackage{lscape} 
\usepackage{apjfonts}
\usepackage{latexsym}
\usepackage{amsmath} 
\usepackage{subfigure}

\shorttitle{The Star Formation History of M32}
\shortauthors{Monachesi et al.}
\begin{document}\title{The Star Formation History of M32
  \footnotemark[1]}  \footnotetext[1]{Based on observations  made with
  the NASA/ESA Hubble Space Telescope, obtained at the Space Telescope
  Science Institute, which is operated by the Association of
  Universities for Research in Astronomy, Inc., under NASA contract
  NAS 5-26555. These observations are associated with GO proposal
  10572.}

\author{Antonela Monachesi} \affil{Kapteyn Astronomical Institute,
  P.O. Box 800, 9700 AV Groningen, The Netherlands} \affil{Department
  of Astronomy, University of Michigan, 830 Dennison Building, 500
  Church Street, Ann Arbor, MI 48109, USA} \email{antonela@umich.edu}
\author{Scott C. Trager} \affil{Kapteyn Astronomical Institute,
  P.O. Box 800, 9700 AV Groningen, The Netherlands} \author{Tod
  R. Lauer} \affil{National Optical Astronomy
  Observatory\footnotemark[2], P.O.~Box 26732, Tucson, AZ, 85726, USA}
\footnotetext[2]{The National Optical Astronomy Observatory is
  operated by AURA, Inc., under cooperative agreement with the
  National Science Foundation.}  \author{Sebasti\'an
  L. Hidalgo}\affil{Instituto de Astrof\'isica de Canarias, Via
  L\'actea s/n E38200-La Laguna, Tenerife, Spain} \author{Wendy
  Freedman, Alan Dressler} \affil{The Observatories of the Carnegie
  Institution of Washington, 813 Santa Barbara Street, Pasadena, CA,
  91101, USA} \author{Carl Grillmair} \affil{Spitzer Science Center,
  1200 E. California Blvd., Pasadena, CA 91125, USA} \and
\author{Kenneth J. Mighell} \affil{National Optical Astronomy
  Observatory\footnotemark[2], P.O.~Box 26732, Tucson, AZ, 85726, USA}

\begin{abstract}
  We use deep HST ACS/HRC observations of a field within M32 (F1) and
  an M31 background field (F2) to determine the star formation history
  (SFH) of M32 from its resolved stellar population.  We find that
  2--5 Gyr old stars contribute $\sim 40\% \pm 17\%$ of M32's mass,
  while $\sim 55\% \pm 21\%$ of M32's mass comes from stars older than
  5 Gyr. The mass-weighted mean age and metallicity of M32 at F1 are
  $\langle\mathrm{Age}\rangle=6.8\pm1.5\,\mathrm{Gyr}$ and
  $\langle\mathrm{[M/H]}\rangle=-0.01\pm0.08\,\mathrm{dex}.$ The SFH
  additionally indicates the presence of young ($<2$ Gyr old),
  metal-poor ($\mathrm{[M/H]} \sim -0.7$) stars, suggesting that
  blue straggler stars (BSS) contribute $\sim2$\% of the mass at F1;
  the remaining $\sim3$\% of the mass is in young metal-rich
  stars. Line-strength indices computed from the SFH imply a
  light-weighted mean age and metallicity of 4.9 Gyr and
  $\mathrm{[M/H]=-0.12}$ dex, and single-stellar-population-equivalent
  parameters of $2.9 \pm 0.2$ Gyr and $\mathrm{[M/H]=0.02} \pm
    0.01$ dex at F1 ($\sim 2.7\,r_e$). This contradicts spectroscopic
  studies that show a steep age gradient from M32's center to
  $1\,r_e$.

  The inferred SFH of the M31 background field F2 reveals that the
  majority of its stars are old, with $\sim 95\%$ of its mass already
  acquired 5--14 Gyr ago.  It is composed of two dominant populations;
  $\sim 30 \% \pm 7.5\%$ of its mass is in a 5--8 Gyr old population,
  and $\sim 65\% \pm 9\%$ of the mass is in a 8--14 Gyr old
  population. The mass-weighted mean age and metallicity of F2 are
  $\langle\mathrm{Age}\rangle=9.2\pm1.2\,\mathrm{Gyr}$ and
  $\langle\mathrm{[M/H]}\rangle=-0.10\pm0.10\,\mathrm{dex}$,
  respectively. Our results suggest that the inner disk and spheroid
  populations of M31 are indistinguishable from those of the outer
  disk and spheroid. Assuming that the mean age of M31's disk at F2
  ($\sim 1$ disk scale length) to be $\sim5$--9 Gyr, our results agree
  with an inside-out disk formation scenario for M31's disk.
 \end{abstract}


\section{A Brief History of  Star Formation in M32}

M32 (NGC 221) is a compact, low-luminosity elliptical galaxy,
satellite of our neighbor M31. Due to its proximity, we can study M32
with great detail not only from its integrated light but also from its
individual, resolved stars in a way that is impossible for most of the
elliptical galaxies, given their greater distances and high
densities. Thus, M32 is a very important galaxy to understand the
formation and evolution of low-luminosity spheroidal star
systems. However, M32's SFH, and therefore its origin, is still
controversial. The different scenarios proposed to explain its origins
extend from a true elliptical galaxy at the lower extreme of the mass
sequence \citep[e.g.,][] {Faber73, Nieto_prugniel87, Kormendy_etal09}
to a threshed spiral galaxy \citep[e.g.,][]{Bekki_etal01,
  Chilingarian_etal09}.

\begin{figure} \centering
  \includegraphics[width=80mm,clip]{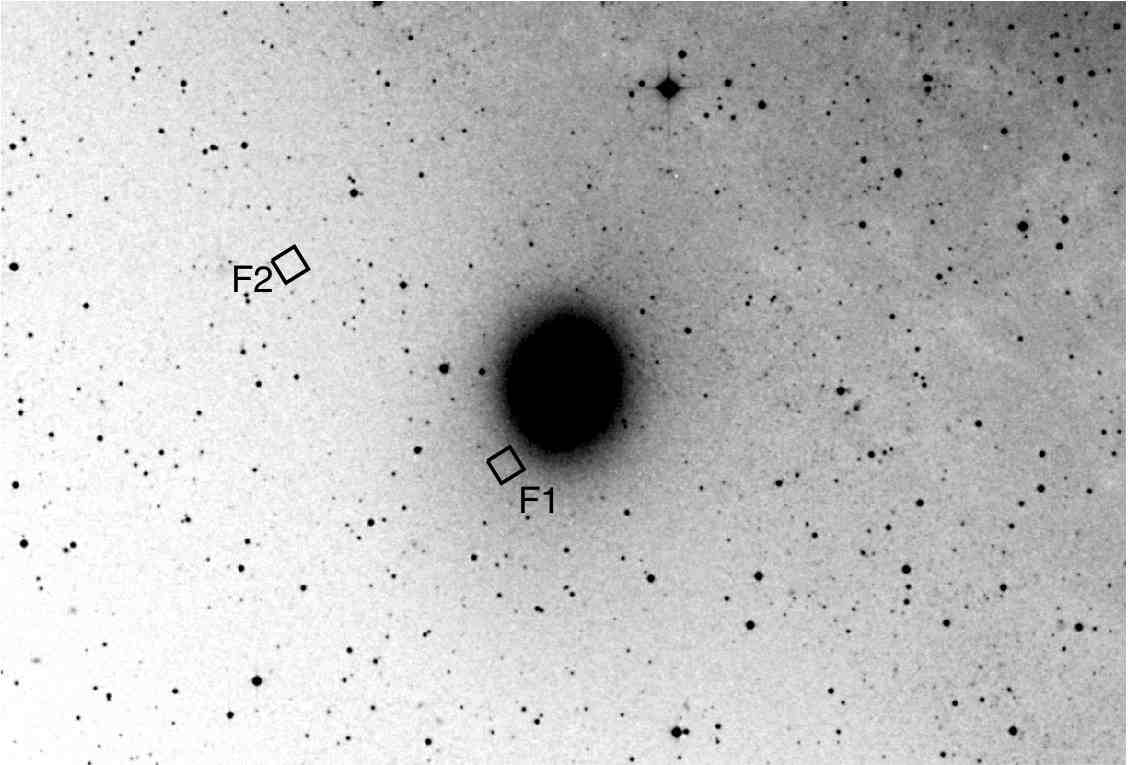}
  \caption{Location of our two HST ACS/HRC pointings: M32 (F1) field
    and M31 background (F2) field, indicated as small black boxes.
    Each field covers a region of $26 \times 29 \,\mathrm{arcsec^{2}}$
    on the sky. The field F1 is located at 110\arcsec\ from the
    nucleus of M32 and represents the best compromise between
    minimizing image crowding and contamination from M31.  The F2
    field is at the same isophotal level in M31 as F1. At the distance
    of M32, each field occupies an area of 11752 pc$^2$.  Thirty-two
    exposures in each of the $F435W$ ($B$) and $F555W$ ($V$) filters
    were taken for each field.  North is up and East is to the left.}
 \label{fig:location}
 \end{figure}

The only way to accurately determine the age, and thus the SFH, of a
galaxy is by directly observing its oldest main-sequence turnoff
(MSTO). With this goal in mind, we were awarded 64 orbits with HST
ACS/HRC to observe two fields near M32, F1 and F2
(Fig.~\ref{fig:location}), in order to detect the oldest MSTOs of this
galaxy.

 \subsection{The deepest HST CMD of M32}

 In \citet[][hereafter Paper I]{Monachesi_etal11} we introduced our
 observations and presented the deepest HST color-magnitude diagram
 (CMD) of M32 yet obtained, reaching more than 2 mag fainter than the
 RC and fully resolving the RGB and the AGB. Paper I significantly
 improved our knowledge on the stellar populations of M32. We have
 found that M32 is dominated by intermediate-age (2--8 Gyr old) and
 old (8--10 Gyr old), metal-rich ($[\mathrm{Fe/H}] \sim -0.2$) stars
 and it contains some ancient ($>10$ Gyr), metal-poor stars
 ($[\mathrm{Fe/H}] \sim -1.6$) as well as possible young populations
 (0.5 -- 2 Gyr old stars).

\begin{figure*}
\centering
 \includegraphics[width=130mm, clip]{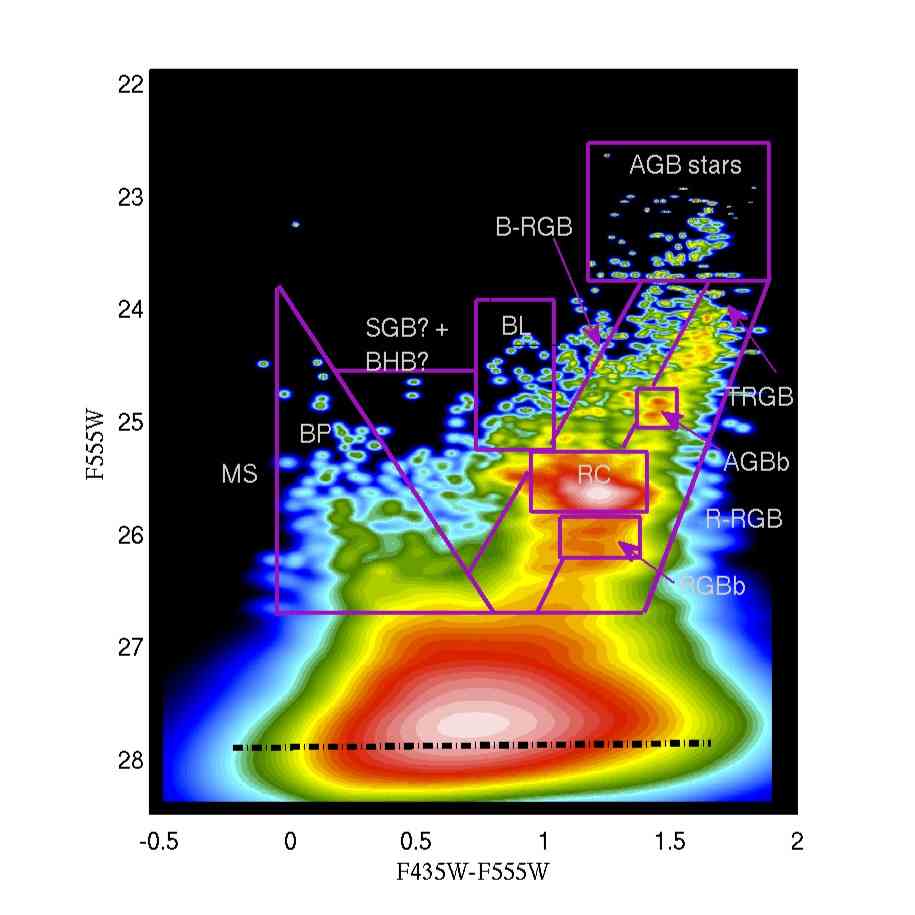}
 \caption{Error-based Hess diagram for M32, corrected for
   contamination from M31 background stars. The boxes indicate various
   features that represent different stellar populations. MS: Main
   Sequence; BP: Blue Plume; SGB: Subgiant branch; BHB: Blue
   Horizontal Branch; BL: Blue Loop; RC: Red Clump; RGBb: Red Giant
   Branch bump; R-RGB: Red-Red Giant Branch; B-RGB: Blue-Red Giant
   Branch; TRGB: Tip of the Red Giant Branch; AGB: Asymptotic Giant
   Branch; and AGBb: Asymptotic Giant Branch bump. The dotted-dashed
   line indicates the 50\% completeness level of our data. Magnitudes
   are calibrated onto the VEGAmag system. (This is Fig.~12 from Paper
   I; we refer the reader to that paper for more details.)}
\label{fig:hess_boxes}
\end{figure*}

 These conclusions were provided by our qualitative analysis of the
 CMD of M32, which shows a red clump (RC), a red giant branch (RGB), a
 RGB bump (RGBb), an AGB bump (AGBb), and a blue plume (BP). Figure 12
 of Paper I, reproduced here as Figure~\ref{fig:hess_boxes}, shows a
 Hess representation of the CMD of M32 decontaminated of M31 stars,
 where the different evolutionary features are highlighted. We
 summarize here the main findings and conclusions of Paper I.

\begin{itemize}

\item The core-helium burning stars are concentrated in a RC and its
  mean color and magnitude suggest a mean age of 8--10 Gyr for a
  metallicity of $[\mathrm{M/H}]\sim -0.2$ in M32.

\item The first detection of the RGB bump and the AGB bump in M32
  permits a constraint on the mean age and metallicity of the
  population.  This gives a mean metallicity of M32 higher than
  $\mathrm{[M/H]} \sim -0.4$ dex and a mean age between 5 and 10 Gyr.

\item The metallicity distribution of M32 inferred from the CMD has a
  peak at $[\mathrm{M/H}]\sim -0.2 \, \mathrm{dex}$. Overall, the
  metallicity distribution function implies that there are more
  metal-rich stars than metal-poor ones. Metal-poor stars with
  $[\mathrm{M/H}] <-1.2$ contribute very little, \emph{at most} 6\% of
  the total $V$-light or 4.5\% of the total mass, to M32 in F1,
  implying that the enrichment process largely avoided the metal-poor
  stage.

\item Bright AGB stars at $F555W < 24$, i.e.  above the TRGB, confirm
  the presence of an intermediate-age population in M32 (ages of 1--7
  Gyr).

\item The observed blue plume is genuine, not an artifact of crowding,
  and contains stars as young as $\sim0.5$ Gyr.  The detected blue
  loop, with stars having masses of $\sim 2$--$3 \,M_{\odot}$ and ages
  between $\sim 0.3$ and $\sim 1$ Gyr, and the possible presence of a
  bright SGB are different manifestations of the presence of a young
  population.  However, in Paper I we suggest that it is likely that
  this young population belongs to the disk of M31 rather than to
  M32. The fainter portion of the blue plume ($F555W > 26$)
  \emph{does} belong to M32 and indicates the presence of stars with
  ages 1--2 Gyr and/or the first direct evidence of blue straggler
  stars (BSS) in M32.

\item The oldest MSTOs were out of reach, given the severe crowding in
  F1, and there is no significant BHB observed in F1, so an ancient,
  metal-poor population cannot be seen directly in our CMD. We have,
  however, a hint of the presence of such a population from a
  2--$\sigma$ detection of RR Lyrae stars found in F1 and associated
  with M32 using our data \citep{Fiorentino_etal10}.
 
\item In general the CMDs of both fields F1 and F2 show an
  unexpectedly similar morphology. By subtracting the normalized F1
  CMD from the F2 one (see Figure 21 in Paper I), one can detect
  subtle differences. M31 has a younger and more metal-poor population
  than M32, and M32 has a more conspicuous intermediate-age population
  (Fig.~\ref{fig:deconvolvedmcmd}).

\item The CMD of our M31 background field F2 exhibits a wide RGB,
  indicative of a metallicity spread with its peak at
  $[\mathrm{M/H}]\sim -0.4 \, \mathrm{dex}$. The presence of a blue
  plume indicates the presence of stars as young as 0.3 Gyr. Bright
  AGB stars in F2 reveal the presence of an intermediate-age
  population in M31.
\end{itemize}

 \subsection{Completing the picture of M32's SFH}

 The analysis presented in Paper I provided initial constraints on the
 ages and metallicities of the stellar populations of M32 at F1 and
 M31 at F2. That work was based on traditional methods of isochrone
 analysis, and was heavily based on the brighter evolved portions of
 the CMDs, such as the RC, RGB, and bump (RGBb and AGBb) features.

 The approach in this paper is independent. Here we use a
 sophisticated method of CMD analysis and decomposition that digs into
 the fainter, severely-crowded portions of the CMDs near the MSTO and
 sub-giant branches. We recover information from the brighter MSTOs
 present in the CMDs, thus providing quantitative information about
 the younger populations of M32.  In this paper, we derive the
 detailed young and intermediate-age SFH of M32 at $\sim 2\arcmin$
 from its center and of M31 at our background field's location, which
 was not possible from the analysis in Paper I. We find that our field
 in M32 has a substantial population of 2--5 Gyr old stars
 contributing to $\sim 40\% \pm 17\%$ of its mass, an unexpectedly
 large population of young stars at such a large distance from the
 center of an elliptical galaxy.

 The paper is organized as follows. In Section~\ref{sec:photo} we
 briefly describe our observations and
 photometry. Section~\ref{sec:method} describes the method used to
 derive the SFH. We present the results of the SFH analysis obtained
 for F1, F2 and M32 in Section~\ref{sec:results}. In
 Section~\ref{sec:sfhm32} we provide a detailed and complete SFH of
 M32 and discuss its implications on M32's origins, synthesizing a
 complete picture based on both the present and Paper I analyses. In
 Section~\ref{sec:m31} we discuss the SFH of the inner regions of
 M31. Finally, we summarize our results and present our conclusions in
 Section~\ref{sec:conclusions}.

\section{Observations and Photometry}
\label{sec:photo}

The field selection and observational strategy as well as the image
reduction are described in Paper I and we refer the reader to that
paper for details.  Briefly, HST ACS/HRC images of two fields near M32
were observed during Cycle 14 (Program GO-10572, PI: Lauer). The M32
HRC field (F1) was centered on a location $110\arcsec$ south (the
anti-M31 direction) of the M32 nucleus. The background field (F2) was
located $327\arcsec$ from the M32 nucleus, roughly along its minor
axis, at the same isophotal level in M31 as F1. The field locations
are shown in Figure~\ref{fig:location}.

\begin{figure*}
\vspace{-0.001cm}
\includegraphics[width=185mm,clip]{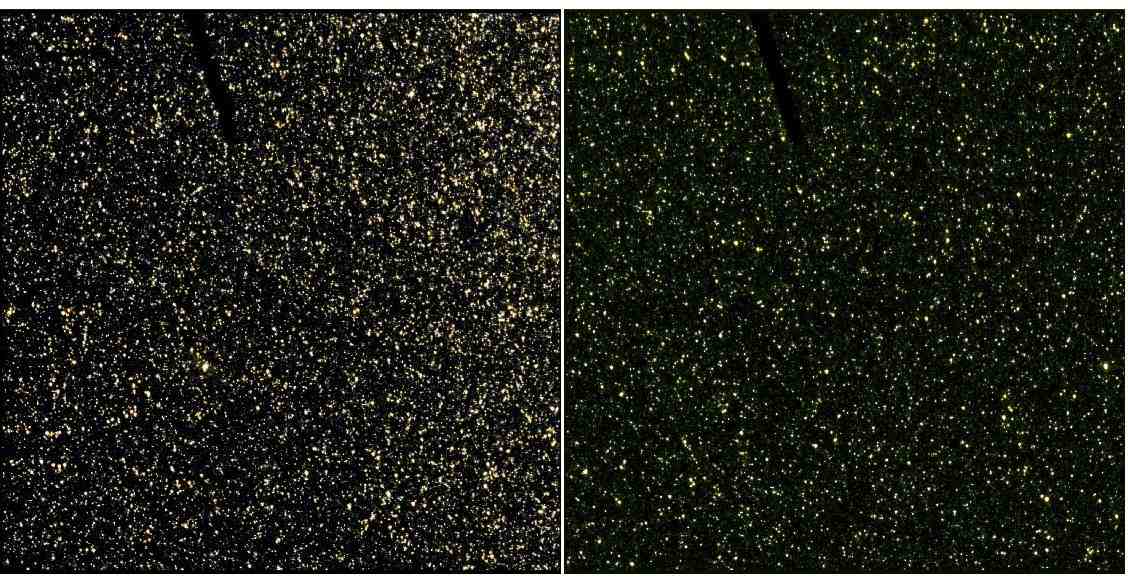}
\includegraphics[width=185mm,clip]{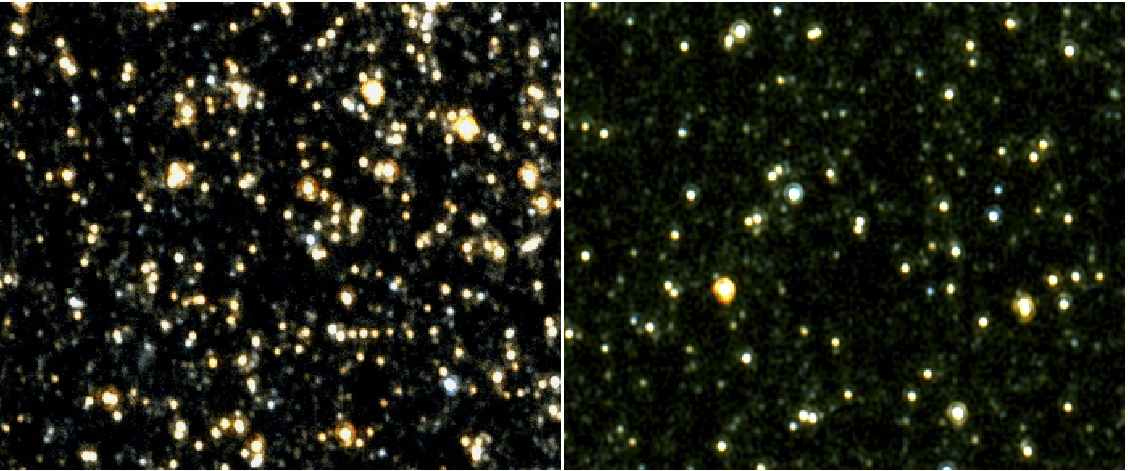}
\caption{Combined color images of the 32 exposures in the F1 (top left
  panel) and F2 (top right panel) fields displayed with the same
  logarithmic stretch.  Each image has a size of 2048$\times$2048
  pixels with a $0\farcs0125$ pixel scale. There is a clear difference
  in stellar density between the images, indicating that crowding is
  more severe in F1 than in F2.  We also note a stellar density
  gradient in the F1 image, becoming higher when approaching the
  center of M32. The long black spot in the top center of each image
  is the occulting finger of the ACS/HRC coronagraph.  The bottom
  panels are zoomed-in images of the centers of the top images for F1
  (left) and F2 (right) fields, where we can better see individual
  stars and the different crowding levels are also evident. Note the
  blue stars in these images. Each zoomed-in image represents an area
  of $\sim15\,\mathrm{arcsec}^2$ on the sky.}
\label{fig:fields}
\end{figure*}

Each field was observed for 16 orbits in each of the $F435W$ ($\sim
B$) and $F555W$ ($\sim V$) filters. All of the images were combined in
an iterative procedure designed to detect and repair cosmic-ray
events, hot pixels, and other defects, with a Nyquist-sampled summed
image as the final product \citep{lauer99a}. Color images of F1 and F2
are shown in the left and right panels of Figure~\ref{fig:fields},
respectively, where the strong crowding in these fields is clearly
visible.  There is however a difference between the stellar density in
F1 and F2: the crowding is more severe in F1 than in F2.  This can
also be seen from the bottom panels of Figure~\ref{fig:fields}, where
zoomed-in images of the top panels are shown.
 
Stellar photometry was performed on deconvolved combined images.  A
detailed description of the deconvolution process is explained in
Paper I. In short, deconvolution was performed on the final images
using the Lucy-Richardson algorithm \citep{Lucy_74, Richardson_72} and
empirically constructed PSFs, one for each image. Stars were
identified in the deconvolved images and their fluxes were
measured. Change transfer efficiency (CTE) and aperture corrections
were applied to the magnitudes, which transform the instrumental
magnitudes into calibrated, apparent magnitudes.

Figure~\ref{fig:deconvolvedmcmd} shows the CMDs derived for F1 (left
panel) and F2 (right panel) from the deconvolved photometry,
calibrated onto the VEGAmag system. They contain 58143 and 27963
stars, respectively, as indicated in Table~\ref{table:photometry}. A
qualitative analysis of these CMDs allowed us to gain some insights
into its stellar populations. This was discussed in detail in Paper I
and we have summarized our conclusions above.

Note the difference between the CMD of F1 and F2 at magnitudes between
$F555W \sim 27$ and 28 (cyan boxes in
Figure~\ref{fig:deconvolvedmcmd}) . The number of stars in this
region, where the brighter MSTOs are located, is larger in F1 than
F2. This suggests that there is a bigger contribution of
intermediate-age stars in F1 than in F2. We can better appreciate this
difference in Figure 21 of Paper I, where we showed a Hess subtraction
of the normalized F1 CMD to the F2 CMD.

\begin{deluxetable}{lcccccc}
  \tabletypesize{\scriptsize}
  \tablecaption{Deconvolved photometry\label{table:photometry}}
  \tablewidth{0pt}  
  \tablehead{\colhead{Field}&
    \colhead{Detections\tablenotemark{a}}&
    \colhead{$R^{F435W}_{\rm{PSF}}$\tablenotemark{b}}&
    \colhead{$R^{F555W}_{\rm{PSF}}$\tablenotemark{b}}&
    \colhead{AC$_{F435W}$\tablenotemark{c}}&
    \colhead{AC$_{F555W}$\tablenotemark{c}}} 
   \startdata 
      F1&58,143&5&5&$-0.25$&$-0.22$ \\
      F2&27,963&6&16&$-0.22$&$-0.10$
      \enddata
      \tablenotetext{a}{Final number of stars detected and used to derive
   CMDs}
\tablenotetext{b}{PSF radius in HRC original pixels}
\tablenotetext{c}{Aperture correction}
\end{deluxetable}

\subsection{Crowding tests}\label{ast}

We performed artificial star tests (ASTs) to assess the completeness
level and quantify the photometric errors of our data. This is a
crucial step for the derivation of the SFH.  The distribution of stars
in the observed CMD is modified from the actual distribution due to
the observational effects, particularly at the fainter magnitudes
where most of the information from the older star formation is
encoded. The ASTs are used to simulate the observational effects in
the synthetic CMDs that are then compared with the observed CMDs in
the analysis described below.

The procedure and results of the ASTs are presented in Paper I and we
refer to that paper for further details; we give a brief description
here in order to provide guidance for later sections of this paper. We
used IAC-STAR \citep{Aparicio_gallart04} to generate $5\times10^5$
artificial stars with realistic colors and magnitudes covering not
only the entire color and magnitude range of the observed stars
  but also $\sim 2$ magnitudes fainter. We injected the artificial
stars into the real images after transforming their magnitudes into
instrumental ACS/HRC fluxes. The number of stars injected per
experiment is 2000, to avoid increasing the already severe crowding of
the real images. We performed 250 ASTs per field/filter combination
for a total of 1000 ASTs.  The images containing real and artificial
stars are photometered exactly in the same way as the original images.
A comparison of the known injected magnitudes and colors of the
artificial stars to those obtained from their photometry allows us to
quantify the photometric errors. The completeness of our data at a
given color and magnitude is calculated as the ratio of
recovered-to-injected artificial stars on that color and magnitude
bin.

\begin{figure*}\centering
 \includegraphics[width=185mm, clip]{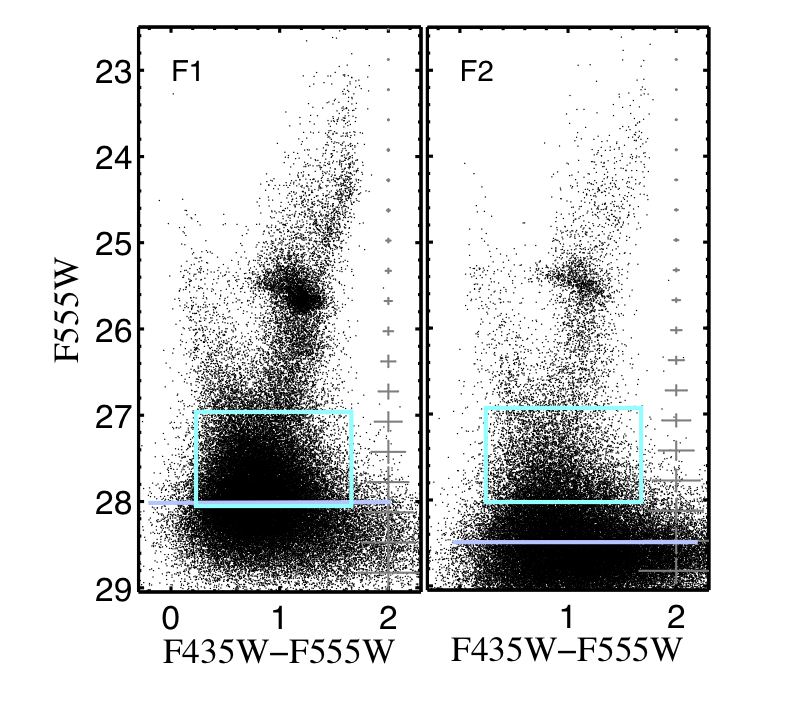}
 \caption{($F435W - F555W$, $F555W$) CMDs of field F1 (left-hand
   panel) and F2 (right-hand panel) obtained using deconvolved images.
   These contain 58143 and 27963 stars respectively, and are
   calibrated onto the VEGAmag HST system. Note the difference between
   the CMDs in the region highlighted with dark blue boxes. The larger
   number of stars in F1 indicate the presence of a more significant
   intermediate-age population in this field compared to F2. The
   region in the box is not an actual ``bundle'' used in the
   derivation of the SFH but a similar one is used to obtain most of
   the information about the SFH of both fields: see
   Section~\ref{sec:method} for more details. The light blue line
   indicates the 50\% completeness level of our data in each field and
   the photometric errors from ASTs refer to a color of
   $(F435W-F555W)=1$.}
\label{fig:deconvolvedmcmd}
\end{figure*}

The results obtained from these ASTs indicate that the limiting
magnitudes of the F1 and F2 CMDs are $F555W \sim 28$ and $\sim 28.5$,
respectively, nearly independent of color. The CMD of F2 is therefore
slightly deeper than that of F1 (cf.\ Figs.\ 8 and 9 of Paper I). The
50\% completeness level as well as the photometric errors derived from
the ASTs for F1 and F2 are indicated in
Figure~\ref{fig:deconvolvedmcmd}.

\section{The IAC Method to resolve the SFH}
\label{sec:method}

To extract the detailed SFH of F1 and F2 we use the well-known method
of fitting synthetic CMDs to the data \citep[see e.g.,][]{Tosi_etal91,
  Bertelli_etal92, Tolstoy_saha96, Aparicio_etal97}.  There are
currently many approaches to derive detailed SFH of galaxies (e.g.,
StarFISH: \citealt{Harris_zaritsky01}, MATCH: \citealt{Dolphin02},
IAC-pop/MinnIAC: \citealt{Aparicio_hidalgo09, Hidalgo_etal11}) as well
as several different stellar libraries (e.g., BaSTI:
\citealt{Pietrinferni_etal04}, Padova/Girardi:
\citealt{Girardi_etal00, Marigo_etal08}) available to compute the
required synthetic CMDs. We use the IAC-pop/MinnIAC method and adopt
the BaSTI and Padova stellar libraries. The IAC-pop code
\citep{Aparicio_hidalgo09} uses a modified $\chi^2$ merit-function
\citep{Mighell99} to compare the observed and synthetic star counts in
different boxes (see below) of the CMDs. A genetic algorithm
\citep{Charbonneau95} is adopted to minimize $\chi^2$. An important
characteristic of the code is that it solves the SFH simultaneously
for age and metallicity distributions. It thus provides the SFH of a
stellar system as a linear combination of simple populations,
i.e. within small ranges of age and metallicity. We refer the reader
to \citet{Aparicio_hidalgo09} and \citet{Hidalgo_etal11} for more
details.

It is important to emphasize that, for the current analysis, we have
mainly used information from the extended MS, MSTO and SGB regions of
the CMDs, as we will see below. We have excluded the RC and most of
the RGB regions, which were the main features analyzed in Paper I and
from which we obtained estimates on the age and metallicity of
M32. This is because the physics governing these phases are more
uncertain than those on the MS and SGB, and differences between
stellar libraries are more severe \citep{Gallart_etal05}. For
instance, the morphology and number of stars occupying the HB/RC
evolutionary phases depend on unknown issues, like mass loss on the
RGB or He-core mass. Small differences in the adopted physics can
significantly alter the number of stars and morphology of these CMD
regions.  The CMD regions that we probe in this paper allow us to
obtain detailed information about the young and intermediate-age
populations of M32, something that was not possible in Paper I, but
conversely, we cannot make a quantitative analysis of the older
populations and so we must rely on the qualitative results of Paper I.

\begin{figure*}
\includegraphics[width=90mm, clip]{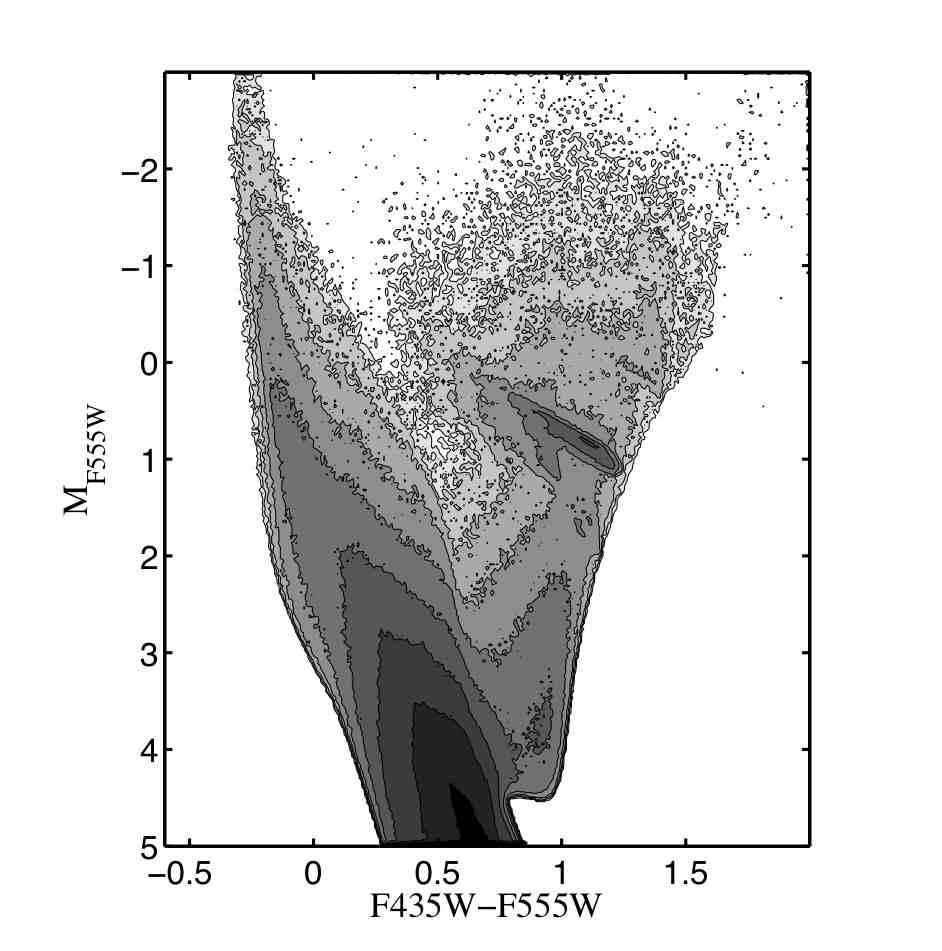}
 \includegraphics[width=90mm, clip]{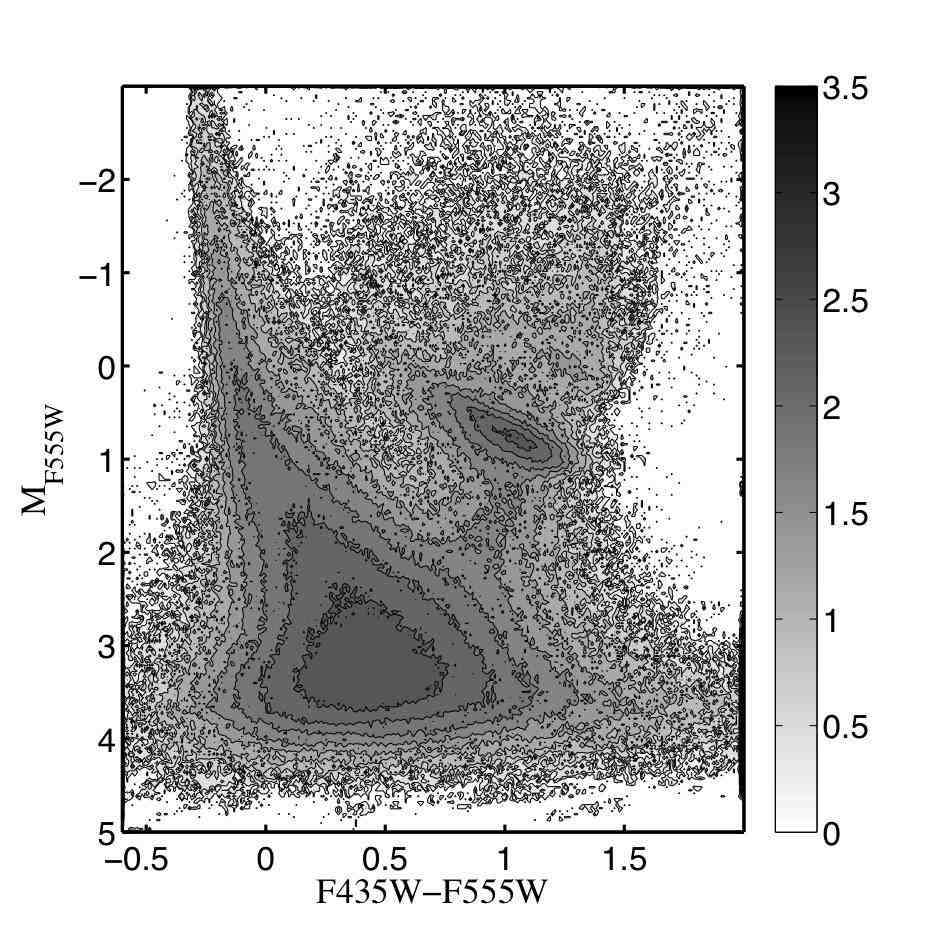}
 \caption{Left-hand panel: Hess representation with a logarithmic
   stretch of the synthetic CMD generated using IAC-STAR for a range
   of age between 0 and 14 Gyr and metallicities uniformly distributed
   at all ages between 0.0001 and 0.04. It contains $5\times10^6$
   stars. Right-hand panel: Hess representation of the model CMD,
   i.e. the synthetic CMD after the observational effects have been
   simulated. It contains $\sim 2\times10^6$ stars. The color bar
   indicates the number of stars per color-magnitude bin in
   logarithmic scale. This model CMD is the one to be compared with
   the observed CMD to derive the SFH.}
\label{fig:scmd}
\end{figure*}

\subsection{Steps carried out to obtain the SFH }

1) \emph {Synthetic CMD}. We first generate a synthetic CMD using
IAC-STAR \citep{Aparicio_gallart04}. The bolometric corrections
applied to both libraries are those of \citet{Origlia_leitherer00}
which transform the theoretical tracks into the ACS/HRC photometric
system. We assume a constant star formation rate (SFR) from 0 to 14
Gyr, and metallicities from $Z = 0.0001$ ($[\mathrm{M/H}] = -2.3$) to
$Z = 0.04$ ($[\mathrm{M/H}] = 0.3$) uniformly distributed at all
ages. Note that there is no assumed age--metallicity relation as
input, and the selected age and $[\mathrm{M/H}]$ ranges are broader
than those expected for the solution. This allows the code to find the
SFH solution with minimum constraints and ensures no lost
information. We adopted a \citet{Kroupa02} initial mass function
(IMF)\footnote{Given that the range of masses we have in our observed
  CMD is rather small and centered around $1\,M_\odot$, where the IMF
  is not especially sensitive to changes, we do not expect the
  effective SFH solution to significantly change if we assume another
  IMF. However, since the large number of lower mass stars are not
  well constrained by the data, different assumptions of the IMF will
  affect the normalization of the SFH, i.e. the total mass of the
  system.} from 0.1 to $100\,M_{\odot}$. The IMF has a slope of 1.3
for stars with masses lower than $0.5\,M_{\odot}$ and 2.3 for stars
with higher masses. We assume a 35\% binary fraction with a relative
mass ratio randomly distributed between 0.5 and 1 (the impact of
different binary fractions on the solution is discussed in the
Appendix). The synthesized CMD, shown in the left panel of
Figure~\ref{fig:scmd}, contains $5\times10^6$ stars and its faintest
magnitude is $\sim$ 2 magnitudes fainter than the 50\% completeness
level of our data. Observational effects (incompleteness and
photometric errors) are simulated using information obtained from the
ASTs described above \citep[see][and references therein for a detailed
  description of this procedure]{Hidalgo_etal11}. The right panel of
Figure~\ref{fig:scmd} shows the synthetic CMD after observational
effects are simulated. We call it a ``model CMD'' following
\citet{Aparicio_etal97}'s notation. The model CMD is the one to be
compared with the observed CMD for the derivation of the SFH of our
fields.
 
2) \emph{Parametrization of the CMDs}.  This is the main input of the
IAC-pop code and was performed using MinnIAC \citep{Hidalgo_etal11}, a
set of routines specially designed for this purpose.  We first define
the ``simple populations'', the age and metallicity bins in which the
model CMD is to be divided. These simple populations represent the
bins in which the SFH is to be determined. The boundaries of the bins
that we used are [0, 0.5, 1, 2, 5, 14]$\times10^9$ years in age and
[0.02, 0.40, 0.80, 1.00, 2.00, 4.00]$\times10^{-2}$ in $Z$,
corresponding to $[\mathrm{M/H}]\approx[-2.0$, $-0.67$, $-0.35$,
  $-0.25$, $0.02$, $0.32$], assuming $Z_{\odot}=0.019$.  These
constitute $5\times5 = 25$ simple populations. The resolution in age
and metallicity was selected, after several experiments on mock
stellar populations, as the optimal choice for our data given the
observational uncertainties\footnote{Experiments on mock stellar
  populations were conducted as follows: we generated synthetic CMDs
  from arbitrary SFHs to which we apply the observational effects of
  our observed CMDs obtained from the ASTs. We then solved for their
  SFHs using the IAC-POP/MinnIAC method as if they were real data,
  adopting different age and metallicity resolutions as first
  reasonable guesses according to the limiting magnitude of our
  data. The SFH solutions were compared with the input SFHs, and when
  these agreed to within $1-\sigma$, we assumed the corresponding age
  and metallicity resolutions were optimal.}. Note that the bin width
in age increases significantly for older populations. This is due to
the limits imposed by the crowding; we cannot extract more detailed
information about the oldest stars.

\begin{figure*}\centering
 \includegraphics[width=80mm, clip]{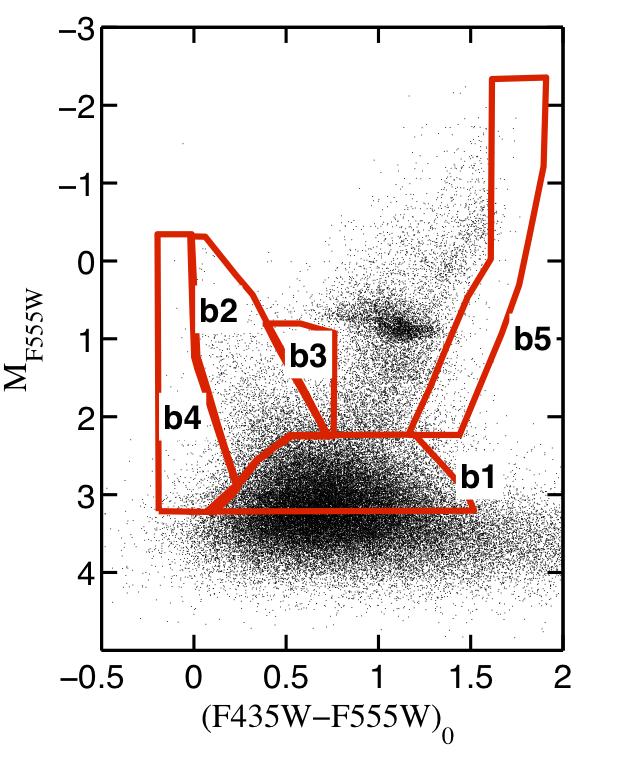}
\includegraphics[width=80mm, clip]{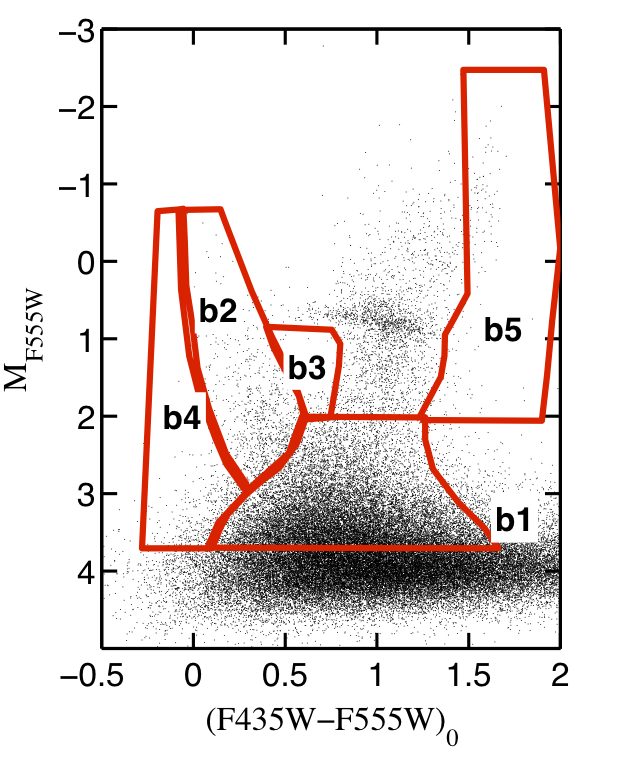}
 \caption{Left-hand panel: CMD of field F1 in absolute magnitudes,
   assuming a distance $\mu_0=24.53$ and $E(B-V) = 0.08$, with the
   location of the bundles superimposed. Right-hand panel: As in the
   left-hand panel but for F2, assuming a distance $\mu_0=24.45$ and
   $E(B-V) = 0.08$. Each bundle in both CMDs is subdivided into boxes
   with sizes that vary from one bundle to another (see
   Table~\ref{table:bundles}). This allows each CMD region used for
   the analysis to have different weights on the extracted SFH. Note
   that no bundles were added below the 50\% completeness level of
   each CMD. Also note that most of the RGB and RC regions are
   likewise excluded of our SFH analysis: uncertainties in the physics
   governing these evolutionary phases are larger than those in the MS
   and SGB region.}
\label{fig:bundles}
\end{figure*}

\begin{figure*}
 \includegraphics[width=180mm]{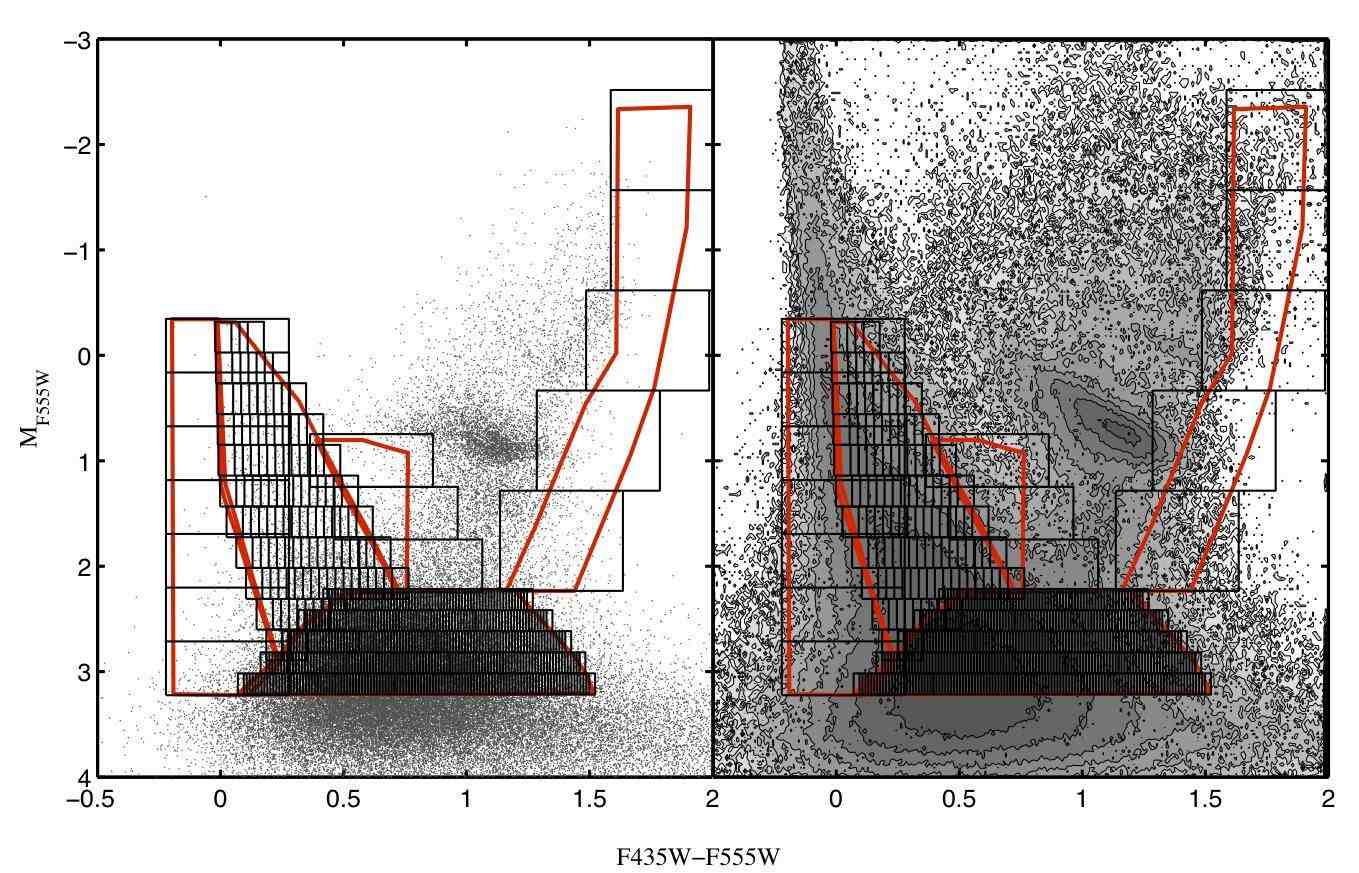}
 \caption{CMD of field F1 (left-hand panel) and model CMD (right-hand
   panel) with the actual bundles and boxes used for the fitting
   superimposed. We show here one sample of boxes in which the stars
   are counted and compared between the observed and model CMDs. Note
   that the boxes inside the bundles are shifted during the dithering
   process, as explained in the text. Also note that only stars inside
   a bundle are considered in the analysis, no matter how big the box
   is.}
\label{fig:boxes}
\end{figure*}

We then define five ``bundles'', macro-regions of the CMDs used for
the fitting. We show the CMDs of F1 and F2 with the selected bundles
superimposed in Figure~\ref{fig:bundles}. The bundles are subdivided
into boxes, whose sizes vary from bundle to bundle. The bundles and
boxes are equally sampled in the observed and model CMDs. Since the
number of stars in each box is the information provided to the IAC-pop
code, the different bundle subdivisions provide the weights that a
given CMD region has on the derived SFH. For instance, CMD regions
well-populated and/or where the input physics is better understood
(e.g., bundle 1) have smaller boxes than CMD regions where either the
number of stars is smaller or the uncertainties in the input physics
significantly impact stellar interior models (e.g., bundle 5). The
properties of the boxes for each bundle are specified in
Table~\ref{table:bundles} and both the observed and model CMDs with
one sample of the boxes superimposed are shown in
Figure~\ref{fig:boxes}. Note that the boxes inside the bundles are
shifted during the dithering process, as explained below, and only
stars inside a bundle are considered in the analysis, no matter how
big the box is. Also note that only stars brighter than the 50\%
completeness level were used to extract the SFH
(cf.~Fig.~\ref{fig:bundles}). Below this region, most of the
information is lost and results obtained from lower-completeness
regions are unreliable (see also the right panel of
Fig.~\ref{fig:scmd}). Also, as mentioned above, we did not use most of
the RGB and RC. Adding bundles in those regions significantly
increased $\chi^2$ from $\sim 2$ to $\sim 5$. Bundles 4 (bluer than
the observed MS) and 5 (redder than the observed RGB) were adopted to
mainly constrain the lowest and highest metallicity, respectively, in
the observed CMD. There are nearly no observed stars in these regions
(see Fig.~\ref{fig:bundles}), whereas there are stars in the model CMD
(see right panel of Fig.~\ref{fig:scmd}). The fact that stars of
certain ages and metallicities appear in the model CMD but not in the
observed one, indicates that those simple populations are not present
in M32.

3) \emph{Solution}. For a given parametrization, i.e., box sizes and
simple population boundaries, MinnIAC counts the stars in each of the
boxes for both the observed and model CMDs.  The number of stars in
each box is the input information to run IAC-pop. IAC-pop compares the
observed and model star counts in each box using a modified $\chi^2$
merit-function \citep{Mighell99}, calculating which combination of
simple populations best reproduces the observed CMD. A SFH solution is
obtained as a linear combination of the simple populations. Thus,
IAC-pop solves the SFH considering the age and metallicity as
independent variables.

\begin{deluxetable}{lcccc}
  \tabletypesize{\scriptsize}
  \tablecaption{CMD regions used for the fitting\label{table:bundles}} 
  \tablewidth{0pt}
  \tablehead{\colhead{Bundle}&
    \colhead{\# of boxes}&
    \colhead{Size of boxes (color, mag)} &
    \colhead{CMD region sampled}} 
  \startdata
  1 & 500 &(0.01, 0.20) &lower MS\\
  2 & 150 &(0.03, 0.30) &upper MS\\
  3 & ~~3 &(0.50, 0.40) &SGB\\
  4 & ~~7 &(0.50, 0.50) &left of the MS\\
  5 & ~~5 &(0.50, 0.90) &Right of the RGB
  \enddata
\end{deluxetable}

4) \emph{Uncertainties and stability of the solution}.  To minimize
biases in the solution due to the sampling of the CMD, MinnIAC allows
slight changes in the input parameters.  The simple populations
(i.e. the age and metallicity bins) are shifted three times a 30\% of
their corresponding bin sizes for each of the following four different
configurations: (1) shifting the age bin toward increasing age at
fixed metallicity, (2) shifting the metallicity bin toward increasing
metallicity at fixed age, (3) shifting both age and metallicity bins
toward increasing values, and (4) shifting both age and metallicity
bins toward decreasing age and increasing metallicity. Furthermore,
for each of these 12 shifts, the boxes are shifted a fraction of their
[color, magnitude] sizes three times: [80\%, 20\%], [20\%, 80\%] and
[20\%, 0\%], respectively. These 36 sets of parameters are used to
generate 36 individual solutions.  The final SFH solution is the
average of these. This ``dithering'' process significantly reduces
fluctuations in the solution associated with the sampling
\citep{Hidalgo_etal11}.  The standard deviation of the ``dithers''
provides a measurement of the uncertainties on the solution
\citep[see][for further discussion of uncertainties in the
  solution]{Aparicio_hidalgo09}.

To account for uncertainties in the distance modulus ($\pm 0.14$:
Paper I), reddening ($\pm 0.03$: \citealt{Burstein_heiles82}),
aperture corrections (Paper I), and other systematics possibly
affecting the zero points of our photometry, we allow the observed CMD
(not the model) to shift in both color and magnitude. The observed CMD
is shifted four times in magnitude and six times in color. The bundles
are correspondingly shifted. MinnIAC repeats the entire process of
generating the input information and averages the 36 individual
solutions generated by IAC-pop, for each of the positions in a
magnitude--color grid.  The grid has 35 nodes, where the shifts in
magnitude are ($-0.14$, $-0.07$, $0$, $0.07$, $0.14$), and the shifts
in color are ($-0.12$, $-0.09$, $-0.06$, $-0.03$, $0$, $0.03$,
$0.06$).  In total we generate $36\times 35 = 1260$ individual
solutions for each field (F1 and F2) and library (BaSTI and
Padova/Girardi) combination.

5) \emph{Final best solution}. After the observed CMD-shifting and
``dithering'' process, we have 35 averaged solutions, one for each
color-magnitude node. Among the 35 mean solutions, the one with the
lowest $\chi^2_{\nu}$ is chosen to be the final solution that best
reproduces our observed CMD.

\subsection{Best SFH solutions for F1 and F2}
\label{sec:bestsfh}

\begin{figure}
 \includegraphics[width=40mm, clip]{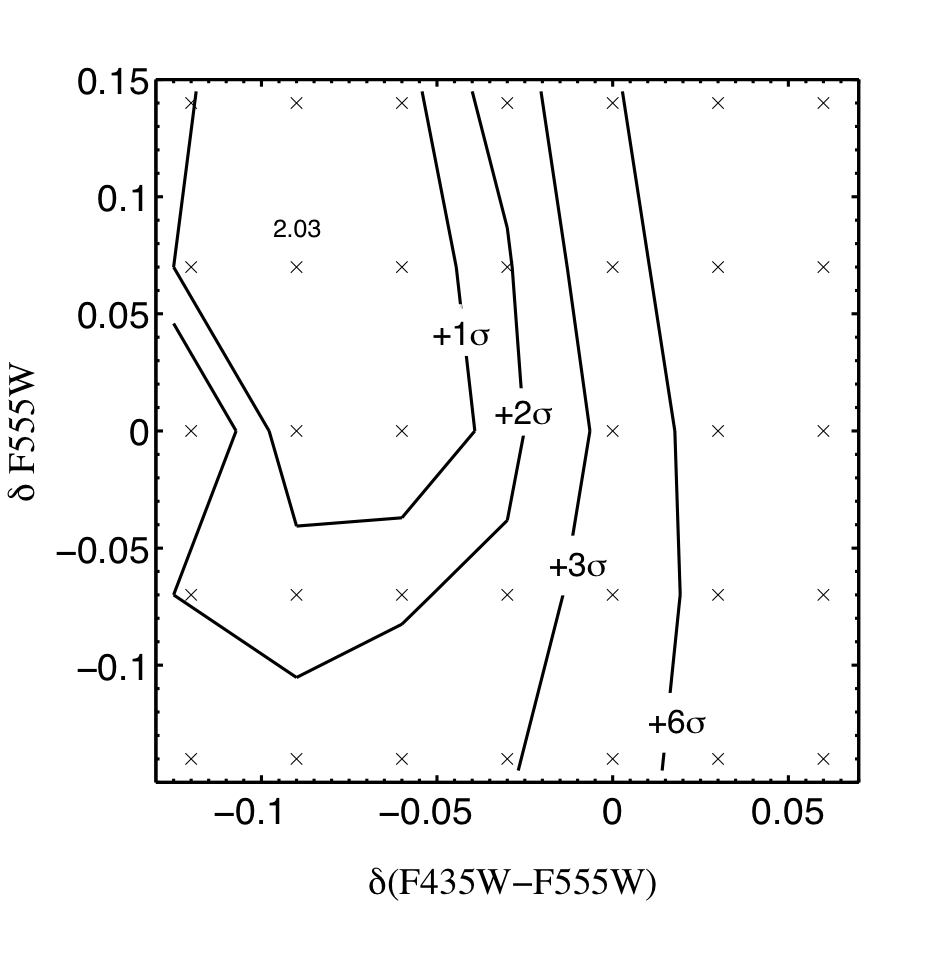}
\includegraphics[width=40mm, clip]{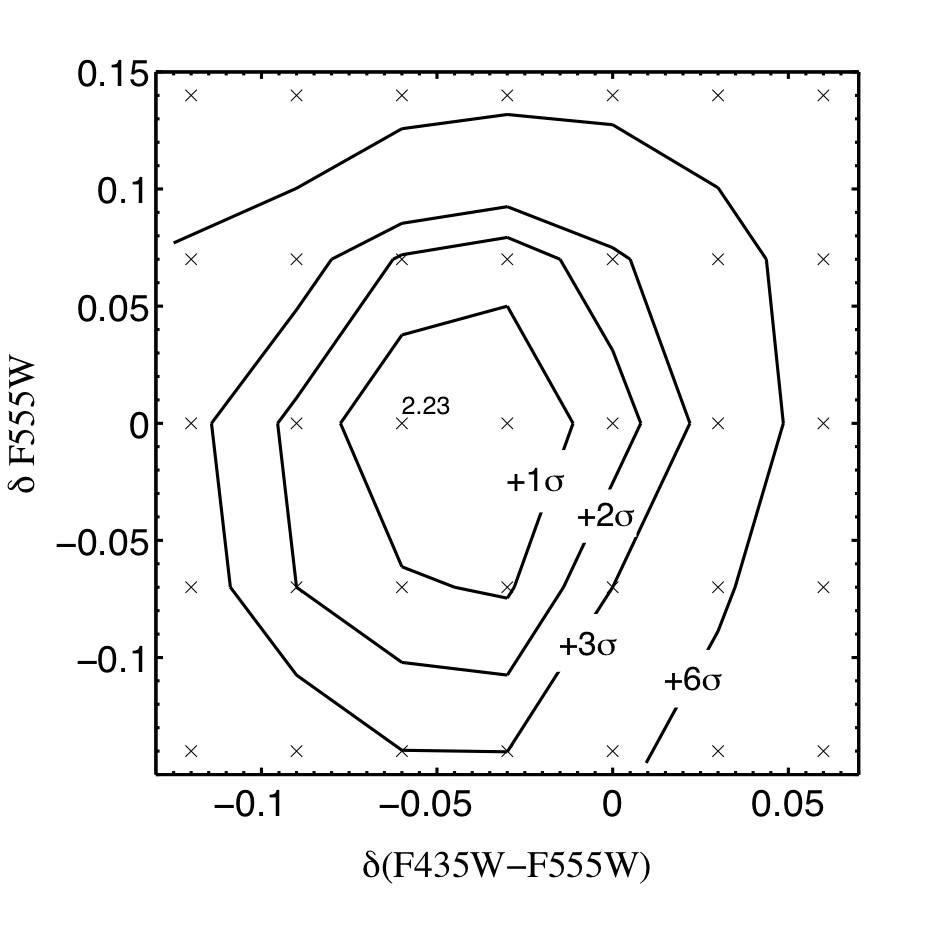}
\caption{Left-hand panel: Grid of color and magnitude shifts applied
  to the observed CMD of field F1. The color-magnitude shift nodes are
  indicated with crosses. For each of these nodes, 36 individual
  solutions were obtained, and the average of its corresponding
  $\chi^2_{\nu}$ was calculated. We consider the solution at which the
  minimum mean $\chi^2_{\nu}$ is reached as the best representation of
  our data. The $\chi^2_{\nu, min}$ value obtained using BaSTI library
  for a binary fraction of 35\% is indicated in the figure, at the
  position where it was found. Contours around this position show the
  1--, 2--, 3-- and 6--$\sigma$ confidence regions, where $\sigma$ is
  defined as the standard deviations of the 36 $\chi^2_{\nu}$
  individual solutions. Right-hand panel: As in the left-hand panel
  but for F2. Note that the $1-\sigma$ confidence area is smaller in
  this case.}
\label{fig:chiplot}
\end{figure}

As previously mentioned, for each shift in color and magnitude of the
observed CMD, we average the 36 individual solutions as well as its
corresponding $\chi^2_{\nu}$. For a 35\% binary fraction, the nodes at
which the mean minimum $\chi^2_{\nu}$, i.e. $\chi^2_{\nu, min}$ was
reached were found at $(\delta (F435W-F555W)_0, \delta M_{F555W}) =
(-0.09, 0.07)$ for F1 with $\chi^2_{\nu, min}=2.03$, and at $(\delta
(F435W-F555W)_0, \delta M_{F555W}) = (-0.06, 0.0)$ for F2 with
$\chi^2_{\nu, min}=2.23$.  We consider the averaged solution
corresponding to $\chi^2_{\nu, min}$ as the one that best reproduces
our observations. Figure~\ref{fig:chiplot} shows how the mean
$\chi^2_{\nu}$ varies as a function of the color and magnitude shifts
applied to the observed CMD in F1 (left panel) and F2 (right panel),
using the BaSTI library and 35\% of binary fraction. The crosses
indicate the 35 nodes at which we calculated a mean $\chi^2_{\nu}$,
average of the 36 $\chi^2_{\nu}$ from the individual solutions. The
contours around the minimum $\chi^2_{\nu}$ (whose value is indicated
in the figure at the position where it was found) show the 1--, 2--,
3-- and 6--$\sigma$ confidence regions, with $\sigma$ defined as the
standard deviation of the mean $\chi^2_{\nu, min}$. We emphasize here
that the shifts in the observed CMD at which we obtained the best
solution do not necessarily represent corrections to the distance or
reddening estimates, since photometric corrections and model
systematics are also present.

The $\chi^2_{\nu, min}$ values suggest that the BaSTI library fits the
data better than the Padova/Girardi isochrones for both fields (see
Table~\ref{table:chivalues} in the Appendix). Nevertheless, the
solutions obtained with both libraries are very similar, with the
Padova/Girardi isochrones generating a best-fit mean solution slightly
more metal-rich than BaSTI. For simplicity, we consider the solutions
obtained using the BaSTI library for most of the following analysis.

\section{Results of the SFH Analysis}
\label{sec:results}
\subsection{The SFH of  F1 and F2} \label{sec:sfh}

\begin{figure*}\centering
\subfigure
{\includegraphics[width=105mm, clip]{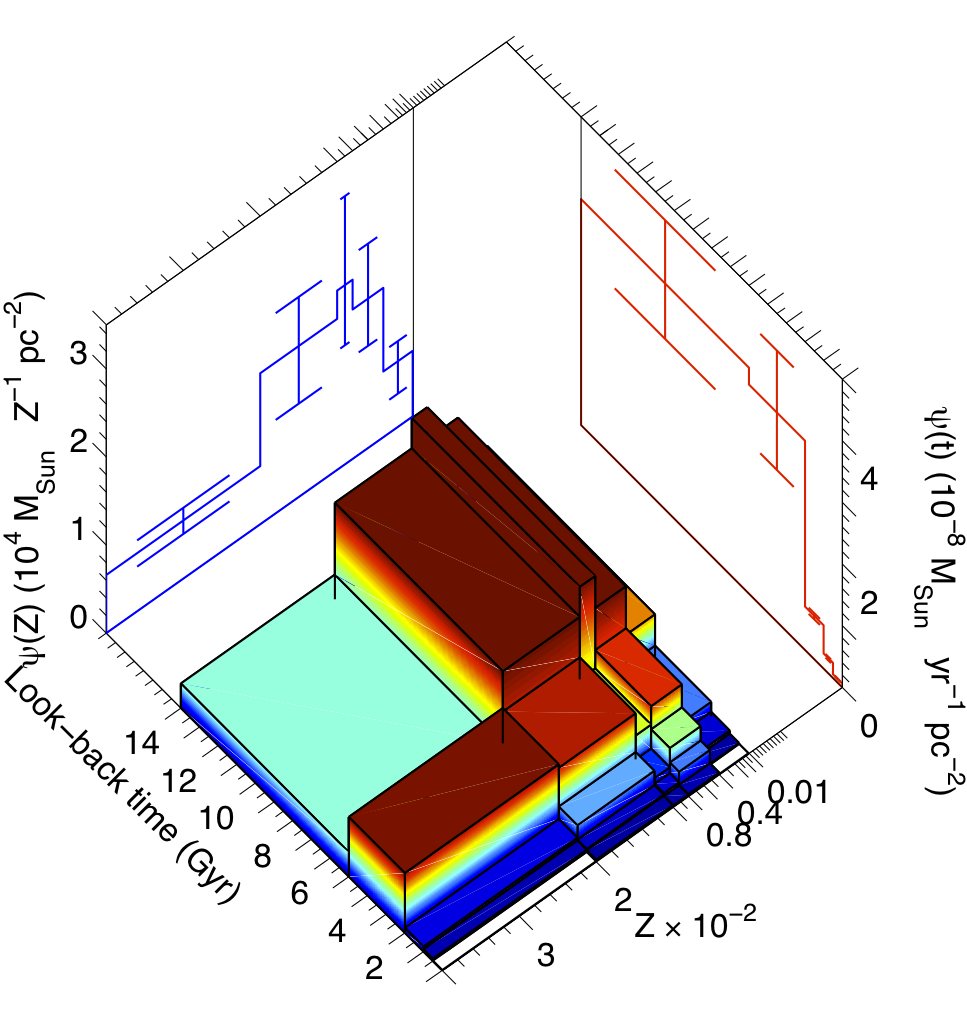}}
\subfigure
 {\includegraphics[width=105mm, clip]{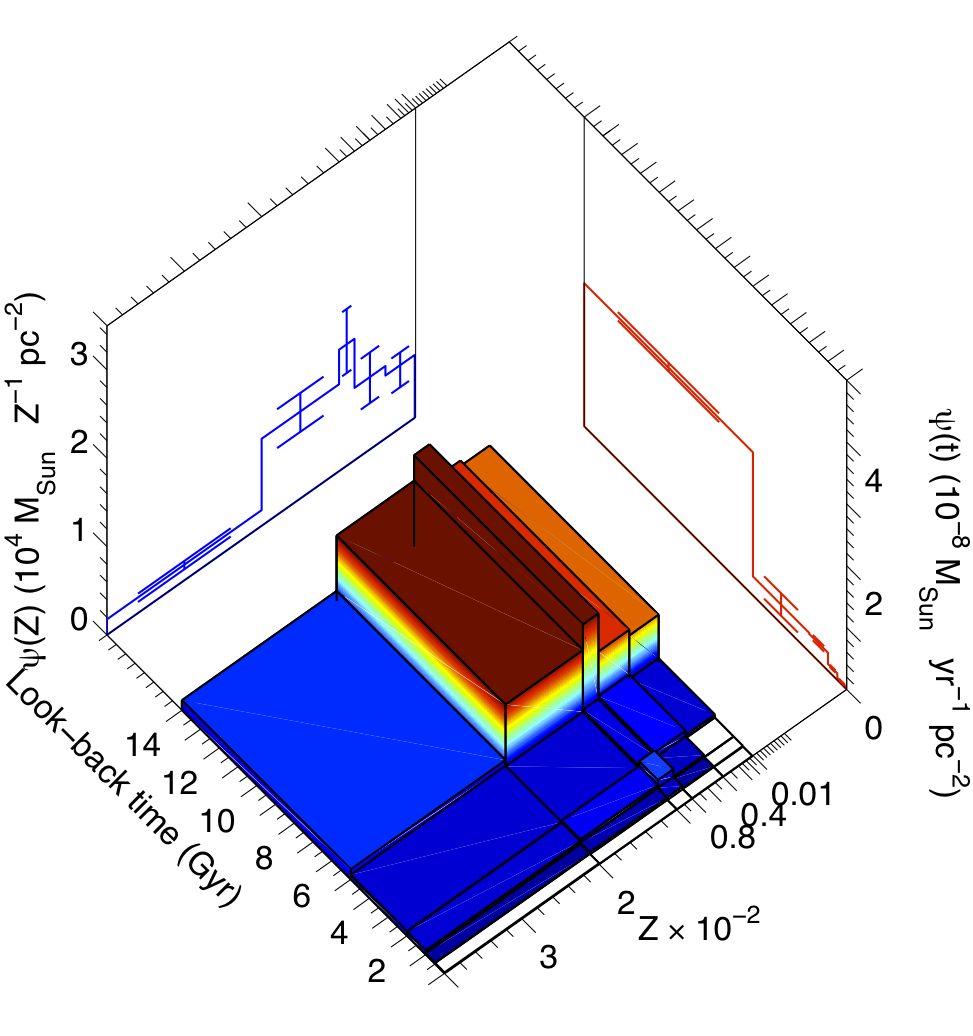}}
 \caption{SFH(t,Z)$=\Psi(t,Z)$ of F1 (top panel) and F2 (bottom panel)
   obtained using BaSTI models and assuming a 35\% binary
   fraction. The blue and red lines are the two SFH projections:
   metallicity distribution $\Psi(Z)$ and age distribution $\Psi(t)$,
   respectively. Note that $\Psi(Z)$ does not represent metallicity
   evolution, as it is integrated over age, and thus should not be
   compared with panel (d) of Figures~\ref{fig:sfhf1}
   and~\ref{fig:sfhf2}, which show Z as a function of age. Each
   solution is calculated by averaging the 36 solutions at the
   $\chi^2_{\nu, min}$ in the $\delta
   \mathrm{mag}-\delta\mathrm{color}$ grid (Sec.~\ref{sec:bestsfh});
   $\chi^2_{\nu,min}=2.03$ for F1 and 2.23 for F2. Recall that the
   number of stars in F2 is $\sim 1/2$ of that in F1. Note the
   prominent stellar population with ages 2--5 Gyr present in F1 but
   nearly absent in F2. Although differences were expected (note the
   different number of stars inside the blue box in
   Figure~\ref{fig:deconvolvedmcmd} and the results in Paper I), the
   significant different SFRs in the 2--5 Gyr bin between the two
   fields is striking.}
\label{fig:sfhf1_f2}
\end{figure*}

In Figure~\ref{fig:sfhf1_f2} we show the best-fit mean SFH$=\Psi(t,Z)$
solution for F1 (top panel) and F2 (bottom panel) in a 3D-histogram
representation, as well as the two projections $\Psi(t)$ (red line)
and $\Psi(Z)$ (blue line).  $\Psi(t)$ is the SFR as a function of time
or age distribution, i.e. $\Psi(t,Z)$ integrated over metallicity, and
$\Psi(Z)$ is the metallicity distribution function, i.e. $\Psi(t,Z)$
integrated over time. Both distributions are normalized by the area in
$\mathrm{pc}^2$. Recall that field F2 has $\sim 1/2$ the number of
stars as in F1.

The most striking feature of Figure~\ref{fig:sfhf1_f2} is the
significant star formation in F1 that occurred 2--5 Gyr ago.  F2 is
predominantly old, with some contribution of young and
intermediate-age stars from 0.5 to 5 Gyr ago, but its 2--5 Gyr old
population is clearly not as prominent as that of F1. We emphasize
here that differences in the intermediate-age population between the
fields were expected (see Paper I and
Fig.~\ref{fig:deconvolvedmcmd}). However, the significant SFR in the
2--5 Gyr bin in F1 compared with F2 is rather surprising.  As F1 has
contributions from both M32 and M31 stars and F2 is expected to have a
negligible contribution from M32, the derived SFHs suggest that the
2--5 Gyr old population in F1 is associated almost entirely with
M32. We discuss this further in the next section. To obtain the mean
age and metallicity of the system, we weight such quantities by the
mass of each bin in age or metallicity, respectively. We call them
hereafter mass-weighted mean age and mass-weighted mean metallicity.

\begin{deluxetable}{lcccc}
  \tabletypesize{\scriptsize} \tablecaption{Integrated quantities
    derived from the SFHs \label{table:sfhs}} \tablewidth{0pt}
  \tablehead{\colhead{Field}&\colhead{$\langle\mathrm{Age}\rangle$
      (Gyr)}&\colhead{$\langle[\mathrm{M/H}]\rangle$ (dex)} &\colhead{int(SFH)
      ($10^6\,M_{\odot}$)}} 
\startdata
\sidehead{BaSTI library}
  F1 & $7.95 ~\pm ~1.35$ & $-0.07 ~\pm ~0.10$ & $5.17 ~\pm ~0.50$\\
  F2 & $9.12 ~\pm ~0.80$ &$-0.19 ~\pm ~0.10$ & $2.59 ~\pm ~0.24$ \\
  M32 (F1-F2) & $6.80 ~\pm ~1.50$ & $-0.01 ~\pm ~0.08$ & $2.60 ~\pm ~0.50$\\
  F2\tablenotemark{a} & $9.15~\pm~1.27$ & $-0.10~\pm~0.10$ & $2.50 ~\pm~ 0.18$\\
\sidehead{Padova/Girardi}
  F1 & $7.99 ~\pm ~1.33$ & $0.01 ~\pm ~0.10$ & $5.88 ~\pm ~0.76$\\
  F2 & $9.03 ~\pm ~0.85$ &$-0.07 ~\pm ~0.10$ & $2.81 ~\pm ~0.29$ \\
  M32 (F1-F2) & $7.03 ~\pm ~1.50$ & $0.06 ~\pm ~0.10$ & $3.07 ~\pm ~0.75$
\enddata
\tablenotetext{a}{SFH of F2 was derived using BaSTI library with an
  extra age bin, from 5--8 Gyr.}
\end{deluxetable}

\begin{figure*}\centering
\subfigure
{\includegraphics[width=85mm, clip]{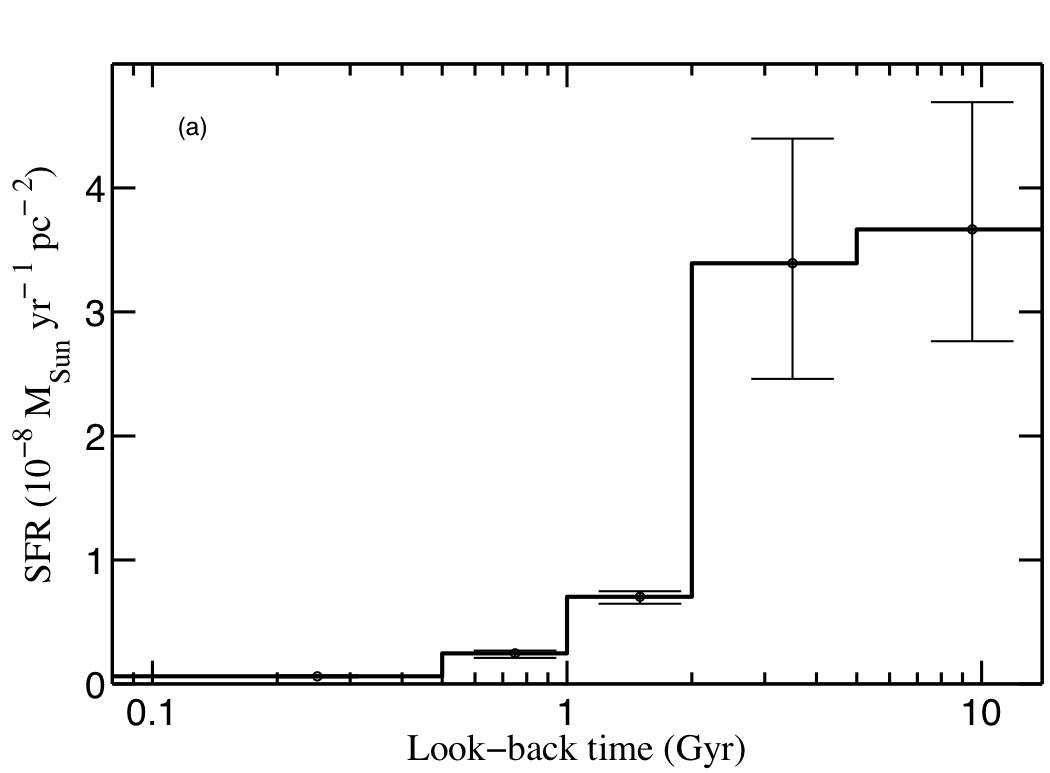}}
\subfigure
{\includegraphics[width=85mm, clip]{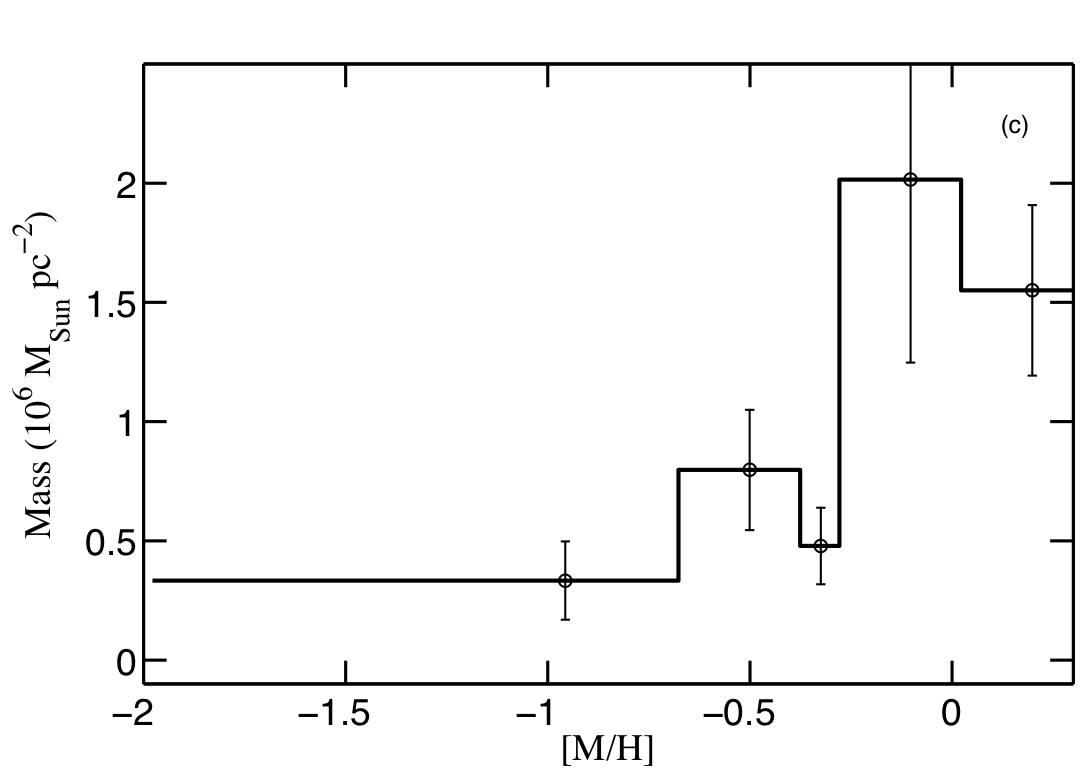}}
\subfigure
{\includegraphics[width=85mm, clip]{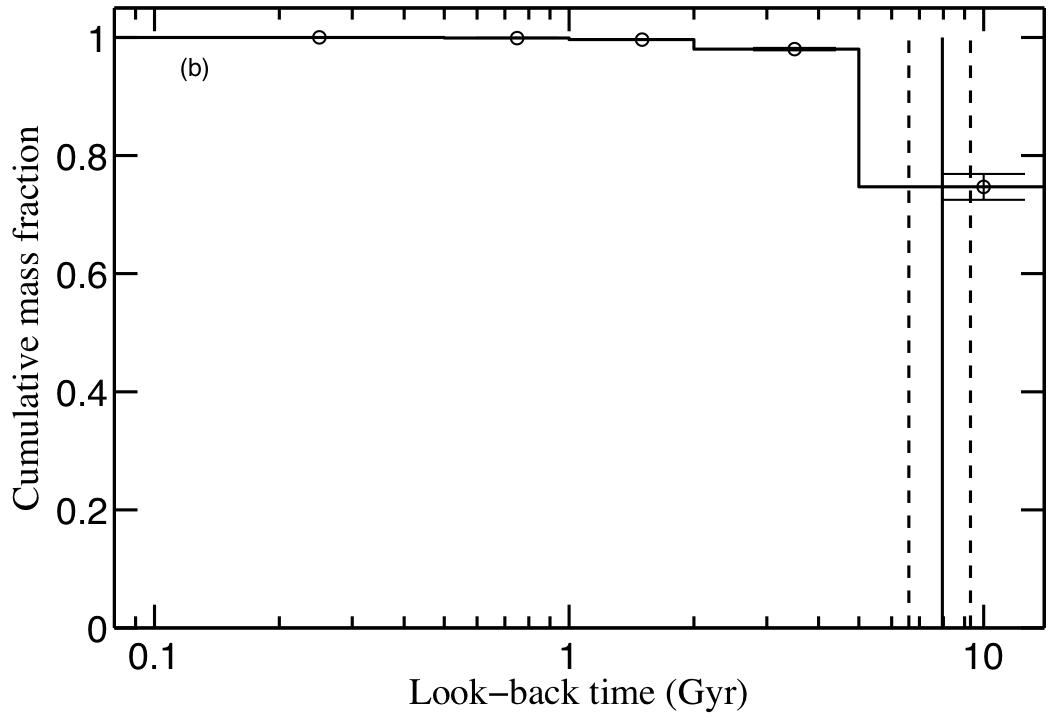}}
\subfigure
{\includegraphics[width=85mm, clip]{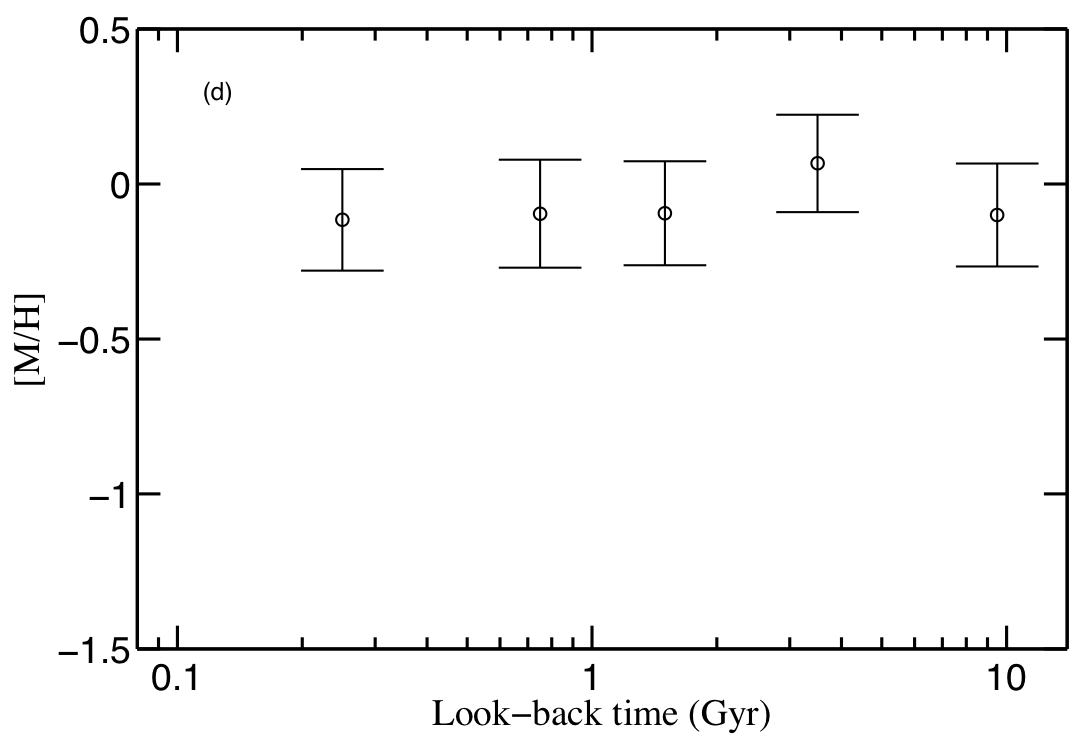}}
\subfigure
{\includegraphics[width=150mm, clip]{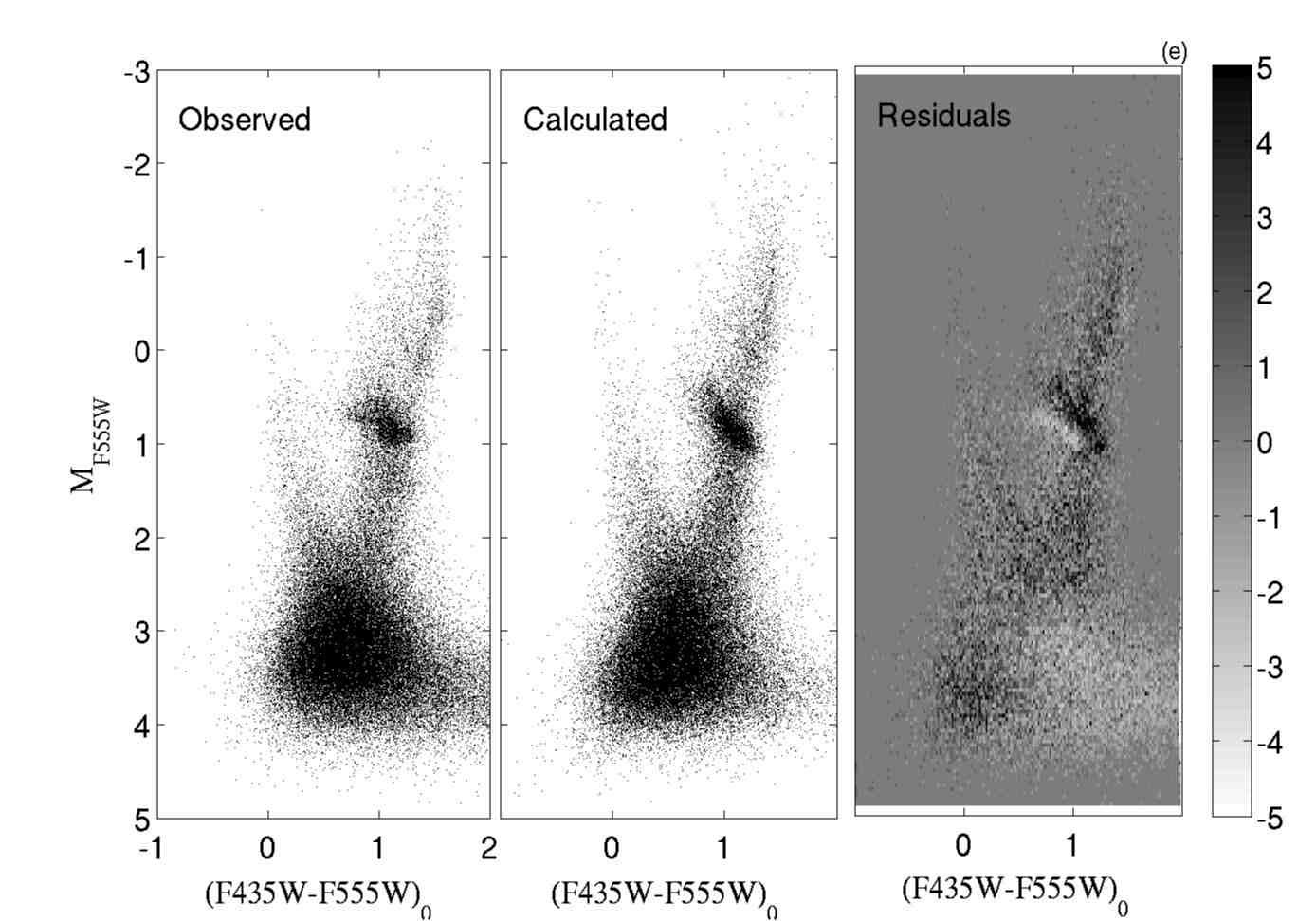}}
\caption{The SFH of F1. (a) SFR as a function of time; (b) cumulative
  mass-weighted age distribution; (c) mass as a function of
  metallicity; (d) age--metallicity relation; and (e) comparison
  between the observed, calculated CMDs and a Hess representation of
  the residuals. The calculated CMD is obtained by randomly extracting
  stars from the model CMD in such a way that the final star
  distribution represents the calculated SFH. Both the observed and
  calculated CMDs were divided into the same $200\times200$ bins. The
  Hess diagram of the residuals in panel (e) shows the subtraction of
  the observed Hess diagram from the calculated one, in units of the
  Poisson uncertainties. The vertical solid line in panel (b)
  represents the mean age ($\sim 8$ Gyr) of the system, and the dashed
  lines indicate the $1\sigma$ deviation of that value.}
\label{fig:sfhf1}
\end{figure*}

\begin{figure*}\centering
\subfigure
{\includegraphics[width=85mm, clip]{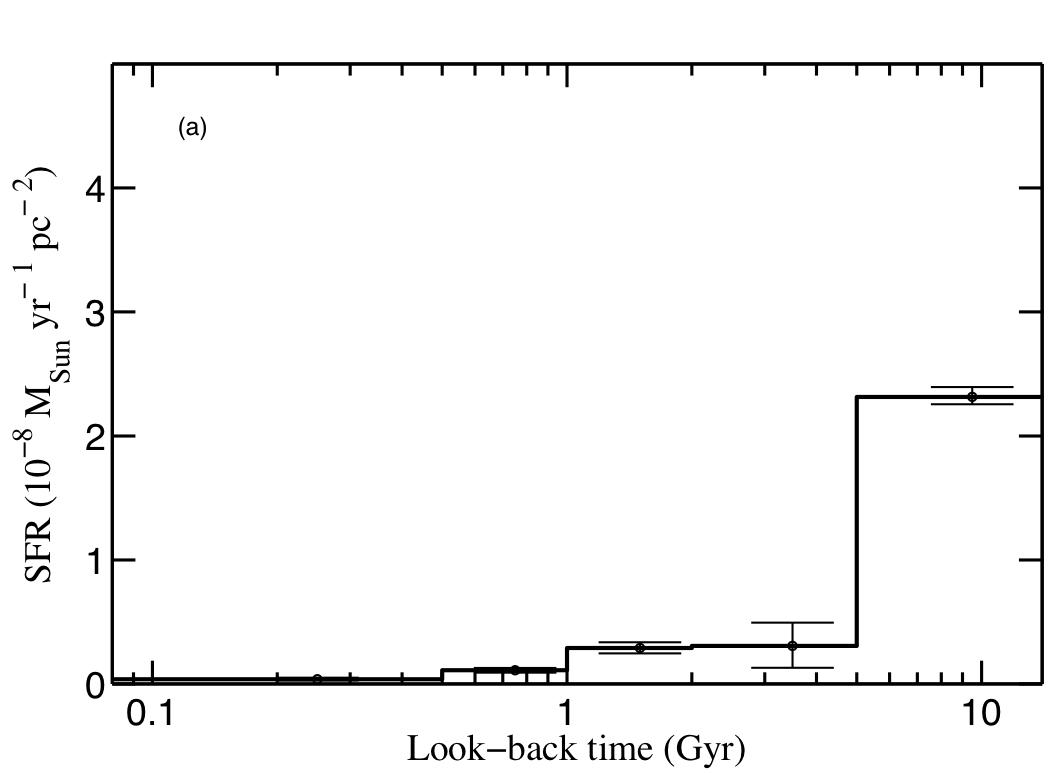}}
\subfigure
{\includegraphics[width=85mm, clip]{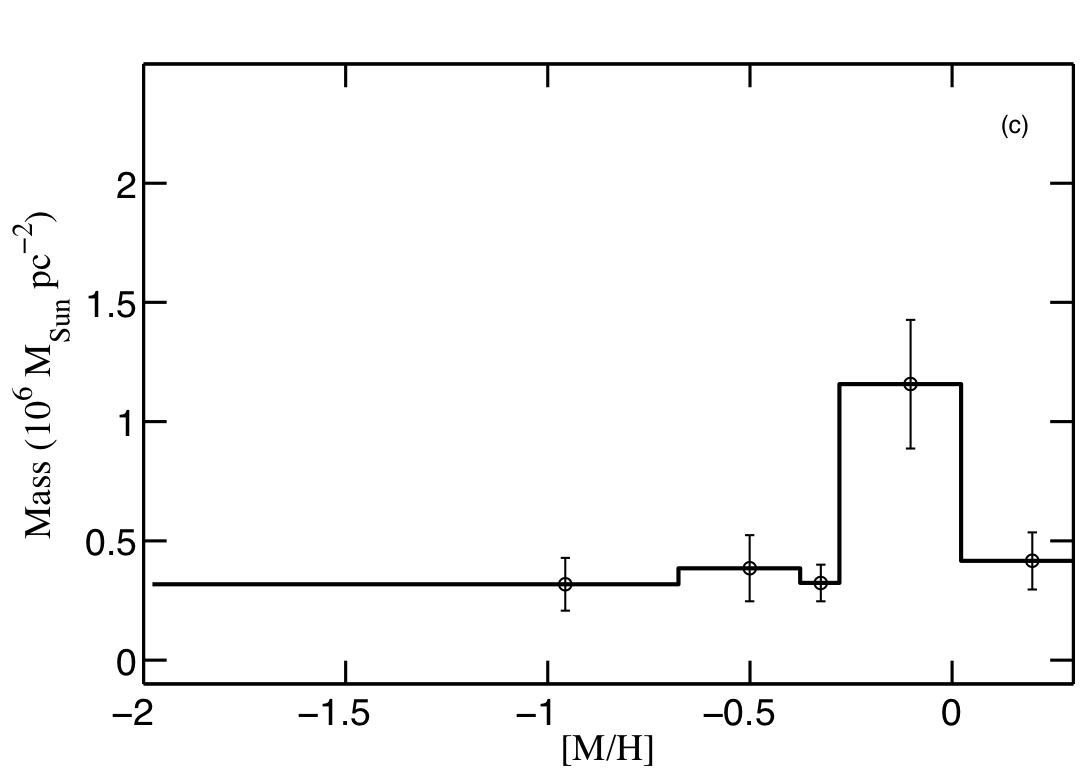}}
\subfigure
{\includegraphics[width=85mm, clip]{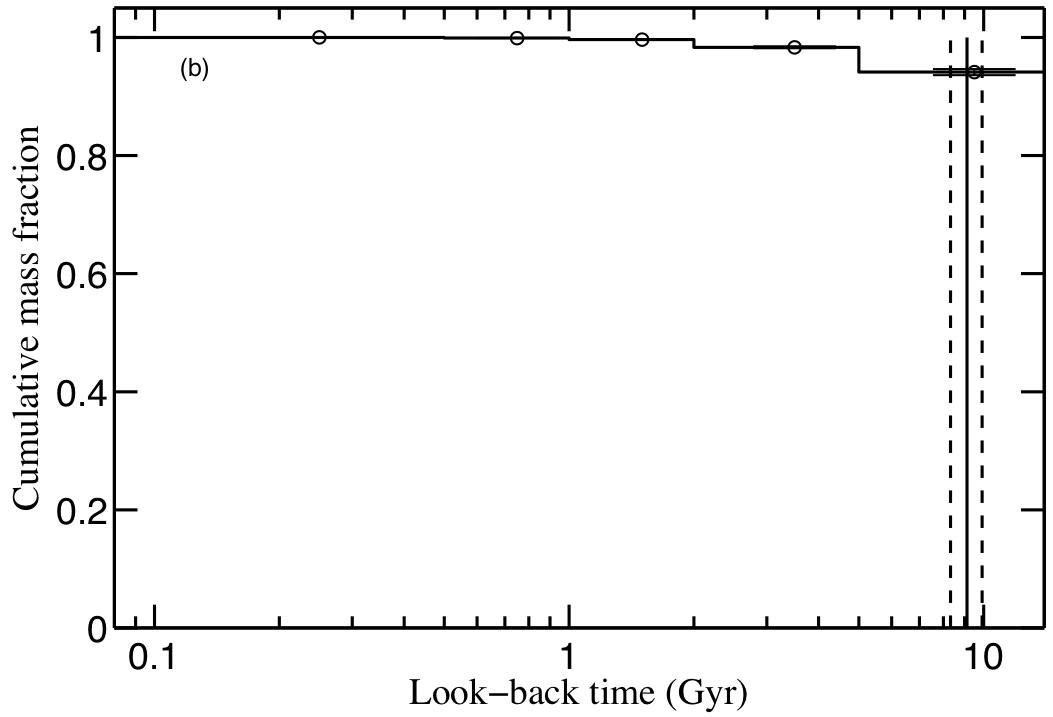}}
\subfigure
{\includegraphics[width=85mm, clip]{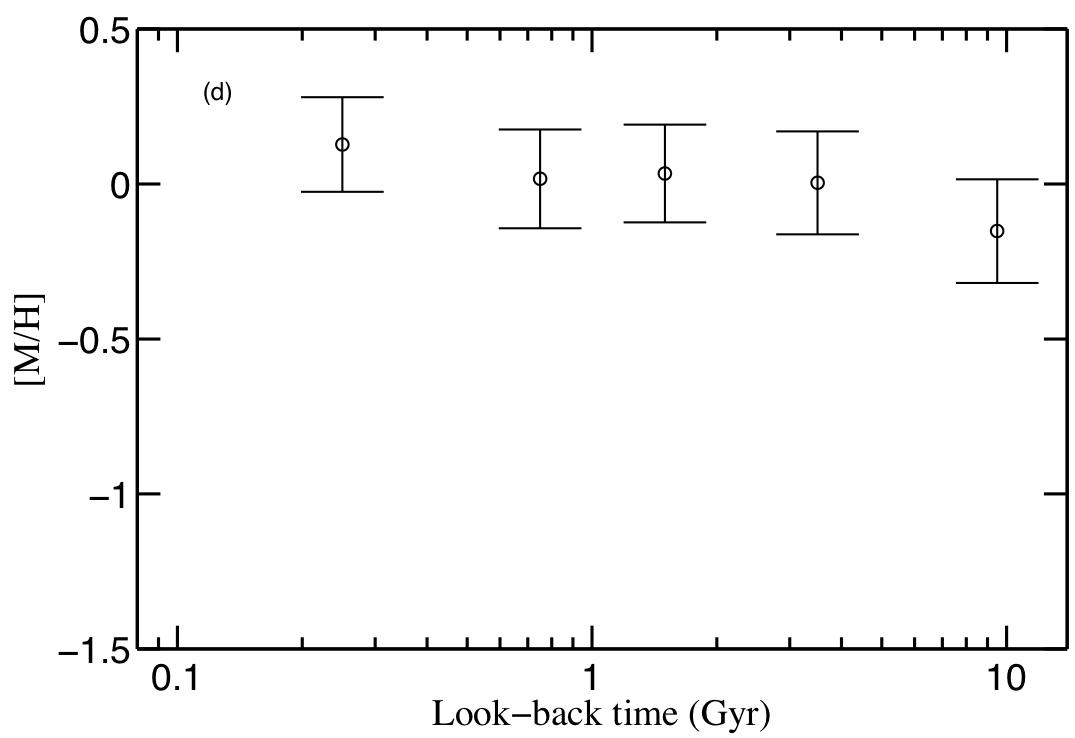}}
\subfigure
{\includegraphics[width=150mm, clip]{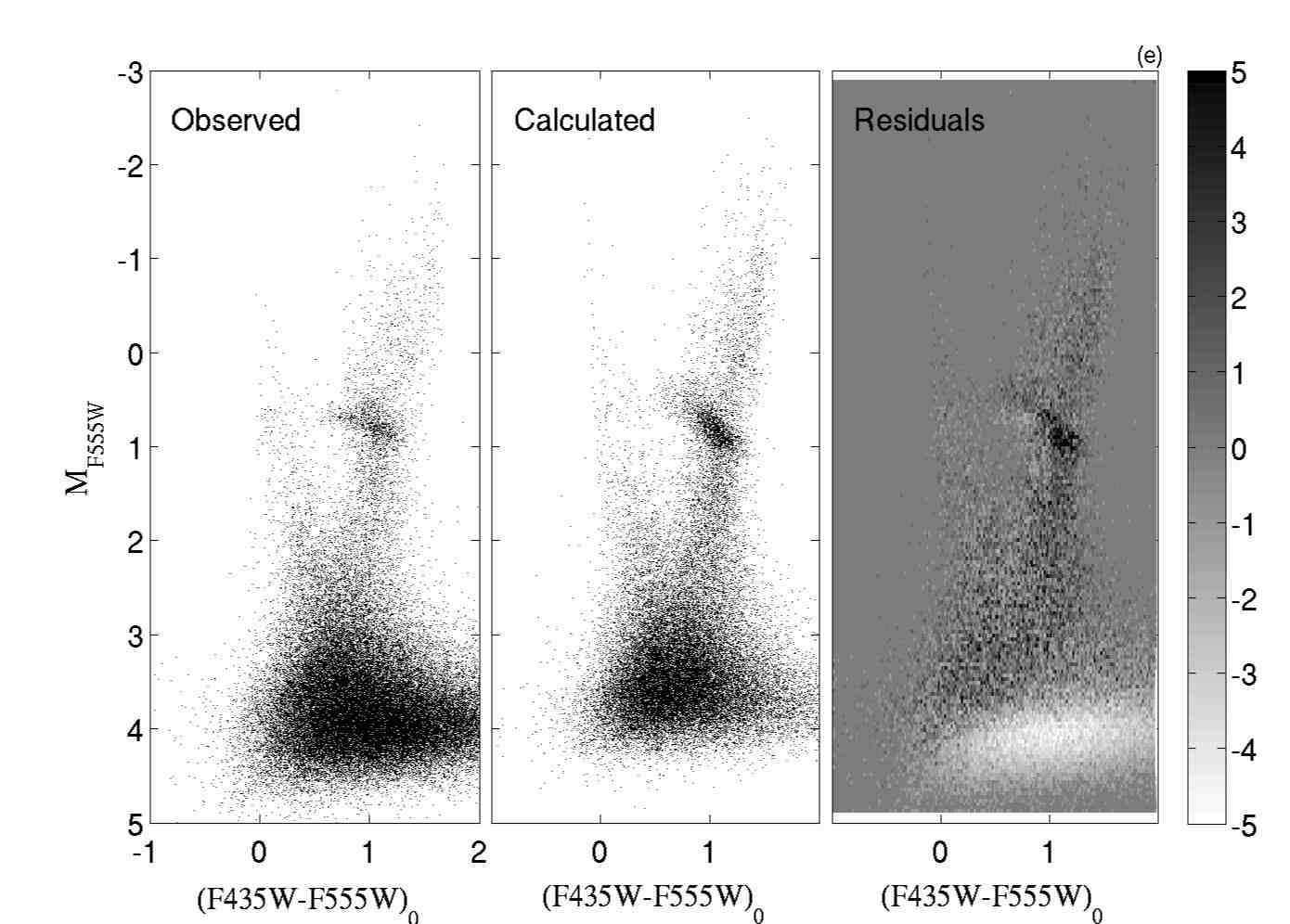}}
\caption{As in Figure~\ref{fig:sfhf1} for F2. The mass in F2 is
  roughly half that of F1.}
\label{fig:sfhf2}
\end{figure*}

Figures~\ref{fig:sfhf1} and \ref{fig:sfhf2} display the main results
projected from the extracted SFHs of F1 and F2, respectively. We find that:

\begin{itemize}
\item F1 acquired 75\% of its stellar mass between 5 and 14 Gyr ago.
  Stars with ages of 2--5 Gyr contribute 23\% of the mass in F1. The
  remaining 2\% of mass in F1 is constituted by stars younger than 2
  Gyr.

\item F1 is metal-rich with an almost constant age--metallicity
  relation, to the limits of the age resolution of the CMD.

\item F1's mass-weighted mean age is $7.95\pm1.35$ Gyr and its
  mass-weighted mean metallicity is
  $[\mathrm{M/H}]=-0.07\pm0.10~\mathrm{dex}$ (Table~\ref{table:sfhs}).

\item F2 is predominantly old, with 95\% of its mass already formed
  5--14 Gyr ago. There is a small contribution of mass to the system
  after that, and it stopped forming stars $\sim$ 0.5 Gyr ago.

\item F2 is also quite metal-rich, but is marginally more metal-poor
  than F1, with a slight age--metallicity relation showing a small
  increase in metallicity at younger ages.

\item F2's mass-weighted mean age is $9.12\pm0.80$ Gyr and its
  mass-weighted mean metallicity is
  $[\mathrm{M/H}]=-0.19\pm0.10~\mathrm{dex}$ (Table~\ref{table:sfhs}).
\end{itemize}
The integrated quantities derived for the SFHs of F1 and F2 using
Padova/Girardi Library are also indicated in Table~\ref{table:sfhs}.

Figures~\ref{fig:sfhf1}e and \ref{fig:sfhf2}e show comparisons between
the observed (left) and calculated CMD (middle) as well as the Hess
diagram of the residuals in units of the Poisson uncertainties
(right), for F1 and F2 respectively. The calculated CMDs have been
obtained by randomly extracting stars from the synthetic CMDs in such
a way that the resulting star distribution follows the best calculated
SFHs. For both F1 and F2, the model CMD shows reasonable agreement
with the observed CMD throughout most evolutionary phases, which is
also reflected in the residual Hess diagrams. The RC regions, however,
show significant discrepancies. This is not surprising; due to
uncertainties in, e.g., the mass loss during the RGB or the He content
of the stars, that particular evolutionary stage is not well-modeled
\citep{Gallart_etal05}---but we have not used this region in deriving
the solutions. There is also some discrepancies for magnitudes fainter
than the 50\% completeness level, but this region was also not used
for the derivation of the SFHs.

\subsection{The SFH of M32 as revealed by the IAC method}

To calculate the SFH of M32, we make use of the derived SFHs of F1 and
F2\footnote{We would ideally need a deep CMD composed solely of M32
  stars to derive the SFH of M32, which we attempted to derive in
  Paper I. Under the assumption that the M31 stellar populations in F1
  and F2 are statistically the same, we subtracted the stars of the F2
  CMD from the CMD of F1 taking into account the difference in
  crowding of the fields. This produced the deepest CMD of M32 yet
  obtained. However, the use of such CMD to extract the SFH of M32
  would introduce uncertainties associated with the decontamination
  process.}. Given the fact that both SFHs have been obtained using
the same stellar population sampling, and assuming that the SFH of M31
in F1 and in F2 is identical and that M32 is not present in F2,
calculating the SFH of M32 is straightforward: we simply subtract the
SFH of F2 from that of F1.

\begin{figure*}\centering
\includegraphics[width=150mm, clip]{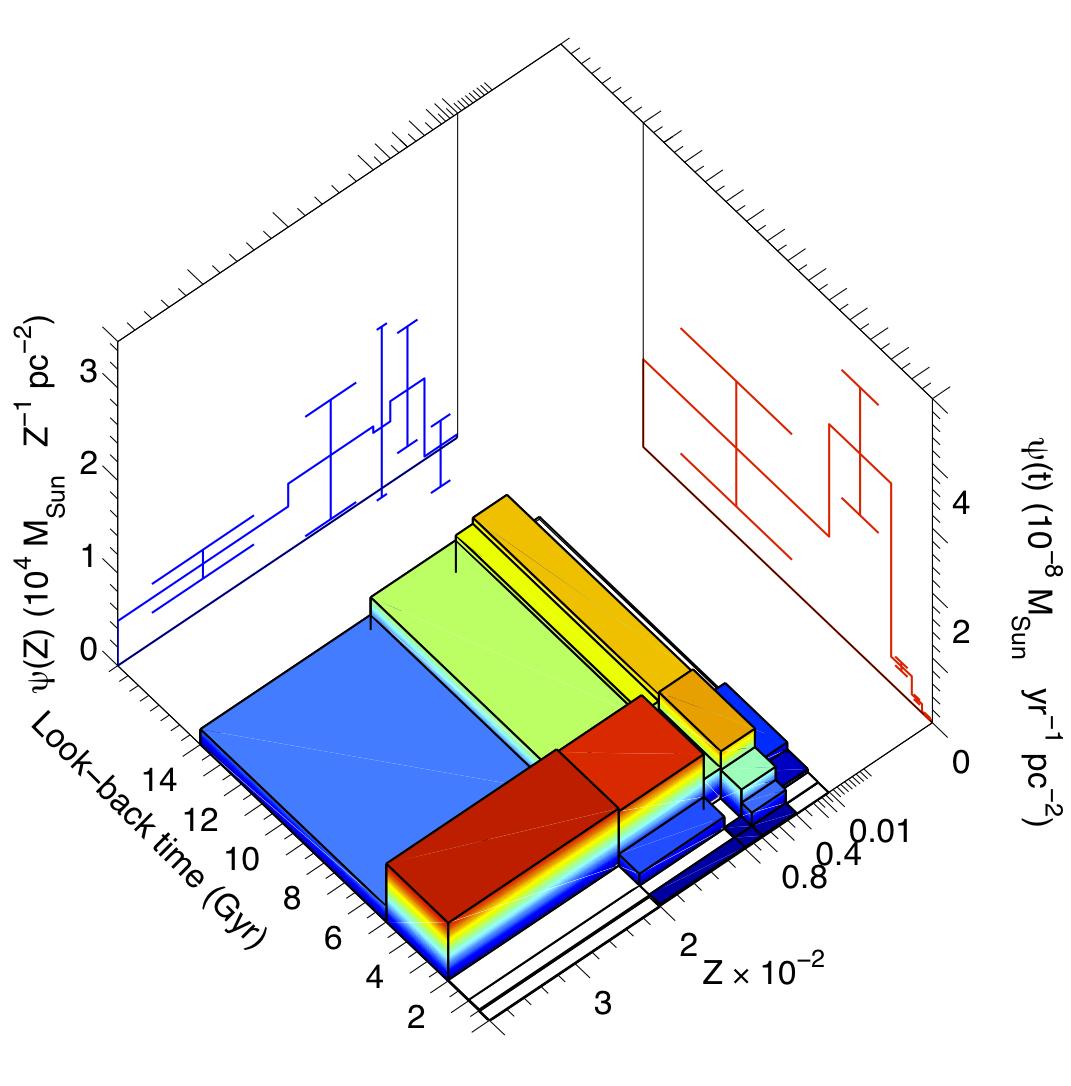}
\caption{SFH of M32 obtained after subtracting the calculated SFH of
  F2 from that of F1. We find two dominant populations contributing to
  the SFH of M32. One is 2--5 Gyr old and contributes $\sim$ 40\% of
  the total mass of M32 at F1. The population older than 5 Gyr
  contributes $\sim$ 55\% of the total M32's mass at F1. Note that
  some of the stars younger than 2 Gyr are quite metal-poor compared
  to the nearly solar mean metallicity of M32. This suggests that
  these are BSS and may be the first direct evidence of such a
  population in M32.}
\label{fig:sfhm32}
\end{figure*}

\begin{figure*}\centering
\subfigure
{\includegraphics[width=85mm, clip]{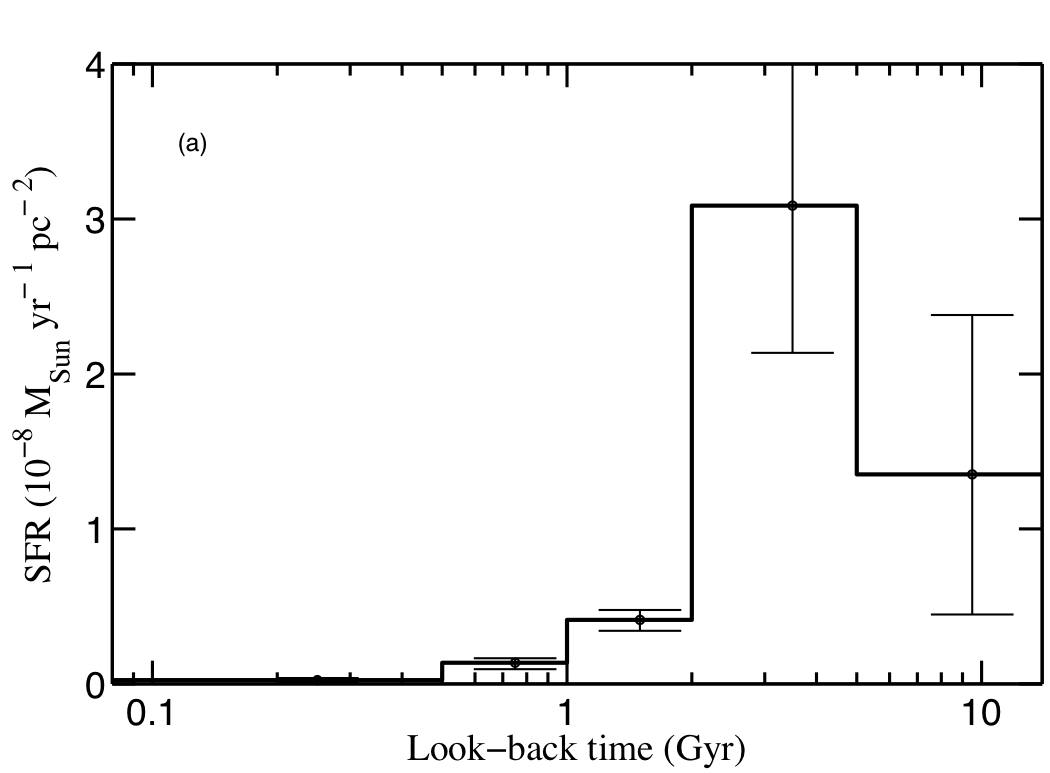}}
\subfigure
{\includegraphics[width=85mm, clip]{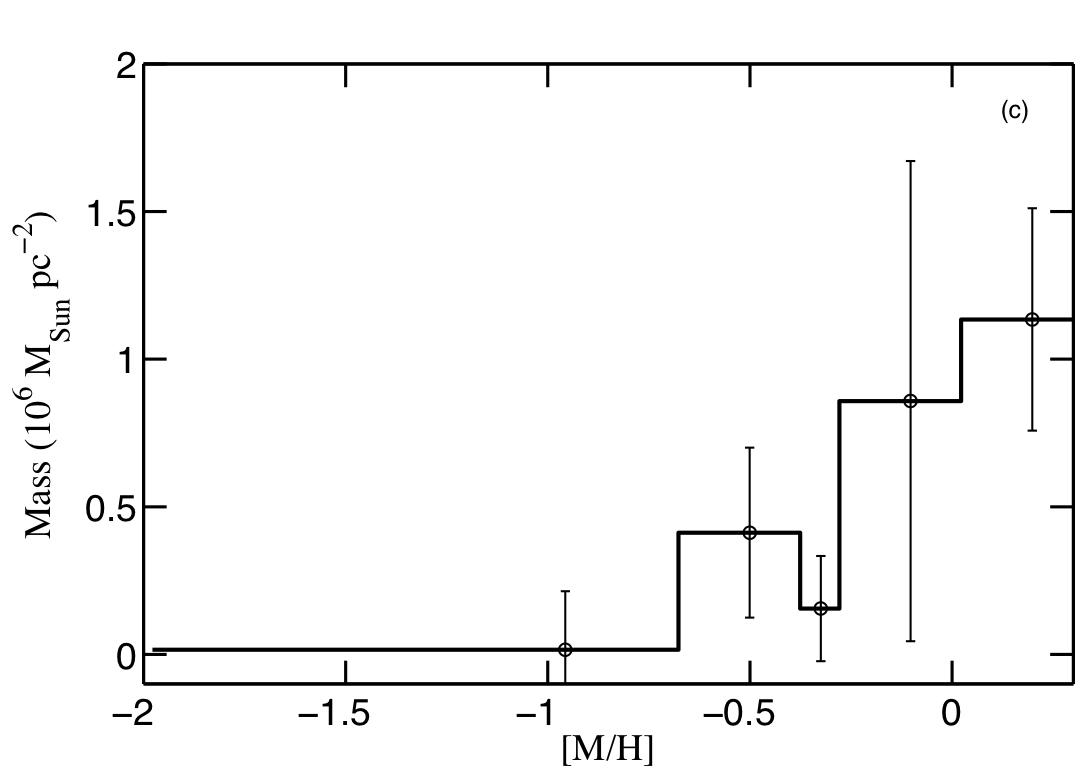}}
\subfigure
{\includegraphics[width=85mm, clip]{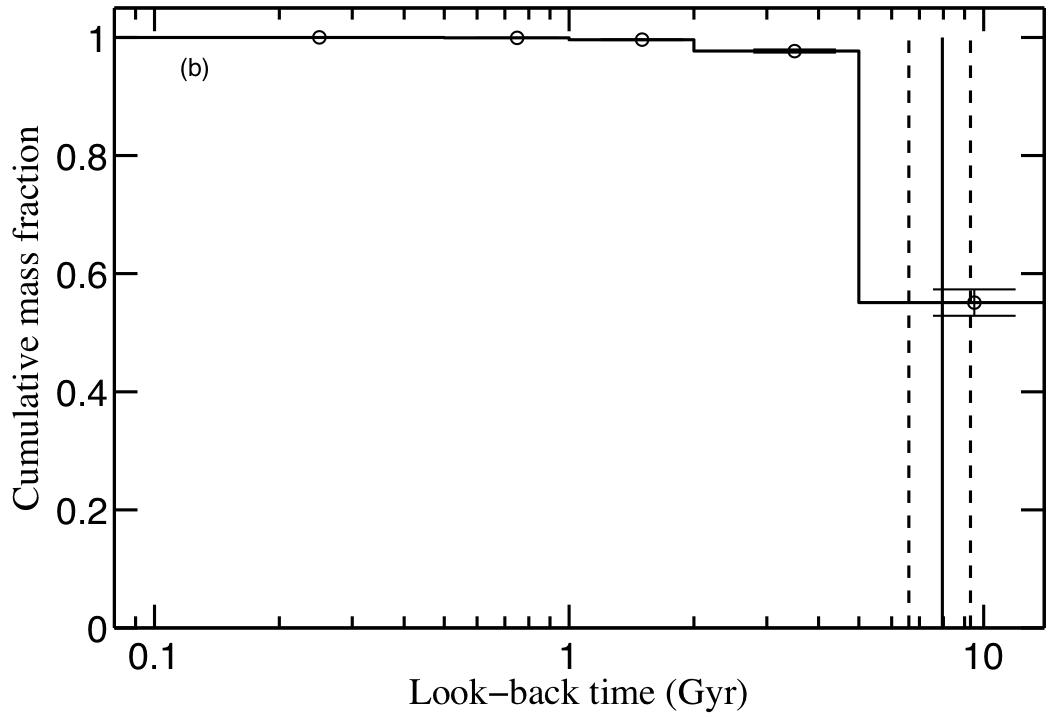}}
\subfigure
{\includegraphics[width=85mm, clip]{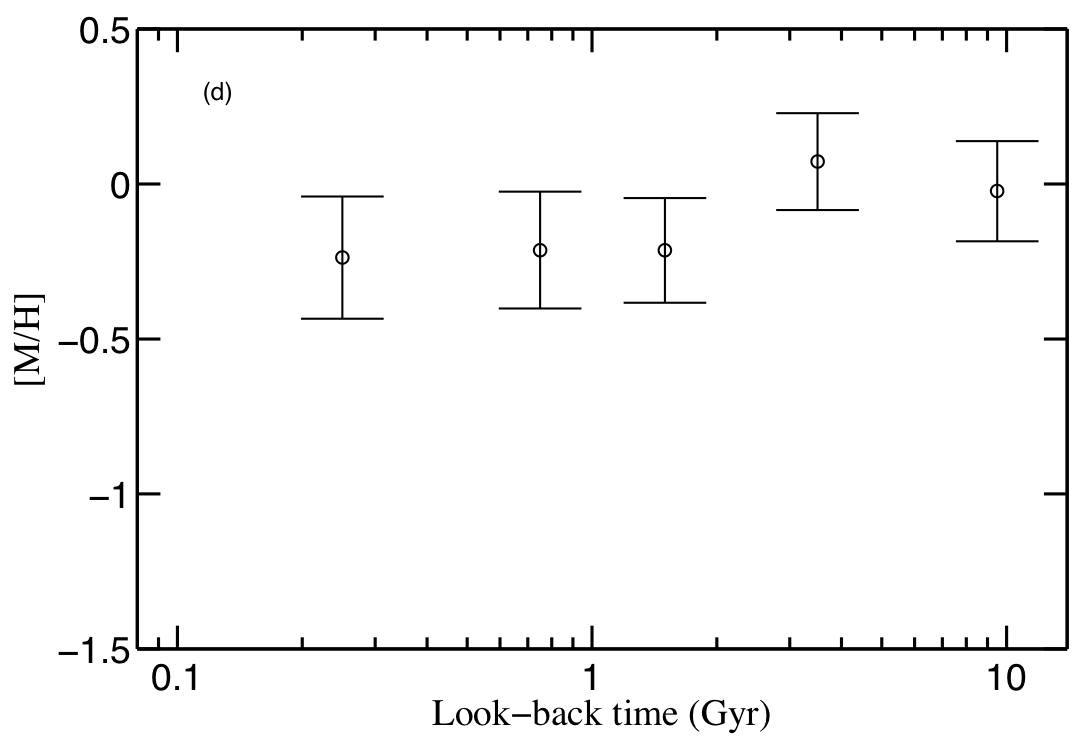}}
\caption{The SFH of M32. (a) SFR as a function of time, clearly
  indicating the two dominant populations: at $\sim 8$ Gyr and $\sim
  4$ Gyr; (b) cumulative mass-weighted age distribution which shows
  how much each population contributes to the total mass of M32 at F2;
  (c) mass as a function of metallicity, indicates the mean
  metallicity of the system, roughly solar; and (d) age--metallicity
  relation, nearly constant. The vertical lines in panel (b) represent
  the mean age ($\sim 6.8$ Gyr) of M32 in F1. The dashed lines
  indicate the $1\sigma$ deviation of this value.}
\label{fig:sfhm32_bis}
\end{figure*}

Figure~\ref{fig:sfhm32} shows the inferred SFH of M32 for the first
time calculated from its resolved stellar population. We used the F1
and F2 SFHs shown in Figure~\ref{fig:sfhf1_f2}, inferred using the
BaSTI library and a 35\% binary fraction. We can see that a major
burst of star formation occurred in M32 2--5 Gyr ago, responsible for
$\sim$ 40\% of M32's current mass at F1's location. This can be seen
from the cumulative mass function, shown in panel (b) of
Figure~\ref{fig:sfhm32_bis}. Stars older than 5 Gyr contribute
$\sim$55\% of the total mass of M32 in this field. From this
CMD-fitting analysis, however, due to the limitations imposed by the
crowding of our fields, we cannot specify when the star formation
started, whether it was constant over the 5--14 Gyr period, or if it
peaked at some age. Integrated quantities derived from the calculated
M32 SFH are indicated in Table~\ref{table:sfhs}. Note that the
estimated mean age and metallicity of M32, $\sim 6.8$ Gyr and $\sim
-0.01~\mathrm{dex}$, respectively, are younger and more metal-rich
that the mean age and metallicity of F1, because M31's mean age and
metallicity in F2 is older and more metal-poor than M32 in F1.

The age--metallicity relation for M32 is nearly constant
(Fig.~\ref{fig:sfhm32_bis}d) although there is apparently a mild
increase at $\sim$ 5 Gyr followed by a small decrease at $\sim$ 2 Gyr.
We note that an almost constant age--metallicity relation appears to
suggest that M32 has not experienced any metal enrichment. However,
the lack of resolution in age means that we cannot extract detailed
information on stars older than 5 Gyr.  Most of the chemical evolution
of the system has likely occurred during that 5--14 Gyr period. M32's
mass-weighted peak in metallicity is at $[\mathrm{M/H}] \sim 0.2$ dex
(Fig.~\ref{fig:sfhm32_bis}c).

\begin{figure*}\centering
\includegraphics[width=120mm, clip]{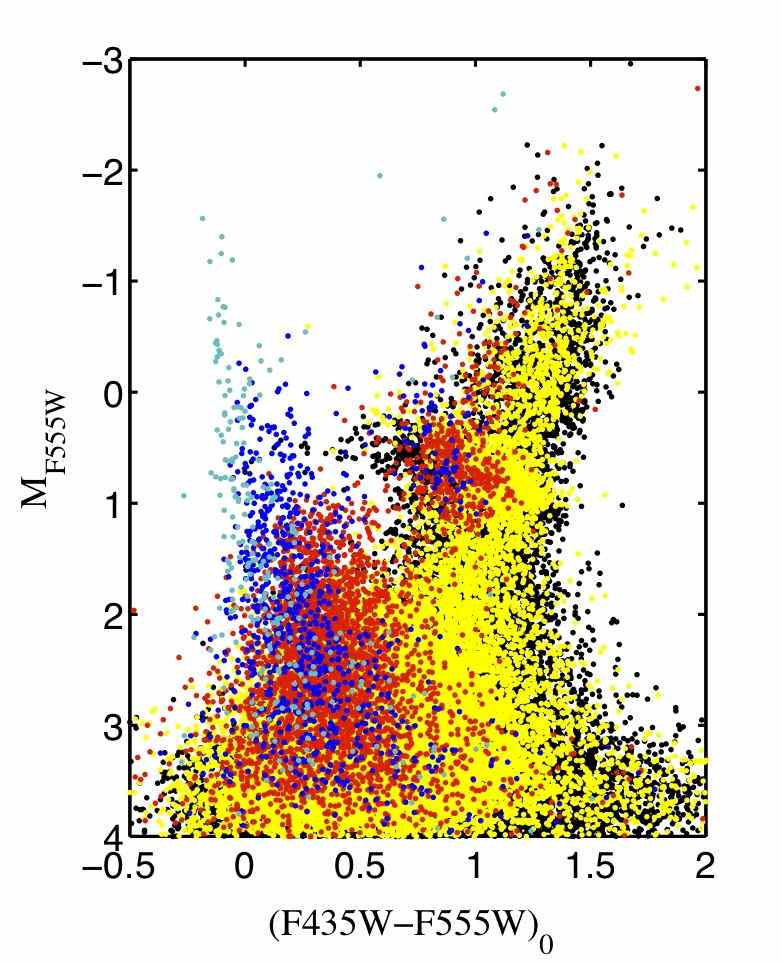}
\includegraphics[width=140mm, clip]{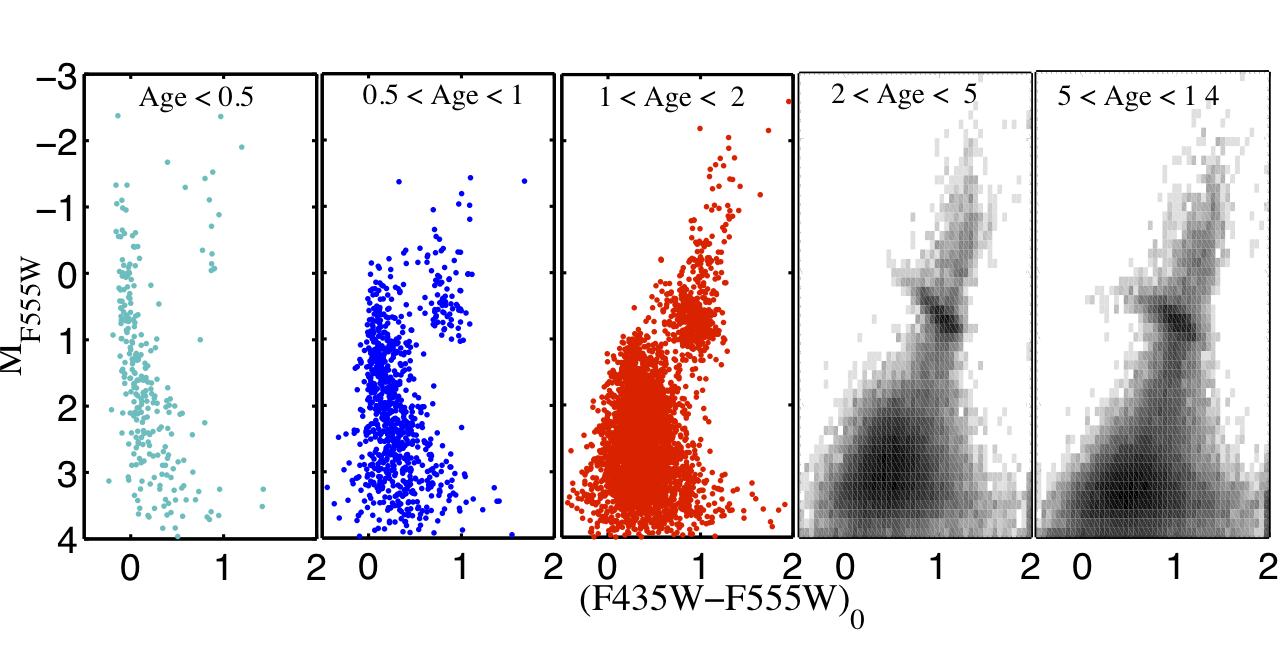}
\caption{Top panel: Calculated CMD of M32, obtained by randomly
  extracting stars from the model CMD in such a way that they follow
  the derived SFH of M32. The stars are color-coded according to age,
  as indicated in the bottom panel, except for 2--5 Gyr old and stars
  older than 5 Gyr, shown as yellow and black dots, respectively, in
  the top panel whereas gray-scale Hess representations of their CMDs
  are shown in the bottom panel. Note how the various ages fit
  together and the age interval that populates each region of the
  CMD. Bottom panel: Each CMD is composed by stars of a different age
  interval, from only an extended main sequence (left panel, ages
  $\sim 0.5$ Gyr) to a CMD with well populated RGB, RC and AGB
  evolutionary phases (right panel, ages of 5--14 Gyr). Note the
  differences in the MSTO region and fainter MS in the last two
  CMDs. The MSTOs for the younger population (2--5 Gyr, yellow dots in
  the top panel and a Hess representation of their CMD in the bottom
  panel) are brighter and bluer than the ones for the 5--14 Gyr old
  population (black dots in the top panel and a Hess representation of
  their CMD in the bottom panel).}
\label{fig:scmd_m32}
\end{figure*}

We show in the top panel of Figure~\ref{fig:scmd_m32} the calculated
CMD of M32, with its stars color-coded according to age. The CMD was
obtained by randomly extracting stars from the model CMD, in such a
manner that their star distribution follows the calculated SFH. This
figure provides explicit information on the age interval that
populates each region of the CMD as well as showing how the various
ages combine. We see, for example, that stars of different ages
contribute to the RC. Younger stars populate the brighter bluer
portion of the RC while older stars populate the fainter, redder
portion of the RC. The BP is only populated by stars younger than 2
Gyr. The bottom panel shows the CMDs produced by each age interval
considered in the extraction of the SFH. We can appreciate in detail
the differences between each CMD as the ages vary, from only an
extended main sequence (bottom left panel, ages $\sim 0.5$ Gyr) to a
CMD with well-populated RGB, RC and AGB evolutionary phases (bottom
right panel, ages of 5--14 Gyr). Note the presence of only few BHB
stars in the bottom right panel, as expected for systems as metal-rich
as M32; in the composite CMD (top panel), these few BHB stars are
mixed with young, blue stars in the extended MS.

Finally, we can qualitatively compare these results with the SFHs
derived for some of the Local Group dwarf satellites, which have been
analyzed by the same or very similar methods\footnote{A more extensive
  and quantitative comparison of our results with other the SFHs of
  Local Group satellites is beyond the scope of this paper.}. It is
interesting to note that the mean age derived for M32 is comparable to
that of dwarf irregular galaxies, such as the Small and Large
Magellanic Clouds and Pegasus \citep{Noel_etal09, Harris_zaritsky09,
  Dolphin_etal05}, whereas it is quite young compared with the mean
ages of dwarf spheroidal galaxies, such as Cetus, Tucana, and Ursa
Minor \citep{Monelli_etal10a, Monelli_etal10b, Dolphin_etal05}. We
also note that a synthetic CMD analysis was performed on archival
WFPC2 data of M32 by \citet{Dolphin_etal05}. Their results suggest an
older mean mass-weighted age for M32, $\sim 8.5$ Gyr. However, the
WFPC2 data not only are shallower than our data but also contain
significant contamination by M31 stars, which were not taken into
account in their SFH analysis.

\subsubsection{Exploring the SFH solution and its robustness}

We now address the robustness and uniqueness of the SFHs derived
here. First, due to the complex parameter space and thus multiple
local minima involved in the process to find a SFH, an algorithm that
guarantees finding the global minimum is strongly desirable. As
mentioned above, we have used a genetic algorithm to find the minimum
$\chi^2$ (see \citealt{Charbonneau95}). This type of algorithm, unlike
the so-called 'downhill' algorithms, is designed such that the
solution found is infinitesimally close to or at the global minimum
independently of the initial seed that started the process, provided
that a sufficiently large number of generations (i.e. variations of
individuals and mutations) is performed. \citet{Aparicio_hidalgo09}
have tested this and found that $\approx 10^5$ generations are enough
to guarantee that the solution will be at or as close as possible to
the global minimum. We have performed $2\times10^5$ generations per
solution, which assures us that we have reached convergence.

In addition, the IAC method does not introduce any systematic error to
the SFH solution, provided that the age and metallicity bins used to
extract the SFH are appropriate to the observed CMD. This has been
verified by several tests on mock stellar populations performed at
Instituto de Astrof\'isica de Canarias in which the SFH of mock
galaxies of known SFH have been recovered rather accurately
\citep[see][]{Aparicio_hidalgo09, Hidalgo_etal09, Hidalgo_etal11}. In
this work, we have also performed several experiments on mock data to
find the appropriate age and metallicity resolution at which,
according to the observational effects of our observed CMDs, the SFH
solutions of mock galaxies are within 1--$\sigma$ to the input ones
(see Sec.~3.1 above). Thus, each SFH solution obtained is ``unique,''
by which we mean that combinations of simple populations within the
error bars of the SFH will produce CMDs indistinguishable from the
best fit CMD. Any other SFH which is combination of simple populations
significantly different that those of the final SFH (i.e., not
possible within the error bars of our solution) will produce a CMD
significantly different than the best-fit CMD and, therefore, than the
observed one.

Finally, even though we only show our best solution, we have explored
other solutions that give similar good fits. We found that the SFH of
the solutions at 1--$\sigma$ confidence area (see
Fig.~\ref{fig:chiplot}) are not significantly different than the best
one, for both F1 and F2. The mass percentages per age range slightly
vary from one solution to the other, but the overall SFH remains the
same. Taking these nearby solutions into account, we find that M32 at
F1 has $\sim 40\% \pm 17\%$ of its mass in a 2--5 Gyr old, metal-rich
population and $\sim55 \% \pm 21\%$ of its mass in stars older than 5
Gyr, with slightly subsolar metallicities. The uncertainties represent
the 1--$\sigma$ error of our best solution and the variations of these
percentages when considering solutions of similar good quality fit.

\subsubsection{Young population (Ages $< 2$ Gyr) vs.\ Blue Stragglers}

In Paper I, we discussed the possibility that the fainter stars in the
BP of M32 could be old BSS rather than a young stellar population with
ages $<2$ Gyr. However, the analysis presented in Paper I did not
allow us to confirm or rule out either case. BSS are stars hotter,
bluer and brighter than the MSTOs in a CMD, thus generating a blue
plume. Given their locations on the CMD, they are burning hydrogen in
their cores with masses larger than the turn-off mass, which indicates
that some sort of mechanism rejuvenated their inner layers. Although
such a mechanism is still a matter of debate, there are currently two
theoretical possible scenarios to explain the BSS origin: they are the
result of either a collision between stars
\citep[e.g.,][]{Sigurdsson_etal94} or mass-transfer in a binary system
\citep[e.g.,][]{Mccrea64, Carney_etal01}. 

We investigate the nature of these stars from the SFH presented
here. Stars younger than 2 Gyr constitute $\sim$ 4\% of the total mass
of M32 at F1. Figure~\ref{fig:sfhm32} shows that some of the young
stars, produced by a very low SFR event at look-back times $<2$ Gyr,
are rather metal poor ($[\mathrm{M/H}]\sim -0.7$) in comparison with
the mean metallicity of M32 ($[\mathrm{M/H}]\sim 0.0$). Given the
almost constant age--metallicity relation for M32 and the presence of
intermediate-age stars (2--5 Gyr old) of solar or even higher
metallicity, it is unlikely that M32 contains at the same time younger
stars with significantly sub-solar metallicities. The most plausible
explanation is that these stars are BSS belonging to an old metal-poor
population.  BSS are found in open and globular clusters
\citep{Ferraro_etal04, Mapelli_etal04, Mapelli_etal06, Piotto_etal04,
  Demarchi_etal06}, dwarf spheroidal galaxies
\citep{Hurleykeller_etal99, Carrera_etal02, Momany_etal07,
  Mapelli_etal09, Monelli_etal10a}, and even in the Milky Way halo
field population \citep{Preston_sneden00}. Therefore, it seems natural
to consider that they can also be found in an elliptical galaxy. These
stars represent $\sim 2\%$ of the mass of M32 in F1 and might be the
first direct evidence of BSS in this galaxy. An alternative
explanation could be that these young and metal-poor stars were
generated by an episode of late infall of metal-poor gas.  However, if
we assume that M32 is interacting with M31, we would not expect M32 to
accrete gas, but instead to lose gas to M31 through stripping.

The other $\sim 2$\% of stars with ages $<2$ Gyr that we find in the
SFH inferred for M32 may indeed represent a young metal-rich
population in M32 at F1.

\section{The Star formation history of M32}
\label{sec:sfhm32}

By combining the results in the present work with the analysis in
Paper I, we can finally provide a detailed and complete SFH of M32. We
conclude that M32 has had an extended SFH and is composed of two main
dominant populations at F1: $\sim 40\% \pm 17\%$ of the mass in a 2--5
Gyr old, metal-rich population and $\sim55 \% \pm 21\%$ of the mass in
stars older than 5 Gyr, with slightly subsolar metallicities. We
confirm the existence of the younger ($< 5$ Gyr) stars through the
presence of bright AGB stars observed in Paper I, with the appropriate
ages. From the RC, RGB bump and AGB bump analyzed in Paper I, the bulk
of the old population is 8--10 Gyr old. We therefore do not expect a
significant contribution from stars older than 10 Gyr in M32 at
F1. Nevertheless, there is a hint of the presence of a few ancient
metal-poor stars present in M32, as revealed by the 2--$\sigma$
detection of RR Lyrae belonging to M32 at F1. The remaining $\sim4$\%
of the mass is roughly equally divided between a young metal-rich
population and a young metal-poor population. We associate the latter
with blue straggler stars belonging to an old (likely metal-poor)
population.

The age--metallicity relation for M32 is nearly constant within our
age resolution, although there is a small increase in metallicity at
at $\sim$ 5 Gyr followed by a small decrease at ages younger than
$\sim$ 2 Gyr. The mass-weighted mean metallicity of M32 is
$[\mathrm{M/H}]\sim -0.01 \, \mathrm{dex}$ with a peak at
$[\mathrm{M/H}]\sim 0.02 \, \mathrm{dex}$. We emphasize here again
that an almost constant age--metallicity relation appears to suggest
that M32 has not experienced metal enrichment; but as in F1, this is
due to the poor age resolution and does not imply the lack of an
age--metallicity relation. Stars with metallicities lower than
$[\mathrm{M/H}]\la -1\, \mathrm{dex}$ only contribute $\sim 5$\% of
the total mass of M32 at $\sim2'$ from its center. This is consistent
with the photometric metallicity function (MDF) of M32 derived in
Paper I, which shows that the majority of the stars has a slightly
sub-solar metallicity at $[\mathrm{M/H}]\sim -0.2 \,
\mathrm{dex}$. The MDF also indicated that metal-poor stars with
$[\mathrm{M/H}] <-1.2$ contribute very little, \emph{at most} 6\% of
the total $V$-light to M32 or 4.5\% of the total mass in F1, implying
that the enrichment process largely avoided the metal poor stage.

\subsection{On the integrated light of M32}

The results obtained in this work are a fundamental step for
understanding the formation and evolution of other low-luminosity
spheroidal systems (elliptical galaxies or bulges). Since integrated
light spectra are, in general, the only means available to study the
stellar populations of these galaxies, we strongly rely on unresolved
stellar population models to learn about their SFHs. These models,
which have become very sophisticated in disentangling the non-trivial
age and metallicity degeneracy \citep{Worthey94, BC03, Thomas_etal03,
  Vazdekis_etal10}, still suffer from several uncertainties: e.g., it
is difficult to distinguish between a young or hot old population
since the latter is not necessarily accounted for in the models
\citep[e.g.][]{Maraston_thomas00}. Calibration of these models, which
requires observations of individual stars in elliptical galaxies, is a
key ingredient that needs to be further developed. As briefly
mentioned in Section 1, M32 is located at a distance such that both
integrated spectroscopy and photometry of its individual stars can be
studied in great detail. A comparison of the stellar parameters
obtained using resolved stars and integrated luminosity is fundamental
to provide a calibration to the unresolved stellar models with an
actual elliptical.

Extensive spectroscopic studies of M32 have been performed, mostly in
the central regions and out to $\sim 1\,r_e$
\citep[e.g.,][]{Oconnell80,Gonzalez93, Worthey04, Rose_etal05}. All
studies agree that the central stellar population has an
SSP-equivalent age of 2.5--5 Gyr and roughly solar metallicity, with
an age gradient that increases the age at $1\,r_e$ by $\sim3$ Gyr and
a mild negative metallicity gradient. Various integrated-light studies
have suggested that M32 underwent a period of significant star
formation in the recent past, i.e., about 5--8 Gyr ago,
\citep[e.g.,][]{Oconnell80, Pickles85, Bica_etal90} based on the
presence of enhanced H$\beta$ absorption in the integrated spectrum of
M32, a signature of an intermediate-age population
\citep[e.g.,][]{Rose94, Trager_etal00a, Worthey04, Schiavon_etal04,
  Rose_etal05, Coelho_etal09}. To date, only Coelho et al. (2009) have
attempted to probe the unresolved stellar populations as far from the
center of M32 as the ACS/HRC field presented in this paper lies,
using longslit observations with GMOS on Gemini. They propose that an
ancient and intermediate-age population are both present in M32 and
that the contribution from the intermediate-age population is larger
at the nuclear region. They claim that a young population is present
at all radii, and they further suggest that there is a strong
component of either very young ($<0.3$ Gyr) and/or very old ($>10$
Gyr), metal-poor stars even in their outermost field.

We can use the inferred SFH of M32 to compute line strength indices
using the models of \citet[][hereafter BC03]{BC03} and to calculate
single stellar population (SSP)-equivalent parameters that can then be
compared with the values obtained from the integrated light of this
galaxy.

Using the BC03 models, we obtain a $B$-band luminosity-weighted mean
age and metallicity of 4.9 Gyr and $[\mathrm{M/H}]=-0.12 \,
\mathrm{dex}$, respectively, for M32 at F1 from its resolved
SFH. \citet{Coelho_etal09} find an average luminosity-weighted age of
$5.7\pm1.5$ Gyr using BC03, which agrees with our result within the
uncertainties, but their inferred mean metallicity is much lower,
$[\mathrm{M/H}]=-0.6\pm 0.1$ (see their Table 3). Moreover, as
mentioned above, they suggest that there is a strong component of
either very young ($<0.3$ Gyr) or very old ($>10$ Gyr), metal-poor
stars in their field at a radius similar to our F1 field, which is
inconsistent with our data.

\begin{figure}
\centering
\includegraphics[width=84mm, clip]{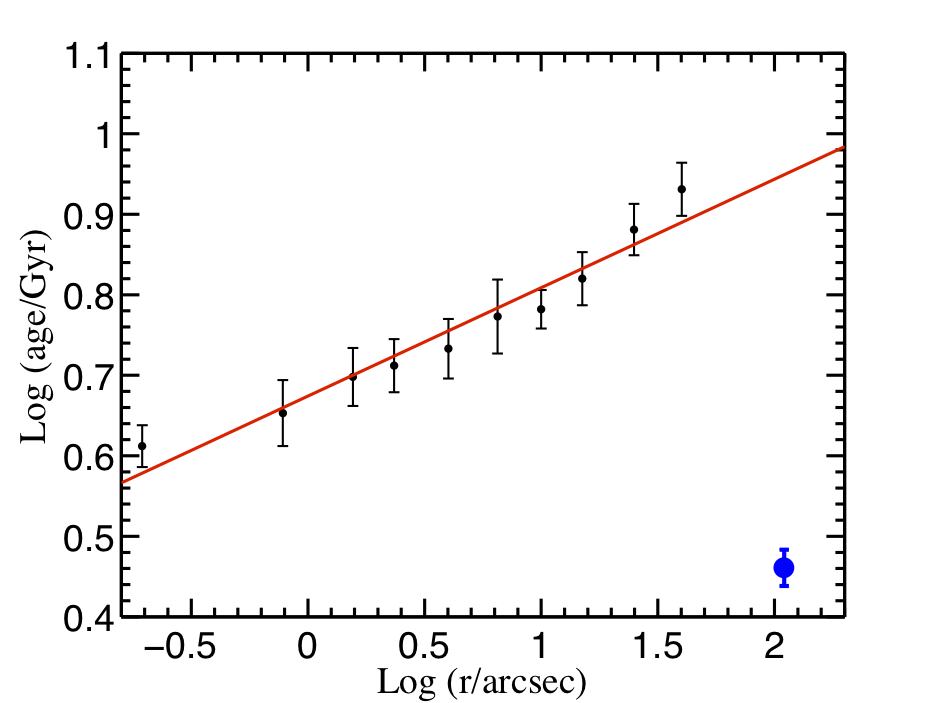}
\includegraphics[width=84mm, clip]{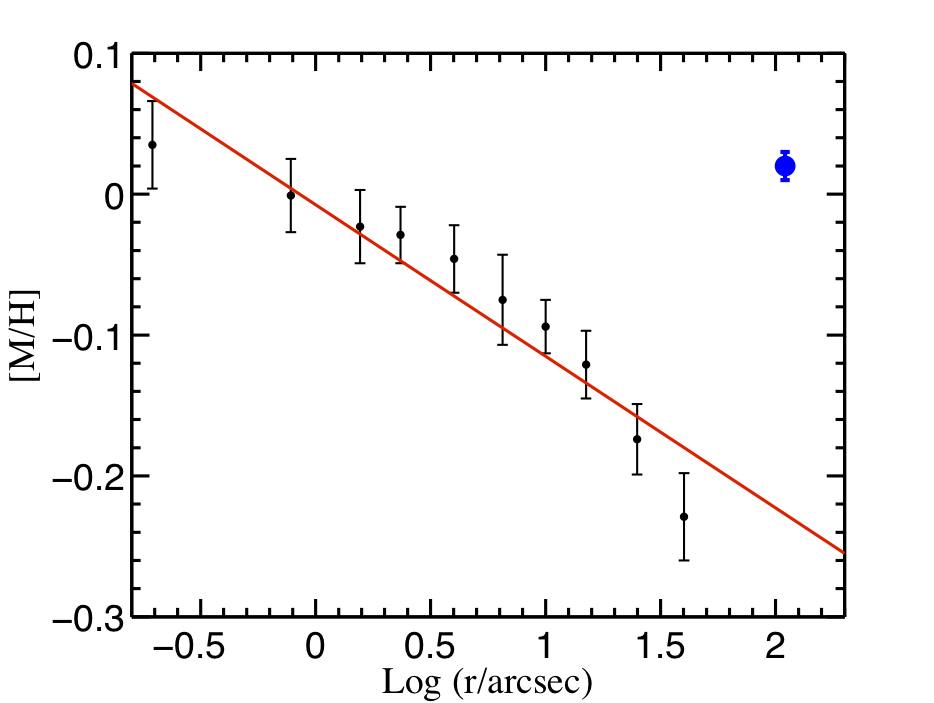}
\caption{SSP-equivalent age (top panel) and metallicity (bottom panel)
  values inferred from line-strength fits given by \citet{Worthey04}
  as a function of radius in M32. The lines are linear fits to the
  data. We can see the steep positive age gradient from M32's center
  to $1\,r_e\sim 40''$. An extrapolation of these fits to $\log(110'')
  \sim 2.04$ gives the SSP-equivalent age and metallicity values at
  F1: 8.9 Gyr and $[\mathrm{M/H}]=-0.23\pm 0.03$ dex,
  respectively. The SSP-equivalent parameters calculated from the
  inferred SFH of M32 (blue dots) are in stark contrast with
  predictions of spectral studies.}
\label{fig:extrapol}
\end{figure}

We have also calculated SSP-equivalent parameters that can be compared
with the values obtained from integrated spectra of this galaxy. Using
the BC03 models, the SSP-equivalent values of M32 obtained from its
inferred SFH at F1 are $2.9 \pm 0.2$ Gyr and $[\mathrm{M/H}]= 0.02 \pm
0.01\,\mathrm{dex}$, respectively. Given the radial line-strength
gradients present in M32 \citep[e.g.,][]{Rose_etal05}, we cannot
directly compare these SSP-equivalent values with those obtained from
central integrated spectra of M32 \citep[by,
  e.g.,][]{Trager_etal00a}. We therefore do the following. We use the
values of the line strength indices from \citet{Worthey04} and compute
the SSP-equivalent parameters from polynomial fits to the
absorption-line strengths as a function of radius (his Table 1) using
the modified BC03 models described in \citet{Trager_etal08}. We then
fit straight-line gradients to the SSP-equivalent parameters as a
function of radius and extrapolate these fits to $110^{\prime\prime}$,
F1's position. The SSP-equivalent age and metallicity of M32 at F1
from this extrapolation are $8.9\pm0.5$ Gyr and
$[\mathrm{M/H}]=-0.23\pm 0.03$ dex, respectively. We note that
Worthey's values for Mg$b$ are low compared with \citet{Gonzalez93}
and \citet{Trager_etal98}, and thus the SSP-equivalent age we have
obtained may be slightly overestimated whereas the SSP-equivalent
metallicity may be underestimated. Taken this into account, we
obtained an SSP-equivalent age of $\sim 8.4$ Gyr and a SSP-equivalent
metallicity of $[\mathrm{M/H}]\sim-0.13$ dex.
Figure~\ref{fig:extrapol} shows the SSP-equivalent parameters from
\citet{Worthey04}. The linear fits to the $\log(\mathrm{age/Gyr})$ and
$[\mathrm{M/H}]$ as a function of $\log(r/'')$ are also shown. The
blue stars indicate the SSP-parameters obtained from the inferred SFH
of M32 from BC03 models.

The predicted age and metallicity at F1 from the extrapolation of the
absorption-line gradients are much older and more metal-poor,
respectively, than those obtained from the inferred SFH of M32. This
suggests that either the extrapolation of line strength indices or the
stellar population models, or both, may be in error, but we are
currently unable to discern which. Color profiles in many colors of
M32 are rather flat \citep[][]{Peletier93}. Since M32 does not contain
dust, integrated colours can be good population indicators and the
fact that there are no gradients in colors agree with the results from
the inferred SFH. \citet{Davidge_jensen07} have also challenged the
radial gradients in mean stellar parameters obtained from spectral
studies. They find no evidence for a radial age gradient in M32, based
on the properties of observed brightest AGB stars, in contrast to the
results by \citet{Worthey04} and \citet{Rose_etal05}, who found (as
described above) a significant radial gradient in the mean
luminosity-weighted age of M32.

To further investigate this apparent contradiction, integral-field
spectroscopic observations with VIRUS-P \citep{Hill_etal08} have been
taken at F1 and F2 and will be analyzed in a future paper. This will
provide spectra of the integrated stellar light of M32 for the
fundamental calibration for the study of stellar populations.

\subsection{On the formation of M32}

Certainly, the most striking result of this work is the substantial
contribution of 2--5 Gyr old metal-rich stars to the total mass of M32
at F1. How has an elliptical galaxy like M32 formed such a young
population of stars? What is the origin of this population? In this
section we attempt to address these questions and discuss, in
particular, the most popular proposed formation scenarios for M32.

A formation scenario for M32 has been proposed by \citet[][hereafter
K09]{Kormendy_etal09}, in which M32 is a normal, low-luminosity
elliptical galaxy. K09 find that both central and global parameter
correlations from recent accurate photometry of galaxies in the Virgo
cluster place M32 as a normal, low-luminosity elliptical galaxy in all
regards.  K09 fit a \citet{Sersic68} profile to the SB of M32 with $n
= 2.8$, in agreement with S\'ersic indices of other low-luminosity
ellipticals studied by K09. They interpret the light at the center of
M32 that was not fit by their S\'ersic profile as a signature of
formation in dissipative mergers \citep[][]{Mihos_hernquist94}. Extra
central light is a general feature of coreless galaxies and is
observed in all the other low-luminosity ellipticals of K09's sample.
Within this scenario, the metal-rich 2--5 Gyr old stars contributing
to $\approx 40\% \pm 17\%$ of M32's mass at F1 could be the result of
such a dissipative merger event. Thus, the progenitors of M32 should
have been very gas-rich spiral galaxies, like M33 for
example. However, such progenitors should have stellar masses of the
order of $10^8\, M_{\odot}$, whereas M33's stellar mass is $\sim
3\times 10^9\, M_{\odot}$. There are in fact no gas-rich spiral
galaxies near M31 of the appropriate stellar mass.

An alternative scenario for the formation of M32 has been proposed by
\citet[][hereafter B01] {Bekki_etal01}, who assumed that M32 is the
result of a low-luminosity spiral galaxy, whose bulge, unlike most of
its outer disk, survived its dynamical interactions with
M31\footnote{The idea that M32, as well as other small high-surface
  brightness galaxies, is a tidally truncated galaxy has been
  discussed several decades before B01 models. In, for example,
  \citet{Faber73}, the original truncated galaxy was a more massive
  elliptical galaxy, from which only the tightly bound core of the
  original elliptical remains after a strong tidal interaction.}.  In
their N-body/smoothed particle hydrodynamics simulations, B01
considered a gas-rich low-mass disk galaxy with a bulge orbiting a
massive disk galaxy like M31. About 0.75 Gyr after the interactions
have started, the outer stellar disk (from 2 kpc to 5 kpc) of the
spiral galaxy is stripped away and only keeps $\approx40$\% of its
initial mass in stars initially located in the central regions, i.e.,
within 2 kpc of the center. New star formation is triggered by the
interaction of the gas-rich spiral with M31 but the outer new stellar
component is also tidally stripped away, and consequently only the
central starburst component survives. On the other hand, the bulge is
only weakly affected by tidal interactions with M31 due to its
compactness, and only $\approx 19$\% of its mass is lost. At the end
of their simulations, there is a fractional disk, bulge, and new stars
mass ratio of $\approx 49$\%, $\approx 42$\%, and $\approx 0.9$\%,
respectively, within 2 Kpc of the remnant compact galaxy. Our field F1
is located at 110$\arcsec$, i.e. $\sim 0.5$ kpc from the galactic
center at M32's distance and, assuming either an inside-out or
outside-in formation scenario for the disk \citep[see e.g.,][and
discussion in Section~\ref{sec:m31}]{Sommer-larsen_etal03} and
considering that we are looking at a $\sim0.5\,R_d$, where $R_d=0.9$
kpc is the scale length radius of B01's disk, the disk stars that we
should be observing there would have mean ages of 8--12 Gyr. Assuming
that the bulge is predominantly old, this scenario is difficult to
reconcile with our results, given the substantial 2--5 Gyr old
intermediate-age population detected in this work. However, there has
been some indications of small bulges which could have extended SFHs,
similar to that of M32, with 10--30\% of their total mass at look-back
times between 0.5 and 5 Gyr \citep{Thomas-davies08}.

Therefore it is unclear from our SFH results what the preferred model
for the origins of M32 is. While specific origin models differ in
detail, the general outlines overlap enough to make choosing a
specific model difficult with the age resolution of our current SFH.
More observations of M32-analog systems and simulations of spheroidal
systems with similar SFHs to that we have presented are needed to shed
light on M32's origins. Furthermore, finding evidence of a stellar
stream in the halo of M31 with the ages and metallicities obtained for
M32, which should be left if a major stripping of M32 by M31 has
occurred, would help to constrain the models and assess the validity
of the proposed scenarios.

\section{The disk and spheroid population of M31 in F2}
\label{sec:m31}

In Paper I, we compared our findings in F2, in particular its
metallicity distribution function (MDF), with several previous works
on the disk and bulge of M31 \citep[e.g.][]{Williams_02,
  Worthey_etal05, Olsen_etal06, Brown_etal06}.  We found a reasonably
good agreement with most studies. In this section we discuss our new,
quantitative results on the stellar populations at F2 and their
implications for the formation of the M31's disk.  M31 seems to have
formed most of its stars between 5--14 Gyr ago at F2. As mentioned
above, we cannot precisely indicate when the star formation started in
either F1 nor F2 but we can see that M31 is older than M32 in F1.

\begin{figure*}\centering
\includegraphics[width=130mm, clip]{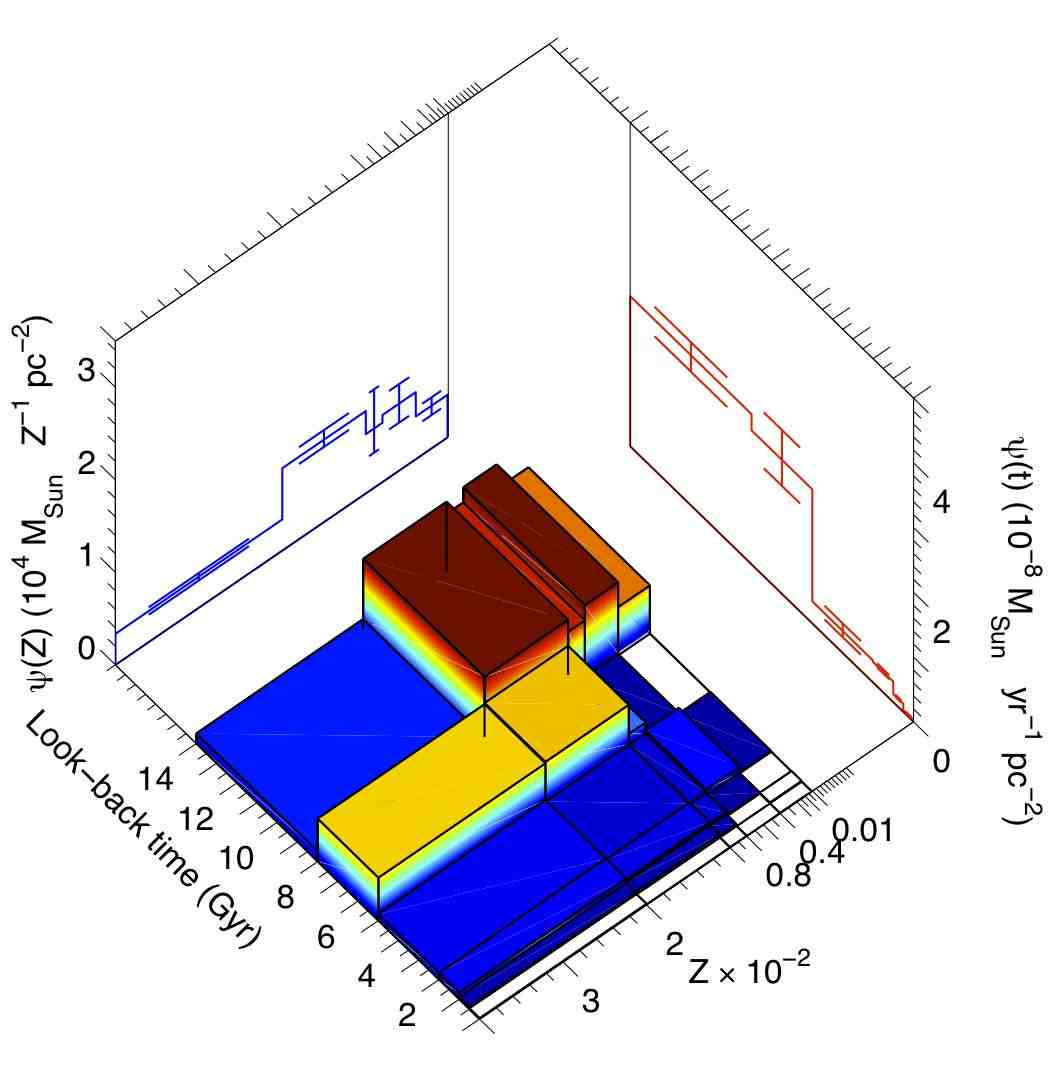}
\caption{A more detailed SFH of F2 in a 3d-histogram representation
  than that shown in Figure~\ref{fig:sfhf1_f2}. This SFH of F2 was
  constructed this time with an extra bin in age covering 5--8 Gyr. We
  now find two dominant populations of M31 at F2: An old more
  metal-poor population, older than 8 Gyr, and an intermediate-age
  more metal-rich population, 5--8 Gyr old. Stars younger than 5 Gyr
  old only contribute $\sim5$\% of the mass of M31 at F2.}
\label{fig:sfhf2extrage}
\end{figure*}

\begin{figure*}\centering
\subfigure
{\includegraphics[width=85mm, clip]{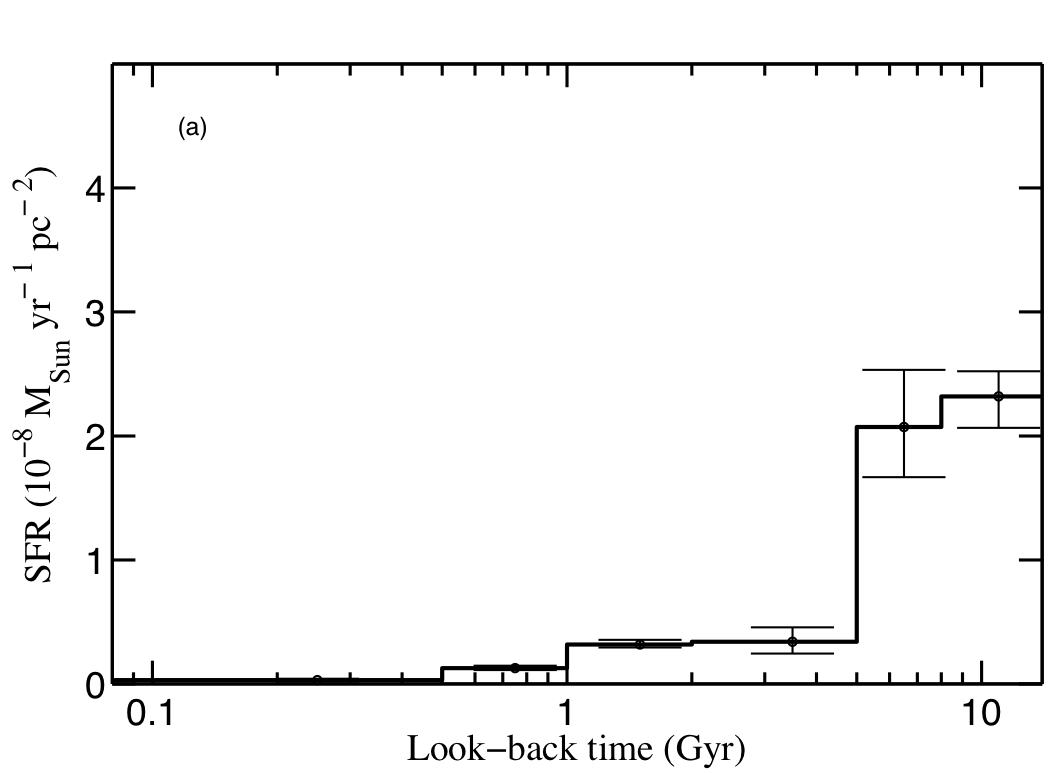}}
\subfigure
{\includegraphics[width=85mm, clip]{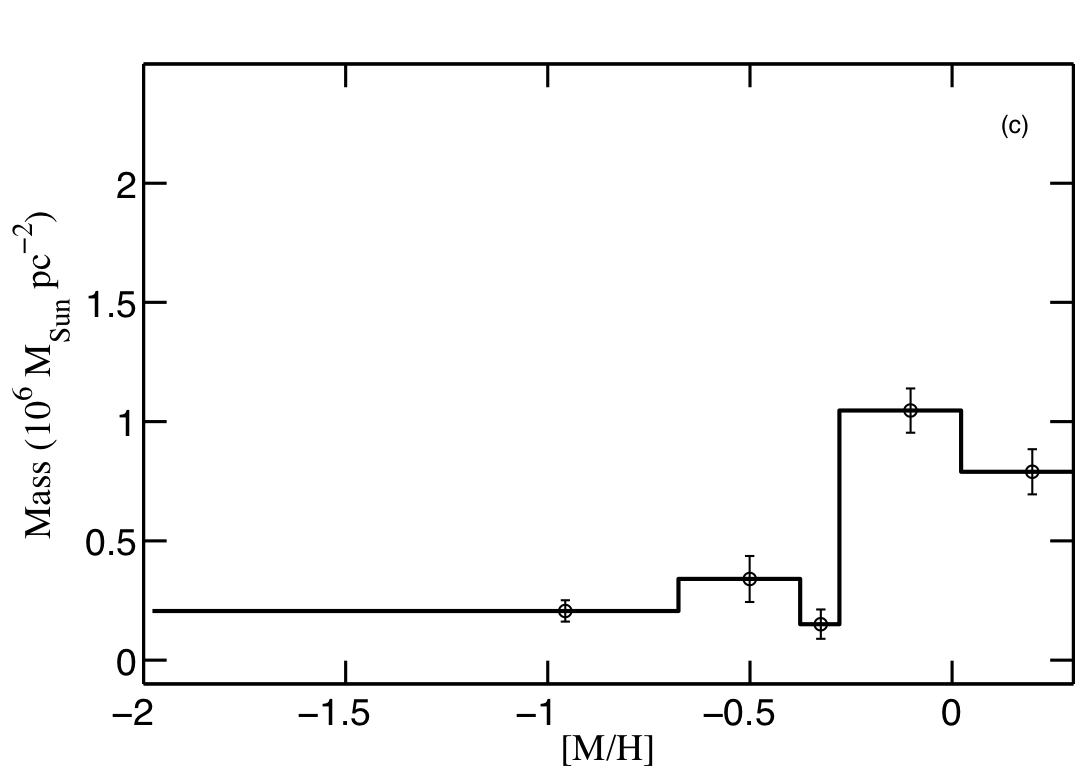}}
\subfigure
{\includegraphics[width=85mm, clip]{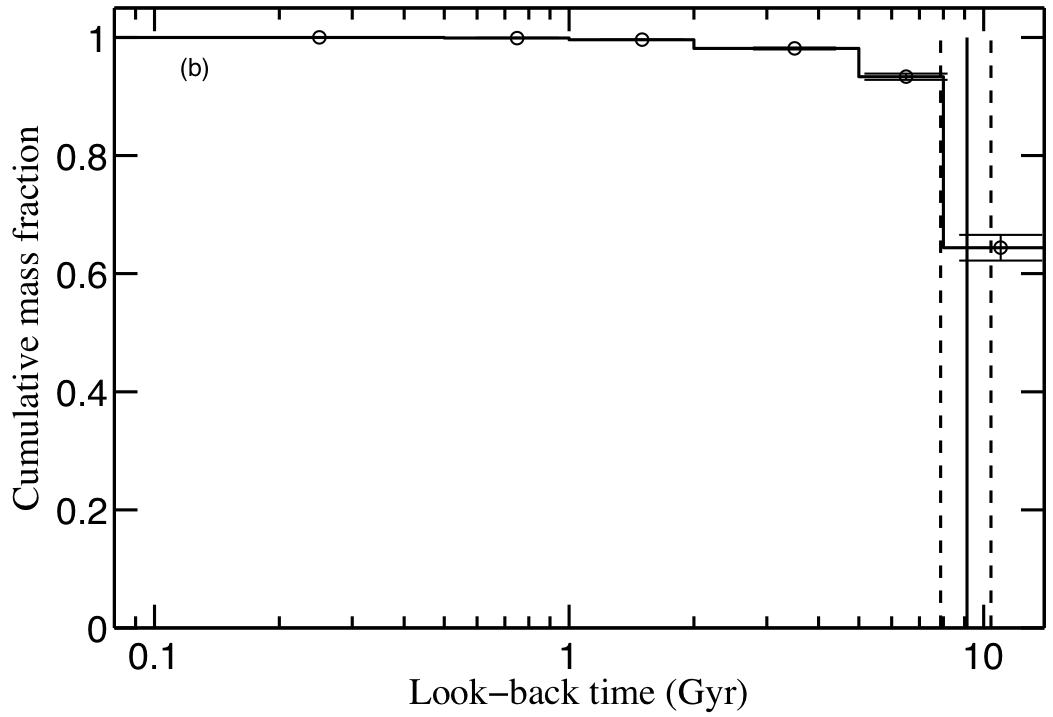}}
\subfigure
{\includegraphics[width=85mm, clip]{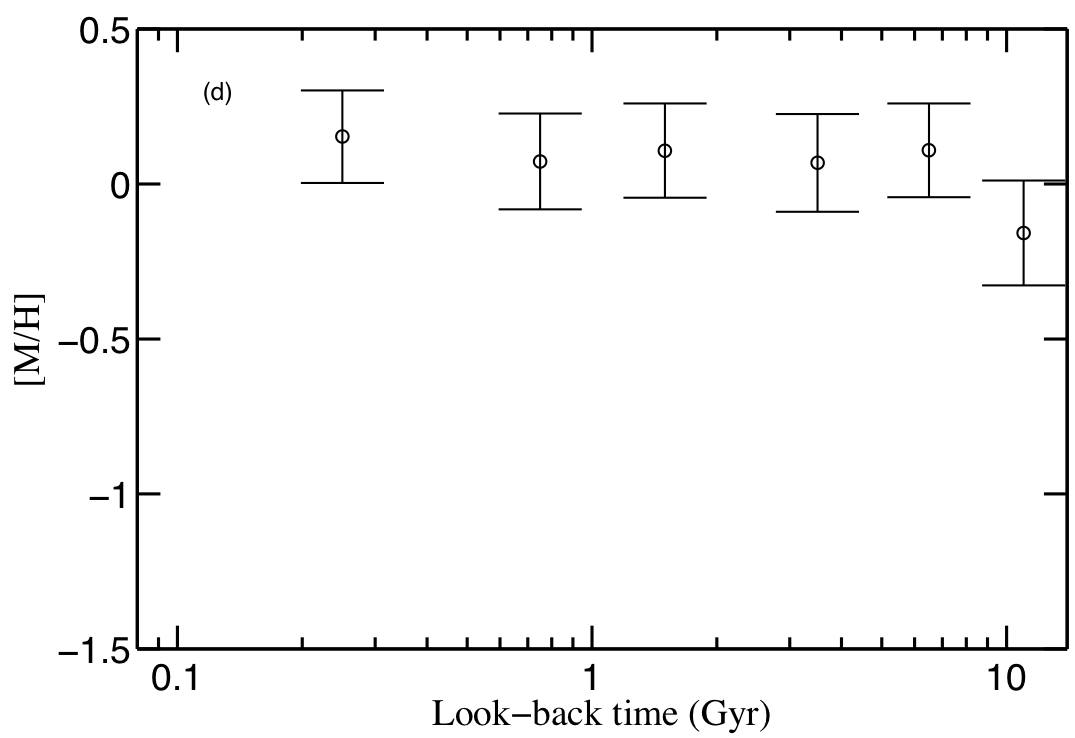}}
\caption{The SFH of F2 inferred using an extra bin, i.e. more
  resolution, in age. (a) SFR as a function of time; (b) cumulative
  mass-weighted age distribution; (c) mass as a function of
  metallicity; (d) age--metallicity relation. The vertical solid and
  dashed lines in panel (b) represent the mean age ($\sim 9.15$ Gyr)
  and $1\sigma$ deviation of that value, respectively.}
\label{fig:sfhf12-age}
\end{figure*}

\citet[][hereafter B06]{Brown_etal06} analyzed three CMDs of different
regions of M31: the spheroid, stream and outer disk. These CMDs
reached well below the oldest MSTOs, and B06 derived SFHs at each
field in great detail. Differences between these SFHs were mainly
found in the age and metallicity distributions of stars older than 5
Gyr. Within this age range (5--14 Gyr) we do not have the resolution
required to inspect different bursts of star formation in F2 in
detail, given the SFH extracted in Section~\ref{sec:results}. We can,
nevertheless study the SFH of F2 in more detail than what is presented
in Section~\ref{sec:results}. As we show in
Figure~\ref{fig:deconvolvedmcmd}, the CMD of F2 is $\sim 0.5$ mag
deeper than the one of F1, which allows us to obtain information of
fainter, i.e. older, MSTOs at F2\footnote{The previous selection of
  age and metallicity bins to derive the SFHs was strictly based on
  the resolution imposed by the CMD of F1. In order to subtract the
  SFH of F2 from that of F1, we required the simple populations
  considered be exactly the same.}. We therefore extracted again the
SFH of F2 following the steps indicated in Section~\ref{sec:method},
but with an extra bin in the age, from 5 to 8 Gyr, for the simple
populations considered.  The boundaries of the bins in age are in this
case [0, 0.5, 1, 2, 5, 8, 14]$\times10^9$ years. The inferred best-fit
mean SFH of F2 with this new resolution in age was found at $(\delta
(F435W-F555W)_0, \delta M_{F555W}) = (-0.03, 0.00)$ with $\chi^2_{\nu,
  min} = 2.12$. Figure~\ref{fig:sfhf2extrage} shows a 3D-histogram
representation of the new SFH solution for F2.  We can now distinguish
two main populations that contribute substantially to our background
field F2, instead of only one old population: $\approx 30\% \pm 7.5\%$
of the total mass in F2 is composed of a 5--8 Gyr old metal-rich
population and $\approx 65\% \pm 9\%$ of the mass is composed of a
8--14 Gyr old, metal-poor population. An age--metallicity relation
shows a slightly steeper slope from an old metal-poorer population to
younger metal-richer ones than before, as shown in
Figure~\ref{fig:sfhf12-age}. We are still not able to answer when the
star formation started in F2.  Nevertheless, our results for the mean
age and metallicity for F2, $9.12\pm 1.21$ Gyr and $-0.10\pm 0.10$ dex
respectively, are in good agreement with B06 results for their outer
disk field, $8.5$ Gyr and $-0.4$ dex, respectively\footnote{The cited
  values correspond to the results obtained by B06 when a 40\% binary
  fraction was assumed.}. In addition, young stars, with ages between
0.3 and 1 Gyr, that populate the BP in the CMD of F2 do not contribute
significantly to the total mass, which is also in agreement with B06's
results.  Interestingly, kinematic data in our field imply that both
the disk and spheroid of M31 contribute to the populations in F2
(K. Howley, 2010, priv. commm). This was also the case for the outer
disk field of B06.  B06, however, attempted to disentangle both
populations assuming that their spheroid field was representative of
the spheroid population present in their outer disk field. By
subtracting the spheroid population, they obtained a younger mean age
for the outer disk of M31---but still older than 5 Gyr.

Given the resolution allowed by the depth of our data, the inner disk
and spheroid populations of M31 (at 5 kpc from its center) seem to be
indistinguishable from the outer disk and spheroid ones (at 25 kpc
from M31 galactic center, B06). Even though we are unable to subtract
the spheroid population that contributes to our field F2, most likely
the mean age of M31's disk at F2 is younger than 8.72 Gyr and older
than 5 Gyr, given the negligible contribution of stars younger than 5
Gyr. This result supports the inside-out disk formation models by
e.g., \citet[][]{Abadi_etal03a, Abadi_etal03b,
  Sommer-larsen_etal03}. Abadi et al. find a mean age of 8--10 at 2
kpc, which radially decreases to 6--8 at 20 kpc. Sommer-Larsen et
al. simulated two spiral galaxies, with two different scenarios of
disk formation: inside-out and outside-in. Our expected mean age for
the disk of M31 at F2 agree with both scenarios within their
uncertainties, assuming a stellar disk scale length of $\approx5$ kpc
\citep[e.g.,][]{Walterbos_kennicutt88, Worthey_etal05}. They find
that, at 1 disk scale length, the mean ages of both simulated disks
are $\sim$6--8 Gyr. However, the significant fraction of stars younger
than 5 Gyr predicted by their outside-in model at F2 is not supported
by our data. Thus, we favor their inside-out model. Furthermore, the
inside-out formation model of \citet{Sommer-larsen_etal03} predicts
that the disk has almost no age gradient which, although surprising,
is also in agreement with the comparison of our and B06 results at
different disk locations. They explain that this prediction is a
consequence of the non linear dependence of the SFR on the cold gas
density, which makes the SFR rather low in the outer disk at late
times, thus the average outer disk stellar age is quite high. An
alternative scenario for the absence of an age gradient, found when
comparing our results with those of B06, is the radial migration of
stars seen in recent simulations of isolated disk formation and
evolution \citep{Roskar_etal08a, Minchev_etal11}. In these
simulations, inside-out disk growth yields a negative age gradient
within the break radius (2--3 disk scale length), after which there is
a positive age gradient due to the secular redistribution of stars,
given their interactions with transient spiral density waves. Of
course, we should keep in mind that what we presented here are the
results of one field in the inner regions of M31 and we need more
observations and statistics to either confirm or rule out what we
suggest. The multi-cycle Panchromatic Hubble Andromeda Treasury (PHAT)
project, which will cover 1/3 of M31 with HST WFC3 and ACS
observations, will resolve the SFH of the disk of M31: our
observations and analysis merely hint at what PHAT is likely to find.

\section{Summary and Conclusions}
\label{sec:conclusions}

We used deep HST ACS/HRC observations to derive the SFH of M32 for the
first time from a detailed modeling of its CMD. The two fields
observed, one closer to M32 (F1) and a background M31 field (F2), were
introduced and used in Paper I to construct deep CMDs of F1 and F2,
and the deepest optical CMD of M32 yet obtained. The IAC-POP/MinnIAC
method was used here to compare the distribution of stars in the
observed CMDs of F1 and F2 with that of a model CMD. We obtained the
SFH of M32 by linearly subtracting the SFHs of F2 from that of F1. The
use of different stellar evolutionary libraries (BaSTI and
Padova/Girardi) and assumptions of binary fractions (0, 0.35, 0.7, and
1) did not significantly modify the solutions obtained, indicating
that our results are robust.

Combining our present results with those of Paper I, we provide an
unprecedented census of the stellar content of M32. The derivation of
the SFH presented in this paper is independent of the analysis
performed in Paper I. In spite of using the same data, the CMD regions
that we have probed in this work are largely different from those used
in Paper I. Our analysis of these regions have allowed us to obtain
detailed information about the young and intermediate-age populations
of M32, whereas only the broadest sketch of these populations were
possible in Paper I. Conversely, detailed information about the older
populations cannot be obtained with our current approach, and
therefore we rely on the qualitative results of Paper I for those
populations.

The main finding of this work is that M32 is composed of two main
dominant populations at F1: $\sim 40\% \pm 17\%$ of the mass in a 2--5
Gyr old, metal-rich population and $\sim 55\% \pm 21\%$ of the total
mass in stars older than 5 Gyr, with slightly subsolar
metallicities. Its mass-weighted mean age and metallicity are
$\langle\mathrm{Age}\rangle=6.8\pm1.5\,\mathrm{Gyr}$ and
$\langle\mathrm{[M/H]}\rangle=-0.1\pm0.08\,\mathrm{dex}$,
respectively. Even though we are unable to specify when the star
formation started in M32 at F1, we make use of the analysis of Paper I
to constrain the older population. We know from the RC, RGb and AGB
bumps that the bulk of the old population is 8--10 Gyr old. Thus, we
do not expect a significant contribution from stars older than 10 Gyr
in M32. There has been, however, a marginal detection of RR Lyrae
belonging to M32 at F1, which reveal the presence of a few ancient
metal-poor stars in M32 \citep{Fiorentino_etal10}. The remaining
$\sim$ 4\% of the mass is distributed in genuine young metal-rich
stars ($\sim 2$\%) and young metal-poor stars ($\sim 2$\%) which we
associate with blue straggler stars belonging to an old metal-poor
population. In addition, we used the inferred SFH of M32 to calculate
the SSP-equivalent age and metallicity parameters from unresolved
stellar population models, which are $2.89 \pm 0.15$ Gyr and
$[\mathrm{M/H}]=0.02 \pm 0.01$ dex, respectively. These values,
however, contradict spectroscopic studies, which show a steep age
gradient from M32's center to $1\,r_e$.

Based on our present results, it is not currently possible to choose a
preferred model for M32's origins between two popular ones: a true
low-luminosity elliptical or a former spiral galaxy whose bulge
survived its dynamical interaction with M31. Future observations to
find M32-analog systems as well as simulations of spheroidal systems
with similar SFHs to M32 may shed light on this issue.

On the other hand, the inferred SFH for F2 shows that the stellar
populations of the inner regions of the disk and spheroidal components
of M31 are older and more metal-poor than M32. Its mass-weighted mean
age and metallicity are
$\langle\mathrm{Age}\rangle=9.15\pm1.2\,\mathrm{Gyr}$ and
$\langle\mathrm{[M/H]}\rangle=-0.10\pm0.10\,\mathrm{dex}$,
respectively. F2 has two main components: $65\% \pm 9\%$ of the mass
composed by a 8--14 Gyr old more metal-poor population and $30\% \pm
7.5\%$ of the mass in more metal-rich stars of 5--8 Gyr old. There is
a small contribution from stars younger to 5 Gyr to the total mass.
The inner disk and spheroidal stellar populations seem to be
indistinguishable from those of the outer disk and spheroid. Assuming
that M31's disk at F2 ($\sim$ 1 disk scale length) has a mean age
between $\sim$ 5 and 9 Gyr, our results are in agreement with
inside-out disk formation models. But of course, we need more
observations and statistics to confirm or rule out this suggestion.

Lastly, while this paper accounts for the SFH history of the bulk of
M32's mass, it does not offer strong constraints on small ``tracer''
populations that may testify to the present level of any very recent
star formation in M32, as well as fossil remnants that may date to
times well before the 10 Gyr mark, after which the bulk of M32 stars
were formed.  Digging down to fainter stars in the M32 CMD to detect
any very old ( $>10$ Gyr) MSTO is not possible with HST, or any
instrument presently under development, given that the M31 background
prevents observation at lower surface brightness levels where the HST
angular resolution would be more effective.  Instead, we believe the
best hope of detecting any ancient, metal-poor, remnant from the very
first stages in the M32 progenitor would come from unambiguously
detecting RR Lyrae stars at higher surface brightness levels than were
observed in F1, where the M31 contamination is relatively much weaker.
Our own image simulations show that RR Lyraes can be detected with HST
at substantially smaller radii than the F1 location.  Likewise, better
constraints on a $< 2$ Gyr population native to M32 may be provided by
bright, if rare, tracers that can still be recognized at much higher
M32 surface brightnesses than were observed in the present work.  As
with RR Lyraes, bright young blue MS stars or AGB stars should be
detectable throughout the body of M32. \\

\acknowledgments We thank Reynier Peletier, Eline Tolstoy and Antonio
Aparicio for their valuable comments on an early version of this
paper. A.M. wishes to thank the hospitality of the Instituto de
Astrof\'isica de Canarias and the Department of Physics and Astronomy
of Michigan State University, where part of this work was carried
out. We thank the anonymous referee for the very careful reading of
the manuscript and comments which helped to improve this paper. This
work has made use of the IAC-STAR synthetic CMD computation
code. IAC-STAR is supported and maintained by the computer division of
the Instituto de Astrof\'isica de Canarias. NOVA is acknowledged for
financial support. Support for program GO-10572 was provided by NASA
through a grant from the Space Telescope Science Institute, which is
operated by the Association of Universities for Research in Astronomy,
Inc., under NASA contract NAS 5-26555.

\emph{Facility:} \facility{HST (ACS)}

\bibliographystyle{apj}
\bibliography{references_sfhm32}

\begin{thebibliography}{80}
\expandafter\ifx\csname natexlab\endcsname\relax\def\natexlab#1{#1}\fi

\bibitem[{{Abadi} {et~al.}(2003{\natexlab{a}}){Abadi}, {Navarro}, {Steinmetz},
  \& {Eke}}]{Abadi_etal03a}
{Abadi}, M.~G., {Navarro}, J.~F., {Steinmetz}, M., \& {Eke}, V.~R.
  2003{\natexlab{a}}, \apj, 591, 499

\bibitem[{{Abadi} {et~al.}(2003{\natexlab{b}}){Abadi}, {Navarro}, {Steinmetz},
  \& {Eke}}]{Abadi_etal03b}
---. 2003{\natexlab{b}}, \apj, 597, 21

\bibitem[{{Aparicio} \& {Gallart}(2004)}]{Aparicio_gallart04}
{Aparicio}, A., \& {Gallart}, C. 2004, \aj, 128, 1465

\bibitem[{{Aparicio} {et~al.}(1997){Aparicio}, {Gallart}, \&
  {Bertelli}}]{Aparicio_etal97}
{Aparicio}, A., {Gallart}, C., \& {Bertelli}, G. 1997, \aj, 114, 669

\bibitem[{{Aparicio} \& {Hidalgo}(2009)}]{Aparicio_hidalgo09}
{Aparicio}, A., \& {Hidalgo}, S.~L. 2009, \aj, 138, 558

\bibitem[{{Bekki} {et~al.}(2001){Bekki}, {Couch}, {Drinkwater}, \&
  {Gregg}}]{Bekki_etal01}
{Bekki}, K., {Couch}, W.~J., {Drinkwater}, M.~J., \& {Gregg}, M.~D. 2001,
  \apjl, 557, L39

\bibitem[{{Bertelli} {et~al.}(1992){Bertelli}, {Mateo}, {Chiosi}, \&
  {Bressan}}]{Bertelli_etal92}
{Bertelli}, G., {Mateo}, M., {Chiosi}, C., \& {Bressan}, A. 1992, \apj, 388,
  400

\bibitem[{{Bica} {et~al.}(1990){Bica}, {Alloin}, \& {Schmidt}}]{Bica_etal90}
{Bica}, E., {Alloin}, D., \& {Schmidt}, A.~A. 1990, \aap, 228, 23

\bibitem[{{Brown} {et~al.}(2006){Brown}, {Smith}, {Ferguson}, {Rich},
  {Guhathakurta}, {Renzini}, {Sweigart}, \& {Kimble}}]{Brown_etal06}
{Brown}, T.~M., {Smith}, E., {Ferguson}, H.~C., {Rich}, R.~M., {Guhathakurta},
  P., {Renzini}, A., {Sweigart}, A.~V., \& {Kimble}, R.~A. 2006, \apj, 652, 323

\bibitem[{{Bruzual} \& {Charlot}(2003)}]{BC03}
{Bruzual}, G., \& {Charlot}, S. 2003, \mnras, 344, 1000

\bibitem[{{Burstein} \& {Heiles}(1982)}]{Burstein_heiles82}
{Burstein}, D., \& {Heiles}, C. 1982, \aj, 87, 1165

\bibitem[{{Carney} {et~al.}(2001){Carney}, {Latham}, {Laird}, {Grant}, \&
  {Morse}}]{Carney_etal01}
{Carney}, B.~W., {Latham}, D.~W., {Laird}, J.~B., {Grant}, C.~E., \& {Morse},
  J.~A. 2001, in Bulletin of the American Astronomical Society, Vol.~33,
  American Astronomical Society Meeting Abstracts, 1311--+

\bibitem[{{Carrera} {et~al.}(2002){Carrera}, {Aparicio},
  {Mart{\'{\i}}nez-Delgado}, \& {Alonso-Garc{\'{\i}}a}}]{Carrera_etal02}
{Carrera}, R., {Aparicio}, A., {Mart{\'{\i}}nez-Delgado}, D., \&
  {Alonso-Garc{\'{\i}}a}, J. 2002, \aj, 123, 3199

\bibitem[{{Charbonneau}(1995)}]{Charbonneau95}
{Charbonneau}, P. 1995, \apjs, 101, 309

\bibitem[{{Chilingarian} {et~al.}(2009){Chilingarian}, {Cayatte}, {Revaz},
  {Dodonov}, {Durand}, {Durret}, {Micol}, \& {Slezak}}]{Chilingarian_etal09}
{Chilingarian}, I., {Cayatte}, V., {Revaz}, Y., {Dodonov}, S., {Durand}, D.,
  {Durret}, F., {Micol}, A., \& {Slezak}, E. 2009, Science, 326, 1379

\bibitem[{{Coelho} {et~al.}(2009){Coelho}, {Mendes de Oliveira}, \& {Cid
  Fernandes}}]{Coelho_etal09}
{Coelho}, P., {Mendes de Oliveira}, C., \& {Cid Fernandes}, R. 2009, \mnras,
  396, 624

\bibitem[{{Davidge} \& {Jensen}(2007)}]{Davidge_jensen07}
{Davidge}, T.~J., \& {Jensen}, J.~B. 2007, \aj, 133, 576

\bibitem[{{de Marchi} {et~al.}(2006){de Marchi}, {de Angeli}, {Piotto},
  {Carraro}, \& {Davies}}]{Demarchi_etal06}
{de Marchi}, F., {de Angeli}, F., {Piotto}, G., {Carraro}, G., \& {Davies},
  M.~B. 2006, \aap, 459, 489

\bibitem[{{Dolphin}(2002)}]{Dolphin02}
{Dolphin}, A.~E. 2002, \mnras, 332, 91

\bibitem[{{Dolphin} {et~al.}(2005){Dolphin}, {Weisz}, {Skillman}, \&
  {Holtzman}}]{Dolphin_etal05}
{Dolphin}, A.~E., {Weisz}, D.~R., {Skillman}, E.~D., \& {Holtzman}, J.~A. 2005,
  ArXiv Astrophysics e-prints

\bibitem[{{Faber}(1973)}]{Faber73}
{Faber}, S.~M. 1973, \apj, 179, 423

\bibitem[{{Ferraro} {et~al.}(2004){Ferraro}, {Beccari}, {Rood}, {Bellazzini},
  {Sills}, \& {Sabbi}}]{Ferraro_etal04}
{Ferraro}, F.~R., {Beccari}, G., {Rood}, R.~T., {Bellazzini}, M., {Sills}, A.,
  \& {Sabbi}, E. 2004, \apj, 603, 127

\bibitem[{{Fiorentino} {et~al.}(2010){Fiorentino}, {Monachesi}, {Trager},
  {Lauer}, {Saha}, {Mighell}, {Freedman}, {Dressler}, {Grillmair}, \&
  {Tolstoy}}]{Fiorentino_etal10}
{Fiorentino}, G., {et~al.} 2010, \apj, 708, 817

\bibitem[{{Gallart} {et~al.}(2005){Gallart}, {Zoccali}, \&
  {Aparicio}}]{Gallart_etal05}
{Gallart}, C., {Zoccali}, M., \& {Aparicio}, A. 2005, \araa, 43, 387

\bibitem[{{Girardi} {et~al.}(2000){Girardi}, {Bressan}, {Bertelli}, \&
  {Chiosi}}]{Girardi_etal00}
{Girardi}, L., {Bressan}, A., {Bertelli}, G., \& {Chiosi}, C. 2000, \aaps, 141,
  371

\bibitem[{{Gonz{\'a}lez}(1993)}]{Gonzalez93}
{Gonz{\'a}lez}, J.~J. 1993, PhD thesis, Thesis (PH.D.)--UNIVERSITY OF
  CALIFORNIA, SANTA CRUZ, 1993.Source: Dissertation Abstracts International,
  Volume: 54-05, Section: B, page: 2551.

\bibitem[{{Harris} \& {Zaritsky}(2001)}]{Harris_zaritsky01}
{Harris}, J., \& {Zaritsky}, D. 2001, \apjs, 136, 25

\bibitem[{{Harris} \& {Zaritsky}(2009)}]{Harris_zaritsky09}
---. 2009, \aj, 138, 1243

\bibitem[{{Hidalgo} {et~al.}(2009){Hidalgo}, {Aparicio},
  {Mart{\'{\i}}nez-Delgado}, \& {Gallart}}]{Hidalgo_etal09}
{Hidalgo}, S.~L., {Aparicio}, A., {Mart{\'{\i}}nez-Delgado}, D., \& {Gallart},
  C. 2009, \apj, 705, 704

\bibitem[{{Hidalgo} {et~al.}(2011){Hidalgo}, {Aparicio}, {Skillman}, {Monelli},
  {Gallart}, {Cole}, {Dolphin}, {Weisz}, {Bernard}, {Cassisi}, {Mayer},
  {Stetson}, {Tolstoy}, \& {Ferguson}}]{Hidalgo_etal11}
{Hidalgo}, S.~L., {et~al.} 2011, \apj, 730, 14

\bibitem[{{Hill} {et~al.}(2008){Hill}, {MacQueen}, {Smith}, {Tufts}, {Roth},
  {Kelz}, {Adams}, {Drory}, {Grupp}, {Barnes}, {Blanc}, {Murphy}, {Altmann},
  {Wesley}, {Segura}, {Good}, {Booth}, {Bauer}, {Popow}, {Goertz}, {Edmonston},
  \& {Wilkinson}}]{Hill_etal08}
{Hill}, G.~J., {et~al.} 2008, in Presented at the Society of Photo-Optical
  Instrumentation Engineers (SPIE) Conference, Vol. 7014, Society of
  Photo-Optical Instrumentation Engineers (SPIE) Conference Series

\bibitem[{{Hurley-Keller} {et~al.}(1999){Hurley-Keller}, {Mateo}, \&
  {Grebel}}]{Hurleykeller_etal99}
{Hurley-Keller}, D., {Mateo}, M., \& {Grebel}, E.~K. 1999, \apjl, 523, L25

\bibitem[{{Kormendy} {et~al.}(2009){Kormendy}, {Fisher}, {Cornell}, \&
  {Bender}}]{Kormendy_etal09}
{Kormendy}, J., {Fisher}, D.~B., {Cornell}, M.~E., \& {Bender}, R. 2009, \apjs,
  182, 216

\bibitem[{{Kroupa}(2002)}]{Kroupa02}
{Kroupa}, P. 2002, Science, 295, 82

\bibitem[{{Lauer}(1999)}]{lauer99a}
{Lauer}, T.~R. 1999, \pasp, 111, 227

\bibitem[{{Lucy}(1974)}]{Lucy_74}
{Lucy}, L.~B. 1974, \aj, 79, 745

\bibitem[{{Mapelli} {et~al.}(2009){Mapelli}, {Ripamonti}, {Battaglia},
  {Tolstoy}, {Irwin}, {Moore}, \& {Sigurdsson}}]{Mapelli_etal09}
{Mapelli}, M., {Ripamonti}, E., {Battaglia}, G., {Tolstoy}, E., {Irwin}, M.~J.,
  {Moore}, B., \& {Sigurdsson}, S. 2009, \mnras, 396, 1771

\bibitem[{{Mapelli} {et~al.}(2004){Mapelli}, {Sigurdsson}, {Colpi}, {Ferraro},
  {Possenti}, {Rood}, {Sills}, \& {Beccari}}]{Mapelli_etal04}
{Mapelli}, M., {Sigurdsson}, S., {Colpi}, M., {Ferraro}, F.~R., {Possenti}, A.,
  {Rood}, R.~T., {Sills}, A., \& {Beccari}, G. 2004, \apjl, 605, L29

\bibitem[{{Mapelli} {et~al.}(2006){Mapelli}, {Sigurdsson}, {Ferraro}, {Colpi},
  {Possenti}, \& {Lanzoni}}]{Mapelli_etal06}
{Mapelli}, M., {Sigurdsson}, S., {Ferraro}, F.~R., {Colpi}, M., {Possenti}, A.,
  \& {Lanzoni}, B. 2006, \mnras, 373, 361

\bibitem[{{Maraston} \& {Thomas}(2000)}]{Maraston_thomas00}
{Maraston}, C., \& {Thomas}, D. 2000, \apj, 541, 126

\bibitem[{{Marigo} {et~al.}(2008){Marigo}, {Girardi}, {Bressan}, {Groenewegen},
  {Silva}, \& {Granato}}]{Marigo_etal08}
{Marigo}, P., {Girardi}, L., {Bressan}, A., {Groenewegen}, M.~A.~T., {Silva},
  L., \& {Granato}, G.~L. 2008, \aap, 482, 883

\bibitem[{{McCrea}(1964)}]{Mccrea64}
{McCrea}, W.~H. 1964, \mnras, 128, 147

\bibitem[{{Mighell}(1999)}]{Mighell99}
{Mighell}, K.~J. 1999, \apj, 518, 380

\bibitem[{{Mihos} \& {Hernquist}(1994)}]{Mihos_hernquist94}
{Mihos}, J.~C., \& {Hernquist}, L. 1994, \apjl, 437, L47

\bibitem[{{Minchev} {et~al.}(2011){Minchev}, {Famaey}, {Combes}, {Di Matteo},
  {Mouhcine}, \& {Wozniak}}]{Minchev_etal11}
{Minchev}, I., {Famaey}, B., {Combes}, F., {Di Matteo}, P., {Mouhcine}, M., \&
  {Wozniak}, H. 2011, \aap, 527, A147+

\bibitem[{{Momany} {et~al.}(2007){Momany}, {Held}, {Saviane}, {Zaggia},
  {Rizzi}, \& {Gullieuszik}}]{Momany_etal07}
{Momany}, Y., {Held}, E.~V., {Saviane}, I., {Zaggia}, S., {Rizzi}, L., \&
  {Gullieuszik}, M. 2007, \aap, 468, 973

\bibitem[{{Monachesi} {et~al.}(2011){Monachesi}, {Trager}, {Lauer}, {Freedman},
  {Dressler}, {Grillmair}, \& {Mighell}}]{Monachesi_etal11}
{Monachesi}, A., {Trager}, S.~C., {Lauer}, T.~R., {Freedman}, W., {Dressler},
  A., {Grillmair}, C., \& {Mighell}, K.~J. 2011, \apj, 727, 55 (Paper I)

\bibitem[{{Monelli} {et~al.}(2010{\natexlab{a}}){Monelli}, {Hidalgo},
  {Stetson}, {Aparicio}, {Gallart}, {Dolphin}, {Cole}, {Weisz}, {Skillman},
  {Bernard}, {Mayer}, {Navarro}, {Cassisi}, {Drozdovsky}, \&
  {Tolstoy}}]{Monelli_etal10a}
{Monelli}, M., {et~al.} 2010{\natexlab{a}}, \apj, 720, 1225

\bibitem[{{Monelli} {et~al.}(2010{\natexlab{b}}){Monelli}, {Gallart},
  {Hidalgo}, {Aparicio}, {Skillman}, {Cole}, {Weisz}, {Mayer}, {Bernard},
  {Cassisi}, {Dolphin}, {Drozdovsky}, \& {Stetson}}]{Monelli_etal10b}
---. 2010{\natexlab{b}}, \apj, 722, 1864

\bibitem[{{Nieto} \& {Prugniel}(1987)}]{Nieto_prugniel87}
{Nieto}, J., \& {Prugniel}, P. 1987, \aap, 186, 30

\bibitem[{{No{\"e}l} {et~al.}(2009){No{\"e}l}, {Aparicio}, {Gallart},
  {Hidalgo}, {Costa}, \& {M{\'e}ndez}}]{Noel_etal09}
{No{\"e}l}, N.~E.~D., {Aparicio}, A., {Gallart}, C., {Hidalgo}, S.~L., {Costa},
  E., \& {M{\'e}ndez}, R.~A. 2009, \apj, 705, 1260

\bibitem[{{O'Connell}(1980)}]{Oconnell80}
{O'Connell}, R.~W. 1980, \apj, 236, 430

\bibitem[{{Olsen} {et~al.}(2006){Olsen}, {Blum}, {Stephens}, {Davidge},
  {Massey}, {Strom}, \& {Rigaut}}]{Olsen_etal06}
{Olsen}, K.~A.~G., {Blum}, R.~D., {Stephens}, A.~W., {Davidge}, T.~J.,
  {Massey}, P., {Strom}, S.~E., \& {Rigaut}, F. 2006, \aj, 132, 271

\bibitem[{{Origlia} \& {Leitherer}(2000)}]{Origlia_leitherer00}
{Origlia}, L., \& {Leitherer}, C. 2000, \aj, 119, 2018

\bibitem[{{Peletier}(1993)}]{Peletier93}
{Peletier}, R.~F. 1993, \aap, 271, 51

\bibitem[{{Pickles}(1985)}]{Pickles85}
{Pickles}, A.~J. 1985, \apj, 296, 340

\bibitem[{{Pietrinferni} {et~al.}(2004){Pietrinferni}, {Cassisi}, {Salaris}, \&
  {Castelli}}]{Pietrinferni_etal04}
{Pietrinferni}, A., {Cassisi}, S., {Salaris}, M., \& {Castelli}, F. 2004, \apj,
  612, 168

\bibitem[{{Piotto} {et~al.}(2004){Piotto}, {De Angeli}, {King}, {Djorgovski},
  {Bono}, {Cassisi}, {Meylan}, {Recio-Blanco}, {Rich}, \&
  {Davies}}]{Piotto_etal04}
{Piotto}, G., {et~al.} 2004, \apjl, 604, L109

\bibitem[{{Preston} \& {Sneden}(2000)}]{Preston_sneden00}
{Preston}, G.~W., \& {Sneden}, C. 2000, \aj, 120, 1014

\bibitem[{{Richardson}(1972)}]{Richardson_72}
{Richardson}, W.~H. 1972, Journal of the Optical Society of America
  (1917-1983), 62, 55

\bibitem[{{Rose}(1994)}]{Rose94}
{Rose}, J.~A. 1994, \aj, 107, 206

\bibitem[{{Rose} {et~al.}(2005){Rose}, {Arimoto}, {Caldwell}, {Schiavon},
  {Vazdekis}, \& {Yamada}}]{Rose_etal05}
{Rose}, J.~A., {Arimoto}, N., {Caldwell}, N., {Schiavon}, R.~P., {Vazdekis},
  A., \& {Yamada}, Y. 2005, \aj, 129, 712

\bibitem[{{Ro{\v s}kar} {et~al.}(2008){Ro{\v s}kar}, {Debattista}, {Stinson},
  {Quinn}, {Kaufmann}, \& {Wadsley}}]{Roskar_etal08a}
{Ro{\v s}kar}, R., {Debattista}, V.~P., {Stinson}, G.~S., {Quinn}, T.~R.,
  {Kaufmann}, T., \& {Wadsley}, J. 2008, \apjl, 675, L65

\bibitem[{{Schiavon} {et~al.}(2004){Schiavon}, {Caldwell}, \&
  {Rose}}]{Schiavon_etal04}
{Schiavon}, R.~P., {Caldwell}, N., \& {Rose}, J.~A. 2004, \aj, 127, 1513

\bibitem[{{S\'ersic}(1968)}]{Sersic68}
{S\'ersic}, J.~L. 1968, {Atlas de galaxias australes}, ed. {Sersic, J.~L.}

\bibitem[{{Sigurdsson} {et~al.}(1994){Sigurdsson}, {Davies}, \&
  {Bolte}}]{Sigurdsson_etal94}
{Sigurdsson}, S., {Davies}, M.~B., \& {Bolte}, M. 1994, \apjl, 431, L115

\bibitem[{{Sommer-Larsen} {et~al.}(2003){Sommer-Larsen}, {G{\"o}tz}, \&
  {Portinari}}]{Sommer-larsen_etal03}
{Sommer-Larsen}, J., {G{\"o}tz}, M., \& {Portinari}, L. 2003, \apj, 596, 47

\bibitem[{{Thomas} \& {Davies}(2008)}]{Thomas-davies08}
{Thomas}, D., \& {Davies}, R.~L. 2008, in IAU Symposium, Vol. 245, IAU
  Symposium, ed. {M.~Bureau, E.~Athanassoula, \& B.~Barbuy}, 289--292

\bibitem[{{Thomas} {et~al.}(2003){Thomas}, {Maraston}, \&
  {Bender}}]{Thomas_etal03}
{Thomas}, D., {Maraston}, C., \& {Bender}, R. 2003, MNRAS, 339, 897

\bibitem[{{Tolstoy} \& {Saha}(1996)}]{Tolstoy_saha96}
{Tolstoy}, E., \& {Saha}, A. 1996, \apj, 462, 672

\bibitem[{{Tosi} {et~al.}(1991){Tosi}, {Greggio}, {Marconi}, \&
  {Focardi}}]{Tosi_etal91}
{Tosi}, M., {Greggio}, L., {Marconi}, G., \& {Focardi}, P. 1991, \aj, 102, 951

\bibitem[{{Trager} {et~al.}(2008){Trager}, {Faber}, \&
  {Dressler}}]{Trager_etal08}
{Trager}, S.~C., {Faber}, S.~M., \& {Dressler}, A. 2008, \mnras, 386, 715

\bibitem[{{Trager} {et~al.}(2000){Trager}, {Faber}, {Worthey}, \&
  {Gonz{\'a}lez}}]{Trager_etal00a}
{Trager}, S.~C., {Faber}, S.~M., {Worthey}, G., \& {Gonz{\'a}lez}, J.~J. 2000,
  AJ, 119, 1645

\bibitem[{{Trager} {et~al.}(1998){Trager}, {Worthey}, {Faber}, {Burstein}, \&
  {Gonzalez}}]{Trager_etal98}
{Trager}, S.~C., {Worthey}, G., {Faber}, S.~M., {Burstein}, D., \& {Gonzalez},
  J.~J. 1998, \apjs, 116, 1

\bibitem[{{Vazdekis} {et~al.}(2010){Vazdekis}, {S{\'a}nchez-Bl{\'a}zquez},
  {Falc{\'o}n-Barroso}, {Cenarro}, {Beasley}, {Cardiel}, {Gorgas}, {Peletier},
  {Beasley}, \& {Cardiel}}]{Vazdekis_etal10}
{Vazdekis}, A., {et~al.} 2010, \mnras, 404, 1639

\bibitem[{{Walterbos} \& {Kennicutt}(1988)}]{Walterbos_kennicutt88}
{Walterbos}, R.~A.~M., \& {Kennicutt}, Jr., R.~C. 1988, \aap, 198, 61

\bibitem[{{Williams}(2002)}]{Williams_02}
{Williams}, B.~F. 2002, \mnras, 331, 293

\bibitem[{{Worthey}(1994)}]{Worthey94}
{Worthey}, G. 1994, ApJS, 95, 107

\bibitem[{{Worthey}(2004)}]{Worthey04}
---. 2004, \aj, 128, 2826

\bibitem[{{Worthey} {et~al.}(2005){Worthey}, {Espa{\~n}a}, {MacArthur}, \&
  {Courteau}}]{Worthey_etal05}
{Worthey}, G., {Espa{\~n}a}, A., {MacArthur}, L.~A., \& {Courteau}, S. 2005,
  \apj, 631, 820

\end{thebibliography}

\appendix

\section{Effect of binaries}
\label{sec:bin}

The results presented in this work were obtained assuming a 35\%
binary fraction in the synthetic CMD.  To investigate how much this
assumption might affect our solution, we have repeated the entire
process of deriving the best mean SFH of F1 and F2 assuming not only
35\% but also 0\%, 70\% and 100\% binary fractions in the synthetic
CMD. The mass ratios between the components of the binaries were set
to be uniformly distributed between 0.5 and 1.

\begin{deluxetable}{@{}lcccc}
  \tabletypesize{\scriptsize} \tablecaption{$\chi^2_{\nu, min}$ values for the different assumptions considered
    \label{table:chivalues}} \tablewidth{0pt}
  \tablehead{\colhead{Field}& \colhead{Binary \%}&
    \colhead{($\delta color$, $\delta mag$)\tablenotemark{a}}&
    \colhead{$\chi^2_{\nu, min}$}}
\startdata
\cutinhead{BaSTI} 
 F1&0&$(-0.09~,~0.14)$& 2.04 \\
  &35&$(-0.09~, ~0.07)$& 2.03\\
  &70&$(-0.09~,~0.07)$& 2.02 \\
  &100&$(-0.09~,~0.14)$&1.98 \\
\hline
F2&0&$(-0.03~,~0.00)$&2.28\\
&35&$(-0.06~,~0.00)$&2.23\\
&70&$(-0.03~,~0.00)$&2.28\\
&100&$(-0.03~,~0.00)$&2.26\\
\cutinhead{Padova/Girardi}
F1&0&$(0.00~,~0.00)$& 4.07\\
&35&$(0.00~,~0.07)$& 3.07\\
&70&$(0.00~,~0.07)$&3.49 \\
&100&$(0.00~,~0.07)$&3.35 \\
\hline
  F2&0&$(-0.09~,-0.07)$&2.62\\
 &35&$(0.03~,~0.00)$&2.82\\
 &70&$(-0.12~,~0.00)$&2.58\\
 &100&$(-0.09~,-0.07)$&2.53
 \enddata
\tablenotetext{a}{Color and magnitude shifts of the observed CMD
        at which the $\chi^2_{\nu, min}$ value is reached.}
\end{deluxetable}

Table~\ref{table:chivalues} shows the values of the $\chi^2_{\nu,min}$
reached for F1 and F2 as a function of the assumed binary fraction,
and using the BaSTI and Padova/Girardi stellar libraries. We can see
that for F1 the goodness of fit does not significantly improve when
varying the binary fraction if we use the stellar library
BaSTI. However, Girardi/Padova models finds the best fit to the
observed CMD in F1 when the fraction of binaries is 35\%. We therefore
choose this fraction as our baseline model. For F2, the
$\chi^2_{\nu,min}$ as a function of binary fraction is nearly
constant, regardless the stellar library used.  Note that BaSTI
library always recovers a better fit, i.e. lower $\chi^2_{\nu,min}$
than Girardi/Padova ones for both F1 and F2 observed CMDs.  The
position in the ($\delta (\mathrm{color})$, $\delta
(\mathrm{magnitude})$) grid at which $\chi^2_{\nu,min}$ is reached for
F1 is nearly insensitive to changes in the model binary fraction.
This is not the case for F2, which reflects the fact that its CMD is
deeper than that of F1.

\begin{figure*}
\includegraphics[width=90mm, clip]{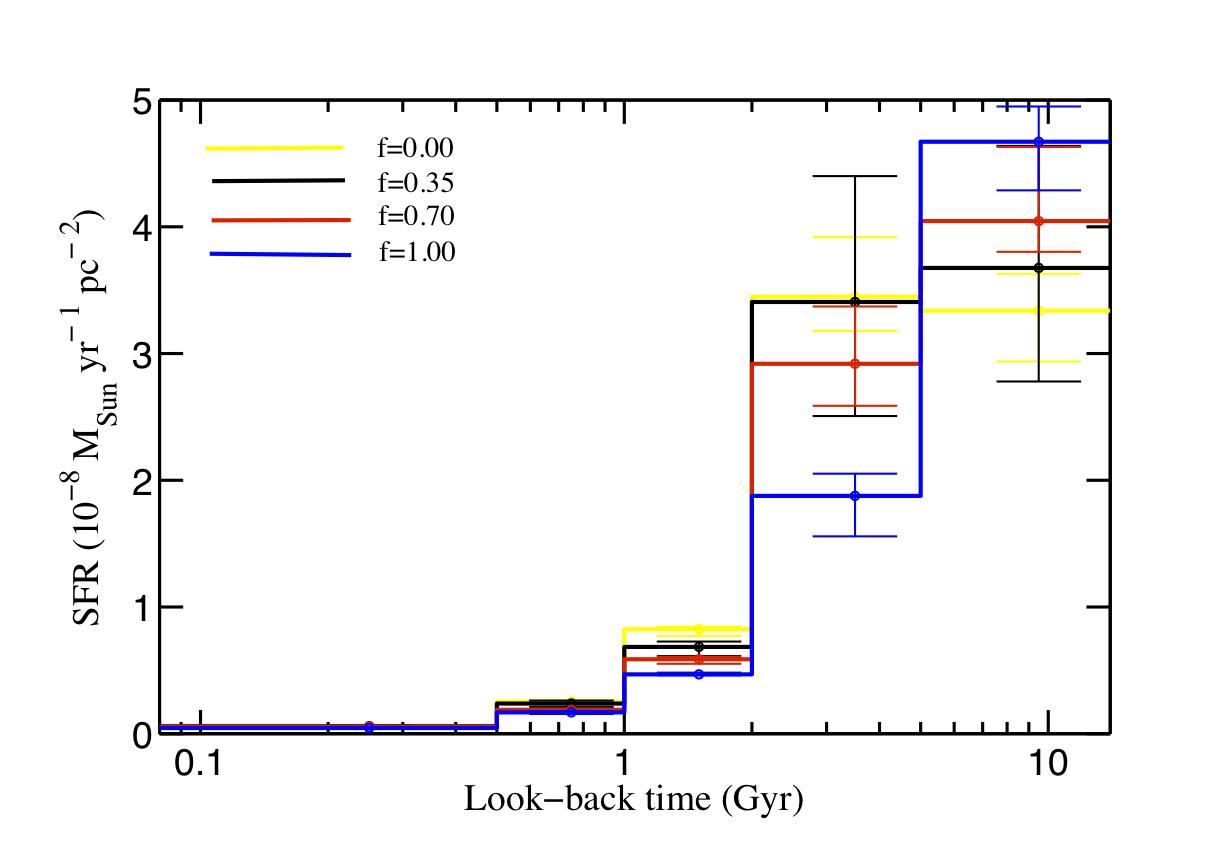}
 \includegraphics[width=90mm, clip]{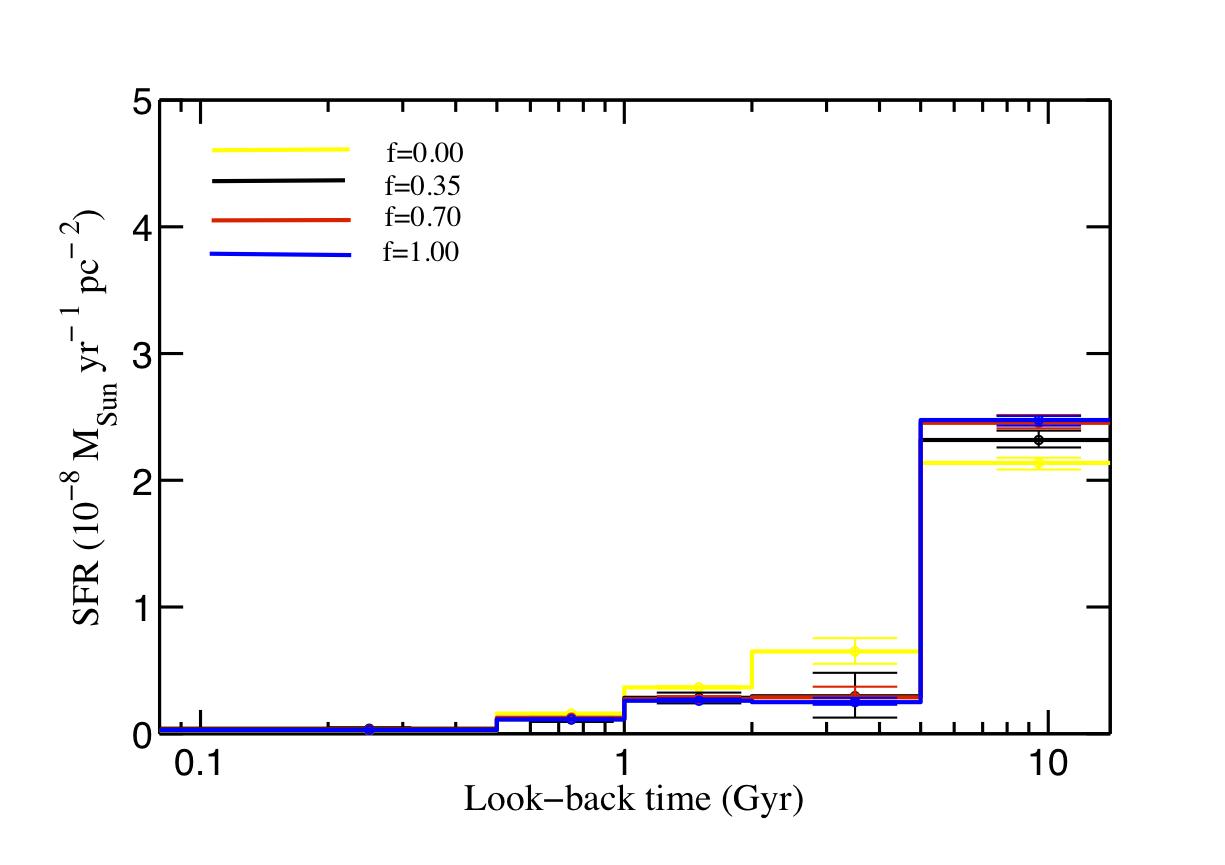}
 \caption{Comparison of the SFRs as a function of age for different
   assumed binary fractions in the synthetic CMD. The left panel shows
   the results for F1 and the right panel shows the results for
   F2. The solution becomes older as we increase the fraction of
   binaries in the model. This can be clearly seen in the first two
   bins of the SFR in F1 (left panel), which represent ages of $\sim
   10$ and $\sim 4$ Gyr, respectively. This reflects the fact that as
   we increase the number of binaries of the model CMD, its
   \emph{effective} MS becomes more luminous and its \emph{effective}
   MSTO becomes brighter and redder.}
\label{fig:binacompa}
\end{figure*}

Figure~\ref{fig:binacompa} shows the comparison of the derived SFHs.
The SFR as a function of time for F1 (left panel) and F2 (right panel)
indicates that the calculated solution does not change significantly
but becomes older as the number of binaries increases in the model
CMD.  This is expected: the larger the number of binaries in a system,
the more luminous the effective (that is, observed) MS and the
brighter and redder the effective MSTO of its CMD.

\end{document}